\documentclass[aps,prl,a4paper,twocolumn,superscriptaddress,citeautoscript,footinbib,floatfix]{revtex4-2}
\usepackage{color}
\usepackage[english]{babel}
\usepackage{lmodern}
\usepackage[T1]{fontenc}
\usepackage{amsmath}
\usepackage{units}
\usepackage{grffile}
\usepackage{color}
\usepackage[bottom]{footmisc}
\usepackage{booktabs}
\usepackage{dcolumn}
\usepackage{graphicx}
\usepackage{xcolor}
\usepackage{lipsum}
\usepackage[normalem]{ulem}
\graphicspath{{figs/}}
\begin{document}
\title{Orbital Selective Mott Transition Effects and Non-Trivial Topology
\\ of Iron Chalcogenide }
\author{Minjae Kim}
\email{garix.minjae.kim@gmail.com}
\affiliation{Korea Institute for Advanced Study, Seoul 02455, South Korea}
\affiliation{Department of Chemistry, Pohang University of Science and Technology (POSTECH), Pohang 37673, Korea}
\affiliation{Department of Physics and Astronomy, Rutgers University, Piscataway, New Jersey 08854, USA}
\author{Sangkook Choi}
\affiliation{Korea Institute for Advanced Study, Seoul 02455, South Korea}
\affiliation{Condensed Matter Physics and Materials Science Department, Brookhaven National Laboratory, Upton, New York 11973, USA}
\author{Walber Hugo Brito}
\affiliation{Departamento de F\'{\i}sica, Universidade Federal de Minas Gerais, C. P. 702, 30123-970 Belo Horizonte, MG, Brazil}
\affiliation{Department of Physics and Astronomy, Rutgers University, Piscataway, New Jersey 08854, USA}
\author{Gabriel Kotliar}
\affiliation{Department of Physics and Astronomy, Rutgers University, Piscataway, New Jersey 08854, USA}
\affiliation{Condensed Matter Physics and Materials Science Department, Brookhaven National Laboratory, Upton, New York 11973, USA}
\date{\today}
\begin{abstract}
The iron-based superconductor FeSe$_{1-x}$Te$_{x}$ (FST) has recently gained significant attention as a host of two distinct physical phenomena: (\textit{i}) Majorana zero modes which can serve as potential topologically protected qubits, and (\textit{ii}) a  realization of the orbital selective Mott transition (OSMT). In this Letter, we connect these two phenomena and provide new insights into the interplay between strong electronic correlations and non-trivial topology in FST. Using linearized quasiparticle self-consistent GW plus dynamical mean-field theory, we show that the topologically protected Dirac surface state has substantial Fe($d_{xy}$) character.
The proximity to the OSMT plays a dual role, it facilitates the appearance of the topological surface state by bringing the Dirac cone close to the chemical potential,
but destroys the Z$_{2}$ topological superconductivity when the system is too close to the orbital selective Mott phase (OSMP). We derive a reduced effective Hamiltonian
that describes the topological band. Its parameters capture all the chemical trends found in the first principles calculation.
Our  findings provide a framework for further study of the interplay between strong electronic correlations and non-trivial topology in other iron-based superconductors.
\end{abstract}
\maketitle

{\it Introduction.}
Quantum information science is a surging frontier of physical
science. By creating quantum states and utilizing them as quantum bits
(qubits) \cite{divincenzo_PhysicalImplementationQuantum_2000}, it
promises vastly improved performance over what we has been  achieved in
computing, sensing, communication, and cryptography in the 20th
century \cite{dowling_QuantumTechnologySecond_2003,ohira_ZeroCorrelationEntanglement_2019}.
Several milestones of quantum technologies, such as universal quantum computers
and the notion of quantum supremacy, have been reached successfully. Today’s quantum technologies are built on a few tens of qubits. They often suffer from computation-destroying noise \cite{preskillQuantumComputingNISQ2018}, spurring the search  for bigger and more robust quantum systems.
Majorana states are emergent quantum states at the boundary of topological superconductors. This quantum system provides a topologically protected route to realize more robust qubits~\cite{kitaev_UnpairedMajoranaFermions_2001,kitaev_FaulttolerantQuantumComputation_2003,read_PairedStatesFermions_2000} against noise than the front runners such as superconducting qubits and trapped-ion qubits.

Among various topological superconductor candidates,
FeSe$_{1-x}$Te$_{x}$ (FST) compounds hold an unique position
\cite{wang_TopologicalNatureMathrmFeSe_2015,xu_TopologicalSuperconductivitySurface_2016,TRS_surface} by
realizing topological superconductivity (TPSC), Majorana states,
and time-reversal symmetry breaking in a single material. These
compounds are correlated quantum materials with a $s$-wave
superconducting gap \cite{yehTelluriumSubstitutionEffect2008,fangSuperconductivityCloseMagnetic2008,salesBulkSuperconductivity142009}. In the normal phase, parity-even and parity-odd bands are inverted
along the $\Gamma-Z$ direction in the first Brillouin zone, and as a result, the spin-orbit coupling (SOC) opens an energy gap at the band crossing
point \cite{wang_TopologicalNatureMathrmFeSe_2015}. This enables non-trivial Z$_{2}$ bulk-band topology and ``spinless'' two-dimensional surface Dirac cone\cite{zhang2018observation,rameauInterplayParamagnetismTopology2019,zakiTimereversalSymmetryBreaking2021}.
This non-trivial bulk-band topology makes the superconductivity at the surface fascinating. When the chemical potential touches the ``spinless'' surface state, the bulk
$s$-wave superconductivity induces topologically non-trivial
superconductivity at the ``spinless'' surface states \cite{fu_SuperconductingProximityEffect_2008}. In contrast, the
surface states are topologically trivial when  the chemical potential is
far from the ``spinless'' surface bands \cite{xu_TopologicalSuperconductivitySurface_2016}.
Following  a  theoretical prediction\cite{fu_SuperconductingProximityEffect_2008,xu_TopologicalSuperconductivitySurface_2016}, signatures of Majorana states were found at the core of the vortices and at antiphase structural domain walls of FST
 \cite{zhang2018observation,wang2018evidence,wang2020evidence}.

FST has also been intensively studied due to the rich physical phenomenon related to its multi-orbital correlated nature, such as  orbital differentiation (which takes place when some orbitals display significant levels of correlation) and its extreme version,
the realization of an orbital selective Mott phase (OSMP)~\cite{yi2015observation,huang2020low,miao2018universal}.
This phase features a localized Fe($d_{xy}$) orbital, whereas the rest of the Fe($d$) orbitals remain itinerant~\cite{yi2015observation,huang2020low,miao2018universal}. Up to now, the concepts of Majorana states and OSMP have been addressed separately as independent phenomena. In this Letter, we show that both are intimately connected.

Density functional theory (DFT) \cite{hohenberg_InhomogeneousElectronGas_1964,kohn_SelfConsistentEquationsIncluding_1965}  is very successful in  predicting the topological properties of weakly correlated materials, and it has been used as a standard method of discovery and screening new topological systems. 
However, it is well known that DFT fails to describe the strong correlation phenomena, such as the OSMP, which occurs in multi-orbital correlated materials. Hence, there are important disagreements between DFT bands and experimental observations on FST.
For instance, in undoped FST, DFT puts the surface Dirac cone excitation energy  $\sim~$100 meV above the Fermi level \cite{wang_TopologicalNatureMathrmFeSe_2015,xu_TopologicalSuperconductivitySurface_2016}. This  implies  that the  surface is topologically trivial in the undoped state, and only becomes non-trivial when the system is sufficiently electron-doped, in stark contrast to the experimental findings~\cite{liElectronicPropertiesBulk2021}.
DFT plus dynamical mean-field theory (DMFT)  flattens  the quasiparticle bands and  brings the Dirac surface bands closer to the chemical potential \cite{Yin2022}.

In this Letter, we demonstrate that the strong orbital-selective correlations and the non-trivial topology of FST are intimately connected.
We use  linearized quasiparticle self-consistent GW~\cite{kutepov_ElectronicStructurePu_2012,kutepov_linearized_2017} combined with DMFT\cite{georges_DynamicalMeanfieldTheory_1996,metzner_CorrelatedLatticeFermions_1989,muller-hartmann_CorrelatedFermionsLattice_1989,brandt_ThermodynamicsCorrelationFunctions_1989,janis_NewConstructionThermodynamic_1991,georges_HubbardModelInfinite_1992,jarrell_HubbardModelInfinite_1992,rozenberg_MottHubbardTransitionInfinite_1992,georges_NumericalSolutionEnsuremath_1992} (LQSGW+DMFT)~\cite{choi_first-principles_2016,choi_comdmft:_2019}
to treat static and dynamic correlations.  Taking into account both electronic correlation and SOC, we successfully reproduces the bulk band topology and surface Dirac cone excitation energy of FST. We then derive an effective Hamiltonian to elucidate the character of the band  which disperse along $k_{z}$ and undergoes band inversion.
This turns out to be our main character, the correlated Fe($d_{xy}$) orbital which can  undergo an orbital selective Mott transition (OSMT), rather than the chalcogen $p_{z}$ orbital as it is usually assumed in the literature ~\cite{fernandes2022iron,Yin2022}.
We use this Hamiltonian to  elucidate the sensitivity of the emerging TPSC of FST  to    the chemical variations in concentration of Se and Te and to the chalcogen height and conclude  that electronic correlations are significant in determining  the region of TPSC, which should be not too far but not to close to the OSMT.

\begin{figure}[t]
\includegraphics[width=\columnwidth]{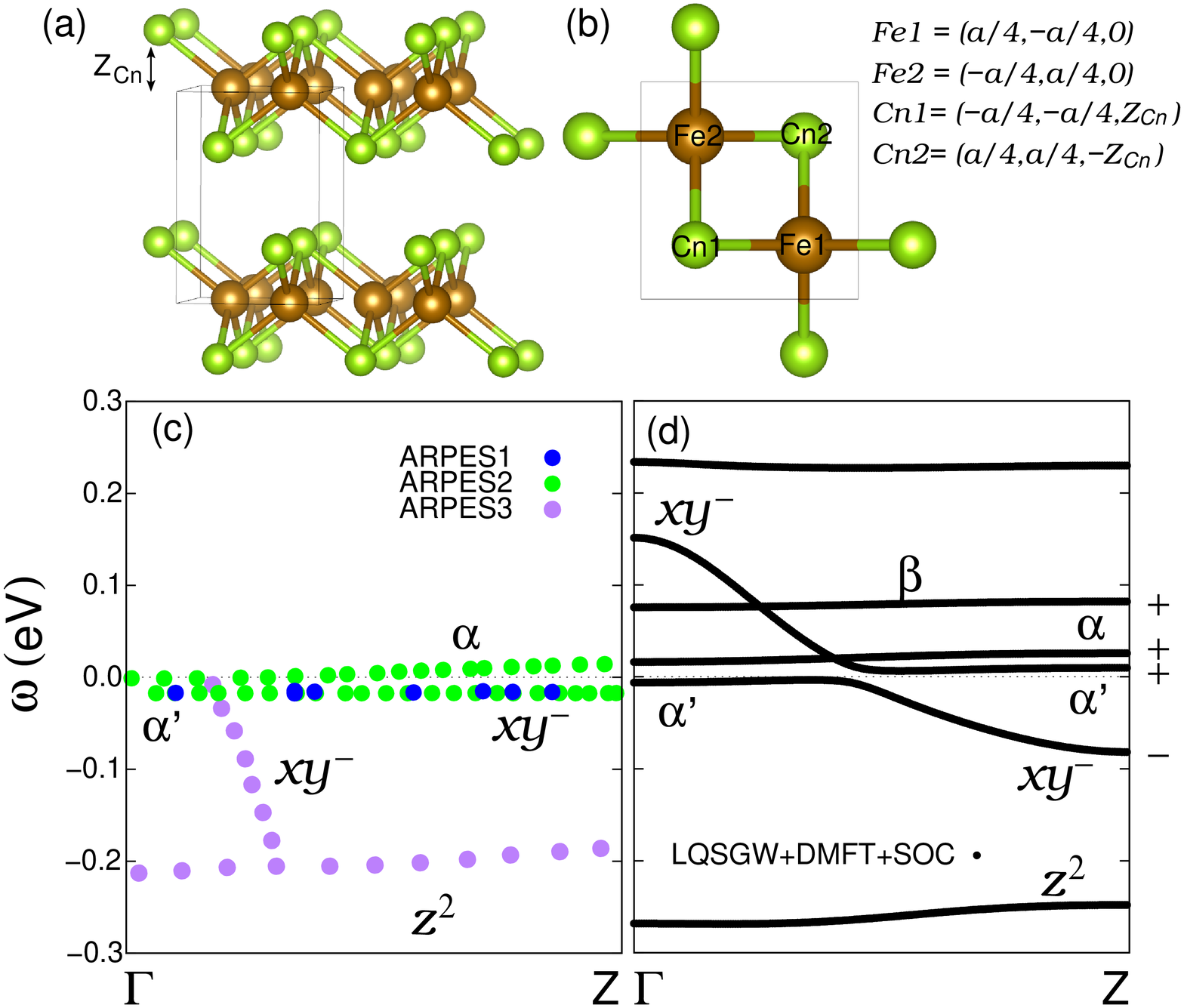}
\caption{Structural inversion symmetry,
and the band inversion
in experiments~\cite{johnson2015spin,lohani2020band,wang_TopologicalNatureMathrmFeSe_2015}
and the present theory.
(a) and (b) The present
structural model for FST and the atomic coordinate in the unit cell.
The Cn indicates a chalcogen atom, and the inversion center is [0,0,0].
(c) Experimental quasiparticle dispersions in the $\Gamma$-Z $k$ point line,
adapted from P. D. Johnson et al.(ARPES1 from Ref.\cite{johnson2015spin}),
H. Lohani et al.(ARPES2 from Ref.\cite{lohani2020band}), and
Z. Wang et al.(ARPES3 from Ref.\cite{wang_TopologicalNatureMathrmFeSe_2015}).
(d) Theoretical quasiparticle dispersions in the $\Gamma$-Z $k$ point line
in the present LQSGW+DMFT+SOC framework.
Parity eigenvalues for each band are denoted in (d),
as $\alpha'$ ($+$), $\alpha$ ($+$), $\beta$ ($+$), and $xy^{-}$ ($-$).
The band has the $z^{2}$ orbital character is also denoted in (c) and (d).
\label{fig:outofplane_main}
}
\end{figure}

\begin{figure}[t]
\includegraphics[width=\columnwidth]{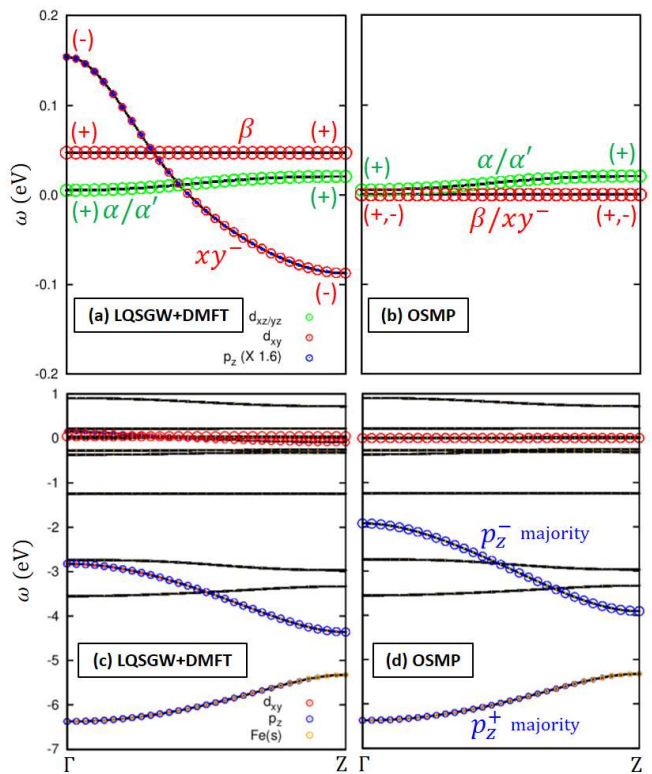}
\caption{Orbital selective Mott transition effects on the electronic structure of FST.
(a) Band structure of FST in the LQSGW+DMFT framework.
(b) Band structure of FST in the orbital selective Mott phase (OSMP) by forcing $Z_{xy}$ to zero from the LQSGW+DMFT result.
For (a) and (b), the size of green, red, and blue circles present Fe($d_{xz/yz}$), Fe($d_{xy}$), and Se($p_{z}$) orbital contributions, respectively. The size of blue circles for the Se($p_{z}$) orbital
is multiplied by the factor of 1.6.
(c) Same as (a) in a wide energy window.
(d) Same as (b) in a wide energy window.
For (c) and (d), the size of red, blue, and orange circles present Fe($d_{xy}$), Se($p_{z}$), and Fe($s$) orbital contributions, respectively.
The parity for $\alpha$, $\alpha'$, $\beta$, and $xy^{-}$ bands are denoted in (a) and (b).
The characterization of $p_{z}^{-}$ majority and $p_{z}^{+}$ majority bands are denoted for (c) and (d) (See Eq.\ref{eq:Hamiltonian_rot} for the $p_{z}^{-}$ and $p_{z}^{+}$ majority bands).
\label{fig:OSMP}
}
\end{figure}

{\it Method.} We model the FeSe$_{0.5}$Te$_{0.5}$ alloy by replacing it by a crystal structure with an averaged chalcogen height, in the spirit of the virtual
crystal approximation (see Figs.~\ref{fig:outofplane_main}(a) and (b)). We  use the lattice constants of FeSe$_{0.49}$Te$_{0.51}$ determined from the neutron powder-diffraction experiments of Ref.~\cite{li2009first}, and the chalcogen height, Z$_{Se}$=1.48 ${\AA}$,  from high-resolution x-ray diffraction data of FeSe$_{0.45}$Te$_{0.55}$~\cite{tegel2010crystal}.
We confirm that this Se chalcogen height of FST ($x\approx0.5$) is an optimized value for the description of angle resolved photoemission (ARPES) experiments along the $\Gamma-M$ line (see the Supplemental Material (SM)~\footnote{The Supplementary Material includes (i) a comparison of ARPES and the present theory, (ii) computational details, (iii) the definition of the projector $f_{Fe-d~or~Se/Te-p}$, (iv) the extraction of the spin-orbit coupling (SOC) constants, (v) the computation of surface electronic structures, (vi) the SOC enhancement and chalcogen heights, (vii) a comparison of the local density approximation (LDA) electronic structures and essential low energy parameters with the FeTe and FeSe chemical formula in the lattice constant of FeSe$_{0.5}$Te$_{0.5}$ with the consideration of the variation in the chalcogen heights, (viii) tight-binding parameters and electronic structures from the local quasi-particle self-consistent GW (LQSGW) and the LQSGW plus dynamical mean-field theory (LQSGW+DMFT), (ix) a detailed discussion on the Z$_{2}$ topology and OSMP, and (x) the effective tight-binding parameters in the Hamiltonian of Eq.\ref{eq:Hamiltonian_rot} and its comparison to the LQSGW+DMFT result in Fig.\ref{fig:OSMP}.} Sections I, II.A, and III).

The quasiparticle bands of FST was computed using the Hamiltonian,
\begin{equation}
  \begin{split}
    &H_{LQSGW+DMFT+SOC}(k) = H_{LQSGW+DMFT}(k) \\
    & +f_{Fe{\text -}d} Z_{imp}(\lambda_{1}+\Delta\lambda_{1})(\mathbf{L}\cdot \mathbf{S})f_{Fe{\text -}d}^\dagger \\
    & +f_{Se/Te{\text -}p}\lambda_{2}(\mathbf{L}\cdot \mathbf{S}) f_{Se/Te{\text -}p}^\dagger ,
\label{eq:Hamiltonian_main}
\end{split}
\end{equation}

where the SOC term was added to the H$_{LQSGW+DMFT}$($k$) from the H$^{nl}_{QP}$($k$) (double counting compensated non-local LQSGW Hamiltonian, see SM\cite{Note1} Section II.B-F). 
In Eq.\ref{eq:Hamiltonian_main}, $f_{Fe{\text -}d/Se{\text -}p/Te{\text -}p}$ is the projection operator to Fe-$d$ / Se-$p$ / Te-$p$ orbitals.
$Z_{imp}(\lambda_{1}$+$\Delta\lambda_{1})$ is the
quasiparticle SOC of Fe($d$) orbital renormalized from electronic correlations~\cite{kim2021spatial,kim2018spin,tamai2019high,linden2020imaginary}.
$\lambda_{2}$ is the average of the SOC of Se/Te($p$) states.
(See SM~\cite{Note1} Section II.G-H). For the LQSGW+DMFT scheme,  we used ComDMFT \cite{choi_comdmft:_2019}. For more details, see Ref.\footnote{ComDMFT is built on top of FlapwMBPT\cite{kutepov_linearized_2017} for the LQSGW part, and ComCTQMC for the quantum impurity problem solution \cite{melnickAcceleratedImpuritySolver2021}. We employed Wien2k~\cite{blaha2001wien2k} to calculate the DFT band structures.}.


{\it Correlated Electronic Structure and Topological Superconductivity.}
Figs. \ref{fig:outofplane_main}(c) and (d),  displays the ARPES quasiparticle dispersions along the $\Gamma$-Z direction from  several experimental groups ~\cite{wang_TopologicalNatureMathrmFeSe_2015,lohani2020band,johnson2015spin}, and from   LQSGW+DMFT+SOC quasiparticle dispersions, respectively.
Even parity bands of $\alpha'$, $\alpha$, and $\beta$ are shown,
as well as an odd parity band which is the main character in this Letter.  The
odd parity band is the most dispersive band along
the $k_{z}$ axis and  is responsible for  the topological phenomena. We anticipate that
this band will be primarily made of a correlated Fe($d_{xy}$) orbital close to an OSMT  and we anticipate this fact, which will  be demonstrated later in this Letter, by using the notation  $xy^{-}$ (See Eq.\ref{eq:Hamiltonian_init} and Eq.\ref{eq:Hamiltonian_rot}).
As seen in Fig.\ref{fig:outofplane_main}(d), a SOC-induced gap opens at the band crossing point between $\alpha'$ and $xy^{-}$ bands.
Although there are   differences in the energy position of the $xy^{-}$ band at Z
among different experiments, there is consensus  that there is a band inversion between the $\alpha'$ and $xy^{-}$ bands~\cite{johnson2015spin,lohani2020band,wang_TopologicalNatureMathrmFeSe_2015}.
Fig.\ref{fig:outofplane_main}(c) displays that, in Refs.~\cite{johnson2015spin,lohani2020band}, the flat band just beneath the chemical potential undergoes a switch of band character (parity) from $\alpha'$ ($+$) to $xy^{-}$ ($-$) in the $\Gamma$-Z direction. In contrast, in Ref.\cite{wang_TopologicalNatureMathrmFeSe_2015}, the energy position of the $xy^{-}$ band at Z is at least below -0.2 eV (See SM~\cite{Note1} Section I.B).

 Fig.~\ref{fig:outofplane_main}(d) presents  the electronic structure obtained within the LQSGW+DMFT+SOC method along $\Gamma$-Z. The chemical potential lies within the SOC-induced gap, which is in agreement with ARPES experiments~\cite{johnson2015spin,lohani2020band,wang_TopologicalNatureMathrmFeSe_2015}. The even-parity $\alpha'$ band lies below the chemical potential at $\Gamma$ and above it at Z, while the odd-parity $xy^{-}$ band lies above the chemical potential at $\Gamma$ and below it at Z. This band inversion, in the presence of time-reversal and inversion symmetries, leads to a non-trivial Z$_{2}$ invariant at the bulk, resulting in the emergence of a surface state Dirac cone at $\Gamma$ in the (001) surface. Our  calculations  are in good agreement with ARPES experiments~\cite{johnson2015spin,lohani2020band,wang_TopologicalNatureMathrmFeSe_2015}, indicating that it properly describes the topological properties of the material. Interestingly, the energy position of the $xy^{-}$ band at Z obtained by the present theory lies between the experiments of Refs.~\cite{johnson2015spin,lohani2020band} and Ref. \cite{wang_TopologicalNatureMathrmFeSe_2015}. This has been attributed to the sensitivity of the $xy^{-}$ band dispersion to the orbital-selective correlation of the Fe($d_{xy}$) orbital, which is affected by the Se/Te ratio. Additionally, the energy position of the Fe($d_{z^{2}}$) driven bands is in agreement with the experiment of Ref.~\cite{wang_TopologicalNatureMathrmFeSe_2015}. This confirms the effectiveness of the LQSGW+DMFT+SOC framework in treating electronic correlations.

\begin{figure}[t]
\includegraphics[width=\columnwidth]{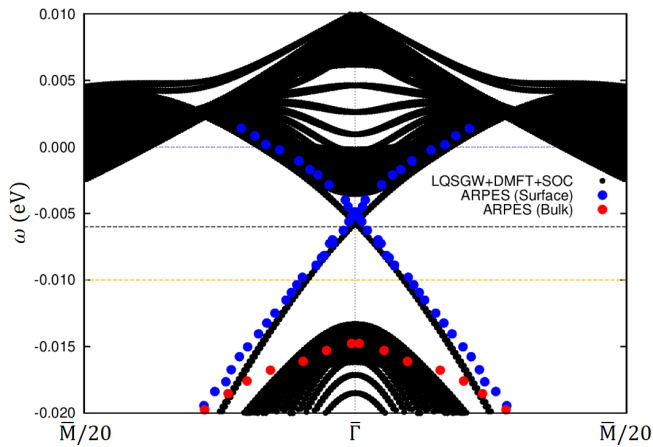}
\caption{Theoretical (001) surface state electronic
structure in the LQSGW+DMFT+SOC near $\overline{\Gamma}$,
compared with the
experimental surface electronic
structure in the ARPES of Ref.\cite{zhang2018observation}.
The black horizontal
dashed line is the original chemical potential in the present theory.
The purple horizontal dashed line is the chemical potential
for the electron doping of 0.035 (electrons/formula unit),
+6 meV in the present theory. The orange horizontal dashed line is the chemical potential
for the hole doping of 0.025 (electrons/formula unit), -4 meV in the present theory.
For the comparison
of the present theory to the ARPES of Ref.\cite{zhang2018observation},
the original chemical potential in the present theory is shifted to the purple horizontal dashed line.
\label{fig:surface}
}
\end{figure}
 Figure~\ref{fig:surface} displays  the surface electronic structure near $\overline{\Gamma}$. It was obtained by constructing a 99-layer slab from the H$_{LQSGW+DMFT+SOC}$($k$) of the bulk in Eq.\ref{eq:Hamiltonian_main} (See SM~\cite{Note1} Section II.I). A comparison of the surface state Dirac cone of this theory with the ARPES data reported in Ref.~\cite{zhang2018observation} for FST reveals an excellent agreement after a small chemical potential shift of +6 meV (0.035 electrons/formula unit). This agreement implies that the present theoretical tools can be used for the quantitative description of the TPSC of FST. This  agreement requires the following important ingredients \textit{(i)} the static self-energy driven lowering of the Fe($d_{xz/yz}$) orbital energy level, \textit{(ii)} the dynamical correlation driven renormalization of bands, and \textit{(iii)} the renormalized SOC from the consideration of the orbitally off-diagonal self-energy (See SM~\cite{Note1} Section I).

{\it Orbital Selective Mott Transition and Non-Trivial Z$_{2}$ Topology.}
We now demonstrate the assertion
that the  $xy^{-}$ odd-parity band,
which is  the most  dispersive band   along
$k_{z}$,  has a dominant  Fe($ d_{xy} $)  orbital contribution
 hybridizing  with the Se($p_{z}$) orbital. The band  inversion in this band drives the  Z$_{2}$ topology. Its sensitivity to the orbital selective correlation derives from its dominant  Fe($d_{xy}$)  orbital character  depicted in Fig.~\ref{fig:OSMP}.
This is not widely recognized in the literature and this band is often labeled as a  $p_{z}$ band in  the literature~\cite{fernandes2022iron,Yin2022} with regards to the band dispersion along $\Gamma$-Z.

 Analysis of Fig.~\ref{fig:OSMP} reveals that the OSMT in FST removes the non-trivial Z$_{2}$ topology of the bulk from the $xy^{-}$ band, as the $\beta$ (even parity) and $xy^{-}$ (odd parity) bands merge to identical flat bands at the chemical potential, with the loss of spectral weights due to the incoherent nature of the Fe($d_{xy}$) orbital in the OSMP. The $p_{z}^{-}$ majority band is also shifted up in the OSMP due to the removed hybridization induced repulsion between $|p_{z}^{-}\rangle$ and $|xy^{-}\rangle$ orbitals (Eq.\ref{eq:Hamiltonian_rot}). This explains the disappearance of the Dirac band from the enhancement of the Te ratio in the experiment\cite{liElectronicPropertiesBulk2021} (Further analysis of the Z$_{2}$ topology is provided in the SM~\cite{Note1} Section IV).

We now construct  an effective Hamiltonian in  Eq.\ref{eq:Hamiltonian_init} and Eq.\ref{eq:Hamiltonian_rot} to analyze the relation between the non-trivial Z$_{2}$ topology and the substantial correlation strength in the   Fe($d_{xy}$) orbital. The Hamiltonian was written in the basis of two Fe($d_{xy}$) orbitals  ($|xy1\rangle$ and $|xy2\rangle$) and two Se($p_{z}$) orbitals ($|p_{z}1\rangle$ and $|p_{z}2\rangle$) in the unit cell, and was transformed to the crystal momentum space with $k_{x,y}$ set to 0 for the $\Gamma$-$Z$ momentum path. The tight-binding parameters in Eq.\ref{eq:Hamiltonian_init}
and Eq.\ref{eq:Hamiltonian_rot} are effective variables that encompass contributions from longer range hoppings and other dispersive orbitals. The effective on-site energy levels for Fe($d_{xy}$) and Se($p_{z}$) orbitals are $\tilde{\epsilon}_{xy}$ and $\tilde{\epsilon}_{p_{z}}$, respectively. The effective nearest neighboring hopping between Fe($d_{xy}$) orbitals is $\tilde{t}_{xy}$, and the effective nearest neighboring hopping between Se($p_{z}$) orbitals is $\tilde{t}_{p_{z}}$. The effective out-of-plane hopping of Se($p_{z}$) is $\tilde{t}_{2}$, and the effective nearest hopping between Fe($d_{xy}$) and Se($p_{z}$) is $\tilde{t}_{1}$. All effective hopping elements are real and positive, thus accounting for the parity of all four orbitals (See SM \cite{Note1} Section V).

\begin{equation}
\begin{split}
&H_{mn}(0,0,k_{z}) =\left[\begin{smallmatrix}
\tilde{\epsilon}_{xy} & -4\tilde{t}_{xy} & 2\tilde{t}_{1} & 2\tilde{t}_{1}\\
-4\tilde{t}_{xy} & \tilde{\epsilon}_{xy} & 2\tilde{t}_{1} & 2\tilde{t}_{1}\\
2\tilde{t}_{1} & 2\tilde{t}_{1} & \tilde{\epsilon}_{p_{z}} & 4\tilde{t}_{p_{z}}+4\tilde{t}_{2}e^{ik_{z}}\\
2\tilde{t}_{1} & 2\tilde{t}_{1} & 4\tilde{t}_{p_{z}}+4\tilde{t}_{2}e^{-ik_{z}} & \tilde{\epsilon}_{p_{z}}
\end{smallmatrix}\right]
\label{eq:Hamiltonian_init}
\end{split}
\end{equation}

We transform the Hamiltonian $H_{mn}(0,0,k_{z})$ in Eq.\ref{eq:Hamiltonian_init}
to the $H_{\tilde{m}\tilde{n}}(0,0,k_{z})$ in Eq.\ref{eq:Hamiltonian_rot}
using the basis transformation as
$|xy^{-}\rangle = \frac{1}{\sqrt{2}}(|xy1\rangle - |xy2\rangle)$,
$|xy^{+}\rangle = \frac{1}{\sqrt{2}}(|xy1\rangle + |xy2\rangle)$,
$|p_{z}^{-}\rangle = \frac{1}{\sqrt{2}}(|p_{z}1\rangle + |p_{z}2\rangle)$,
and $|p_{z}^{+}\rangle = \frac{1}{\sqrt{2}}(|p_{z}1\rangle - |p_{z}2\rangle)$.
The Hamiltonian $H_{\tilde{m}\tilde{n}}(0,0,k_{z})$ in Eq.\ref{eq:Hamiltonian_rot}
is in the order of $|xy^{-}\rangle$, $|xy^{+}\rangle$, $|p_{z}^{-}\rangle$, and $|p_{z}^{+}\rangle$ basis.
In this transformation, from the even parity of the Fe($d_{xy}$) wave function,
$|xy^{-}\rangle$ and $|xy^{+}\rangle$ indicate odd parity Fe($d_{xy}$) and even parity Fe($d_{xy}$) basis, respectively.
Concerning the odd parity of the Se($p_{z}$) wave function,
$|p_{z}^{-}\rangle$ and $|p_{z}^{+}\rangle$ indicate odd parity Se($p_{z}$) and even parity Se($p_{z}$) basis, respectively.

\begin{equation}
\begin{split}
&H_{\tilde{m}\tilde{n}}(0,0,k_{z}) \\
&=\left[\begin{smallmatrix}
\tilde{\epsilon}_{xy}-4\tilde{t}_{xy} & 0 & 4\tilde{t}_{1} & 0\\
0 & \tilde{\epsilon}_{xy}+4\tilde{t}_{xy} & 0 & 0\\
4\tilde{t}_{1} & 0 & \tilde{\epsilon}_{p_{z}}+4\tilde{t}_{p_{z}}+4\tilde{t}_{2}cos k_{z} & -4i\tilde{t}_{2}sin k_{z}\\
0 & 0 & +4i\tilde{t}_{2}sin k_{z} & \tilde{\epsilon}_{p_{z}}-4\tilde{t}_{p_{z}}-4\tilde{t}_{2}cos k_{z}
\end{smallmatrix}\right]
\label{eq:Hamiltonian_rot}
\end{split}
\end{equation}

Analysis of Eq.\ref{eq:Hamiltonian_rot} reveals that the $|xy^{+}\rangle$ basis does not hybridize with any other vector in the  basis at $\Gamma$-Z, and can be regarded as a non-bonding state of Fe($d_{xy}$). The band associated with this orbital character is the $\beta$ band, which is consistent with Eq.\ref{eq:Hamiltonian_rot} as it has an even parity with very weak dispersion in $\Gamma-$Z, as shown in Fig.\ref{fig:outofplane_main}. In contrast, the $|xy^{-}\rangle$ orbital hybridizes with the $|p_{z}^{-}\rangle$ orbital from the $4\tilde{t}_{1}$ term in Eq.\ref{eq:Hamiltonian_rot}, which is enabled by the inversion symmetry of the system~\cite{wang_TopologicalNatureMathrmFeSe_2015}. The band resulting from this hybridization is the $xy^{-}$ band in Fig.\ref{fig:outofplane_main}, possessing an odd parity. It is important to note that the $xy^{-}$ band is primarily composed of the Fe($d_{xy}$) orbital, due to the substantially higher energy level of the $|xy^{-}\rangle$ orbital than that of the $|p_{z}^{-}\rangle$ orbital in Eq.\ref{eq:Hamiltonian_rot}.  The $xy^{-}$ band acquires a band dispersion in $k_{z}$ from the 4$\tilde{t}_{1}$ term.

We now isolate how the parameters in the quasiparticle Hamiltonian, Eq.\ref{eq:Hamiltonian_rot},  vary as the chemistry
and the structure of the compound modifies the strength of the correlations.  Here, the mass renormalization parameters of the Fe($d$) orbital, $Z_{m}$,  are the
key parameters.
This renormalization  is expressed by ${t}_{mn,renormalized}=\sqrt{Z_{m}}t_{mn}\sqrt{Z_{n}}$.
$Z_{xy}$  approaches to zero when the correlation is enhanced with a larger Te ratio~\cite{yi2015observation,huang2020low}.
It can thus be concluded that the reduction of the $\tilde{t}_{1}$ term, Fe($d_{xy}$)-Se($p_{z}$) hopping,
from the OSMT in FST leads to a down shifted energy level of the $xy^{-}$ band, as well as a decrease in the $k_{z}$ dependent dispersion of the $xy^{-}$ band.

\begin{equation}
\begin{split}
&\tilde{t}_{1} \rightarrow \sqrt{Z_{xy}} \tilde{t}_{1},~\tilde{t}_{xy} \rightarrow Z_{xy}\tilde{t}_{xy}
\end{split}
\label{eq:renormalization}
\end{equation}

Eq. \ref{eq:renormalization} demonstrates the renormalization of hopping elements, $\tilde{t}_{1}$ and $\tilde{t}_{xy}$, due to the dynamical correlation of the Fe($d_{xy}$) orbital.  Through Eq. \ref{eq:Hamiltonian_rot} and Eq. \ref{eq:renormalization}, it is evident that in the vicinity of the OSMP, the two Fe($d_{xy}$) dominant bands, $\beta$ and $xy^{-}$, coalesce into a single flat band due to the lack of hybridization with the $|p^{-}_{z}\rangle$ orbital. This implies that close to the OSMP, the Z$_{2}$ topology is trivial, with a removal of band inversion between the $\alpha'$ and $xy^{-}$ bands, which is consistent with the band structure in Fig. \ref{fig:OSMP}. The effective Hamiltonian in Eq. \ref{eq:Hamiltonian_rot} successfully captures the electronic structure in Fig. \ref{fig:OSMP}, having a substantial Fe($s$) orbital contribution in the $p_{z}^{+}$ majority band (See SM~\cite{Note1} Section V).

\begin{figure}[t]
\includegraphics[width=\columnwidth]{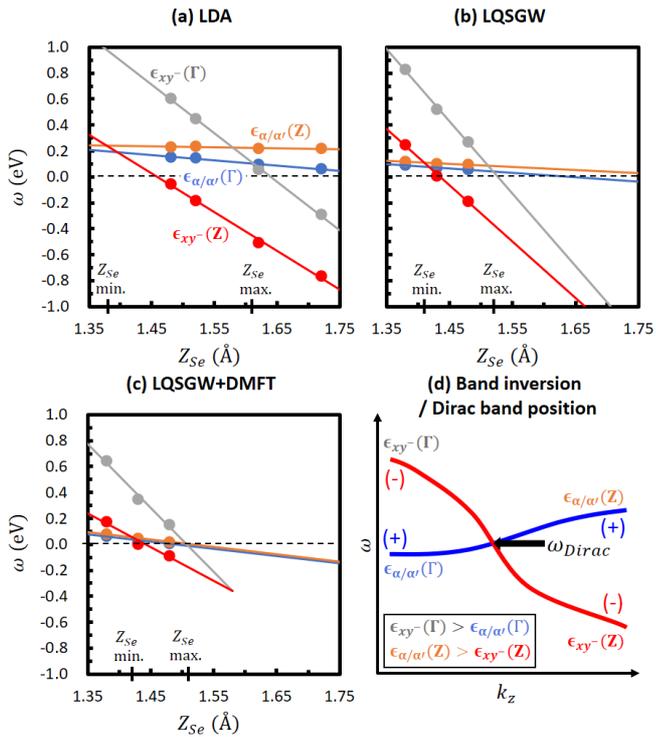}
\caption{
The $Z_{Se}$ dependent phase diagram of the band inversion
condition for the non-trivial Z$_{2}$ bulk topology
and the energy position of the Dirac surface band.
(a) The $Z_{Se}$ dependent variation of
(i) the $\alpha/\alpha'$ bands top and bottom energy positions ($\epsilon_{\alpha/\alpha'}$(Z) and $\epsilon_{\alpha/\alpha'}$($\Gamma$)), and
(ii) the $xy^{-}$ band top and bottom energy positions ($\epsilon_{xy^{-}}$($\Gamma$) and $\epsilon_{xy^{-}}$(Z)).
The $\alpha/\alpha'$ bands degenerate along the $\Gamma$-Z momentum path in computations without SOC.
Linear lines are for the interpolating from the given data (dots) in the LDA.
The range of the $Z_{Se}$
for the band inversion are denoted, [1.38,1.61] $\AA$.
(b) Same as (a) in the LQSGW with the $Z_{Se}$ interval of [1.41,1.52] $\AA$ for the band inversion.
(c) Same as (a) in the LQSGW+DMFT with the $Z_{Se}$ interval of [1.42,1.51] $\AA$ for the band inversion.
(d) A schematic diagram for the band inversion condition and the energy position for the emergence of the Dirac surface band ($\omega_{Dirac}$).
\label{fig:phasediagram}
}
\end{figure}

In Fig.~\ref{fig:phasediagram}, we present the $Z_{Se}$ dependent
top and bottom energy positions of the $\alpha$/$\alpha'$ band
and the $xy^{-}$ band in $\Gamma-$Z
from the LDA, the LQSGW, and the LQSGW+DMFT frameworks.
The condition for the band inversion is that
the top of the $xy^{-}$ band is above the bottom of the $\alpha$/$\alpha'$ band ($k=\Gamma$),
and the bottom of the $xy^{-}$ band is below the top of the $\alpha$/$\alpha'$ band ($k=Z$).
The range of the $Z_{Se}$ for the band inversion gives rise to the non-trivial Z$_{2}$ topology
is determined to be [1.38,1.61] $\AA$ for the LDA, [1.41,1.52] $\AA$ for the LQSGW,
and [1.42,1.51] $\AA$ for the LQSGW+DMFT, respectively.
The electronic correlation renormalizes the bandwidth of the $xy^{-}$ band, reducing
the range of $Z_{Se}$ for the non-trivial Z$_{2}$ topology.
It is also found that the electronic correlation shifts down the $\alpha$/$\alpha'$ band
with the reduced bandwidth of the band.
The electronic correlation effects on the $xy^{-}$ band
explains the removal of the TPSC of FST
upon enhancing Te or Se ratio which changes the chalcogen heights~\cite{liElectronicPropertiesBulk2021}.
Furthermore, in the DFT framework, the substitution of Te for Se
brings a minor enhancement of the $\tilde{t}_{1}$ term while substantially enhances the $\tilde{t}_{2}$ term,
leading to the OSMP with vanishing $\tilde{t}_{1}$ contribution (See SM \cite{Note1} Section VI).
These results demonstrate the essential roles of the electronic correlation
for the observation of the TPSC in FST.
As the strength of the correlations  is very sensitive to the chalcogen height,
we suggest that uniaxial strain can be an ideal tool  for controlling the region
where  non-trivial topology is realized.

{\it Conclusion.}
The  new understanding of the  dominant orbital character of the topologically non-trivial band of FST (i.e. the $xy^{-}$ band) from Fe($d_{xy}$) and its relation to the  OSMP in FST  provides  new insights into  puzzling observations in this  compound and calls for further studies.  The surface layers of FST  are more correlated than the bulk layers due to reduced screening and  the  missing hopping matrix elements, which should result in a surface OSMT at a larger  Se concentration than in the bulk.   This provides a qualitative explanation for the puzzling phase diagram of Refs.~\cite{TRS_surface,liElectronicPropertiesBulk2021} and calls for more quantitative studies using  inhomogeneous qsGW+DMFT~\cite{richler2018inhomogeneous},  to explore  in more detail surface properties.  Furthermore, the OSMP leads to the formation of surface local moments  which will be coupled
to  the itinerant states of Fe($d_{xz/yz}$) from the double exchange,  leading to  possible   time reversal symmetry breaking  states, accounting for the recent experimental observations~\cite{TRS_surface,zakiTimereversalSymmetryBreaking2021}.

The strong sensitivity of the parameters controlling the  topological band  to the chalcogen height,
suggests controlled experiments where stress can be used to stabilize the region of the phase diagram where TPSC with its resulting Majorana zero modes are observed.  This has been   shown in  recent experiments on LiFeAs~\cite{li2022ordered}.

Quantitatively, the successful application of the LQSGW+DMFT+SOC method   which enabled the theoretical estimations of the  parameters of TPSC of FST and their dependence on the structure and the chemistry of the compound  suggests that theoretically guided material   design  is feasible in this  area. It could be applied to other iron pnictides and more generally to the normal state of other  correlated  topological materials  which become superconductors at low temperatures.


\acknowledgements
Acknowledgements

We are greatful to  Youngkuk Kim for discussions with  S.C.  and to Tamaghna Hazra
for discussions with GK. G.K. and S.C. both  acknowledge  comments and  discussions with P. Johnson and A. Tsvelik.  This work was supported by the U.S Department of Energy, Office of
Science, Basic Energy Sciences as a part of the Computational
Materials Science Program. \textit{Ab initio} LQSGW+DMFT calculation used resources of the National Energy Research Scientific Computing Center (NERSC), a U.S. Department of Energy Office of Science User Facility operated under Contract No. DE-AC02-05CH11231.
MK was supported by KIAS Individual Grants(CG083501). SC was supported by a KIAS Individual Grant (CG090601) at Korea Institute for Advanced Study. The DFT calculation is supported by the Center for Advanced Computation at Korea Institute for Advanced Study.

\bibliography{refs_FeSeTe,zotero}


\renewcommand{\thetable}{S\arabic{table}}
\renewcommand{\thefigure}{S\arabic{figure}}
\renewcommand{\thetable}{S\arabic{table}}
\renewcommand\theequation{S\arabic{equation}}
\setcounter{table}{0}
\setcounter{figure}{0}
\setcounter{equation}{0}
\renewcommand{\bibnumfmt}[1]{[S#1]}
\renewcommand{\citenumfont}[1]{S#1}

\def\cred{\color{red}}
\def\cblue{\color{blue}}

\onecolumngrid

\clearpage

\begin{center}
{\bf \large
{\it Supplemental Material:}\\
Orbital Selective Mott Transition Effects and Non-Trivial Topology
\\ of Iron Chalcogenide

}

\vspace{0.2 cm}
Minjae Kim$^{1,2,3}$, Sangkook Choi$^{1,4}$, Walber Hugo Brito$^{5,3}$, and Gabriel Kotliar$^{3,4}$
\vspace{0.1 cm}

{\small{\it
$^{1}$Korea Institute for Advanced Study, Seoul 02455, South Korea \\
$^{2}$Department of Chemistry, Pohang University of Science and Technology (POSTECH), Pohang 37673, Korea \\
$^{3}$Department of Physics and Astronomy, Rutgers University, Piscataway, New Jersey 08854, USA \\
$^{4}$Condensed Matter Physics and Materials Science Department,\\ Brookhaven National Laboratory, Upton, New York 11973, USA \\
$^{5}$Departamento de F\'{\i}sica, Universidade Federal de Minas Gerais, C. P. 702, 30123-970 Belo Horizonte, MG, Brazil
}
}
\end{center}


\section{I. ARPES versus LQSGW+DMFT+SOC}

\subsection{A. In-plane $\Gamma$-M line}

\begin{figure}[b]
\includegraphics[width=\columnwidth]{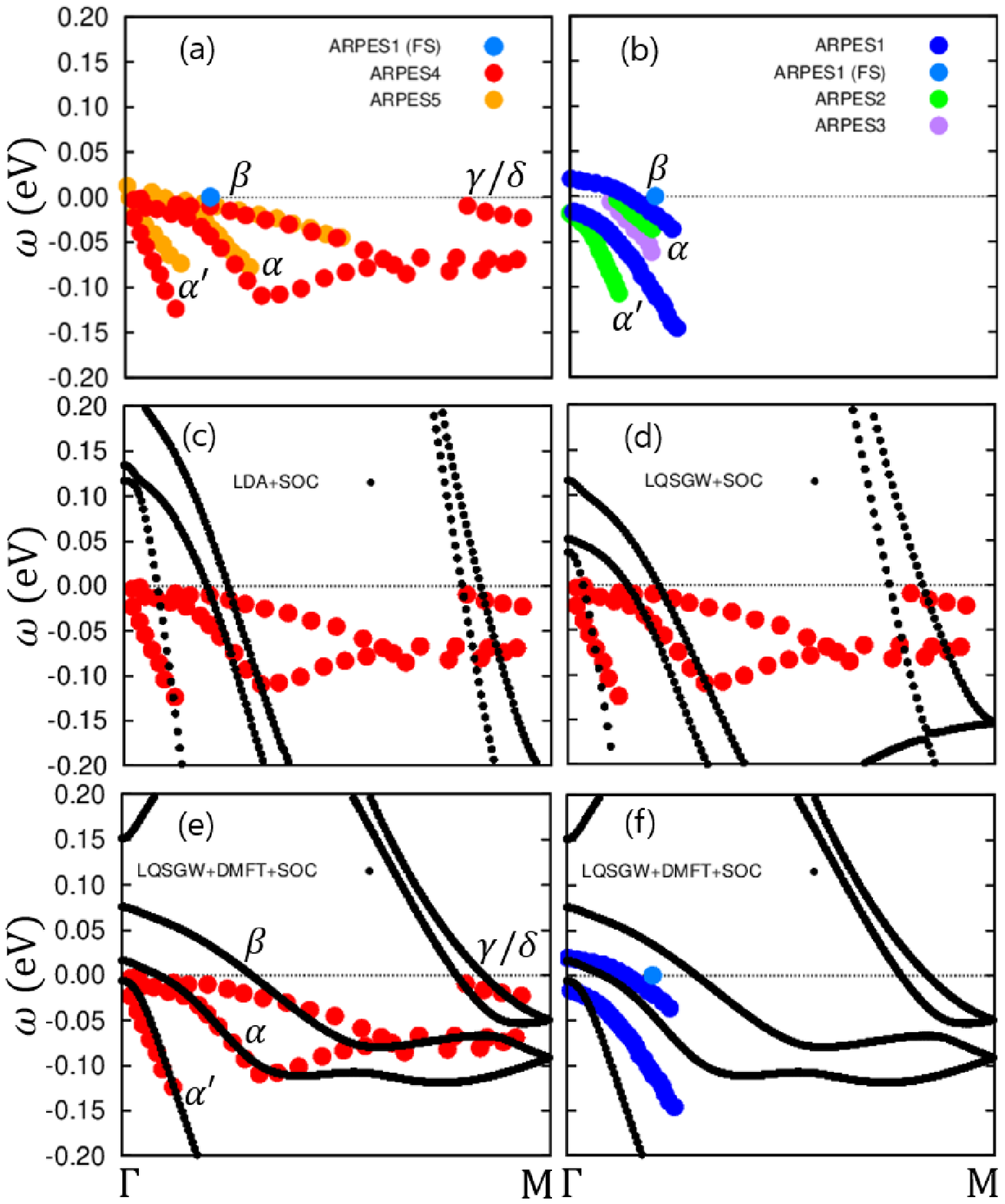}
\caption{
(a) Experimental quasiparticle dispersions in the $\Gamma$-M $k$ point line,
adapted from H. Miao et al.(ARPES4 from Ref.\cite{miao2018universal}),
J. Huang et al.(ARPES5 from Ref.\cite{huang2020low}), and
P. D. Johnson et al.(ARPES1 for the Fermi momentum (FS) of $\beta$ from Ref.\cite{johnson2015spin}).
(b) Refined experimental quasiparticle dispersions near $\Gamma$
in the $\Gamma$-M $k$ point line,
adapted from P. D. Johnson et al.(ARPES1 from Ref.\cite{johnson2015spin}),
H. Lohani et al.(ARPES2 from Ref.\cite{lohani2020band}), and
Z. Wang et al.(ARPES3 from Ref.\cite{wang_TopologicalNatureMathrmFeSe_2015}).
(c), (d), and (e)
Theoretical quasiparticle dispersions in the $\Gamma$-M $k$ point line
are compared to the experiment of ARPES4 from Ref.\cite{miao2018universal},
(c) LDA+SOC, (d) LQSGW+SOC, and (e) LQSGW+DMFT+SOC
theoretical frameworks, respectively.
(f) Theoretical quasiparticle dispersions in the $\Gamma$-M $k$ point line
from the LQSGW+DMFT+SOC is compared to the experiment of ARPES1 from Ref.\cite{johnson2015spin}.
Band indices for hole bands ($\alpha'$, $\alpha$, and $\beta$)
and electron bands ($\gamma$ and $\delta$) are denoted.
\label{fig:inplane}
}
\end{figure}

Fig. \ref{fig:inplane} presents existing quasiparticle dispersions of FST ($x\approx0.5$) from ARPES experiments~\cite{wang_TopologicalNatureMathrmFeSe_2015,lohani2020band,johnson2015spin,huang2020low,miao2018universal} in the $\Gamma$-M $k$ point line, as compared to quasiparticle dispersions obtained in the LDA+SOC, LQSGW+SOC, and LQSGW+DMFT+SOC frameworks. We use the present structural model for FST for computations
of quasiparticle dispersions (See Section II.A). Fig. \ref{fig:inplane}(c) shows an inconsistency in the energy scale of bandwidths between the LDA+SOC and ARPES experiments; the latter is smaller by a factor of 4-5. This result strongly suggests that electronic correlation-induced renormalization beyond the LDA is essential to accurately describe the experimental bulk electronic structures in ARPES~\cite{miao2018universal,johnson2015spin}. While the LQSGW+SOC in Fig. \ref{fig:inplane}(d) improves upon the LDA+SOC with a reduction in bandwidth by a factor of 0.5 and a shrinking of two hole pockets, $\alpha'$ and $\alpha$, these calculations are still not in full agreement with ARPES experiments~\cite{wang_TopologicalNatureMathrmFeSe_2015,lohani2020band,johnson2015spin,huang2020low,miao2018universal}.

Fig.~\ref{fig:inplane}(e and f) presents a comparison between the quasiparticle band structure obtained from LQSGW+DMFT+SOC and ARPES data (along $\Gamma$-M) taken from Refs.~\cite{miao2018universal,johnson2015spin}. The results demonstrate a good agreement, implying the validity of the present structural model for FST and the importance of electronic correlations for the description of bulk electronic structure. In particular, the SOC induced gap at $\Gamma$ for the $\alpha$ and $\alpha'$ bands is consistent with the refined ARPES from Ref.~\cite{johnson2015spin} as seen in Fig.~\ref{fig:inplane}(f), validating the implementation of SOC in the main text and Table~\ref{Parameters}.

Fig.~\ref{fig:inplane}(e) shows good agreement between LQSGW+DMFT+SOC calculations and ARPES results from Ref.~\cite{miao2018universal} for electron bands near the M point ($\delta$/$\gamma$). In comparison, the $\beta$ hole band exhibits a different Fermi momentum and effective mass in the present LQSGW+DMFT+SOC results compared to experiments~\cite{miao2018universal,johnson2015spin}, as depicted in Fig.\ref{fig:inplane}(e and f). As discussed in the main text, the $\beta$ band is primarily composed of $xy$ orbitals and experiences the orbital selective Mott transition in FST. Therefore, (i) its dispersion is sensitive to the chemical composition of the FST alloy
and (ii) the band has a small Fermi velocity.
These factors give rise to the observed variation in its Fermi momentum, as shown in Fig.\ref{fig:inplane}(a and b)~\cite{miao2018universal,johnson2015spin}.

\subsection{B. Out-of-plane $\Gamma$-Z line}

\begin{figure}[t]
\includegraphics[width=\columnwidth]{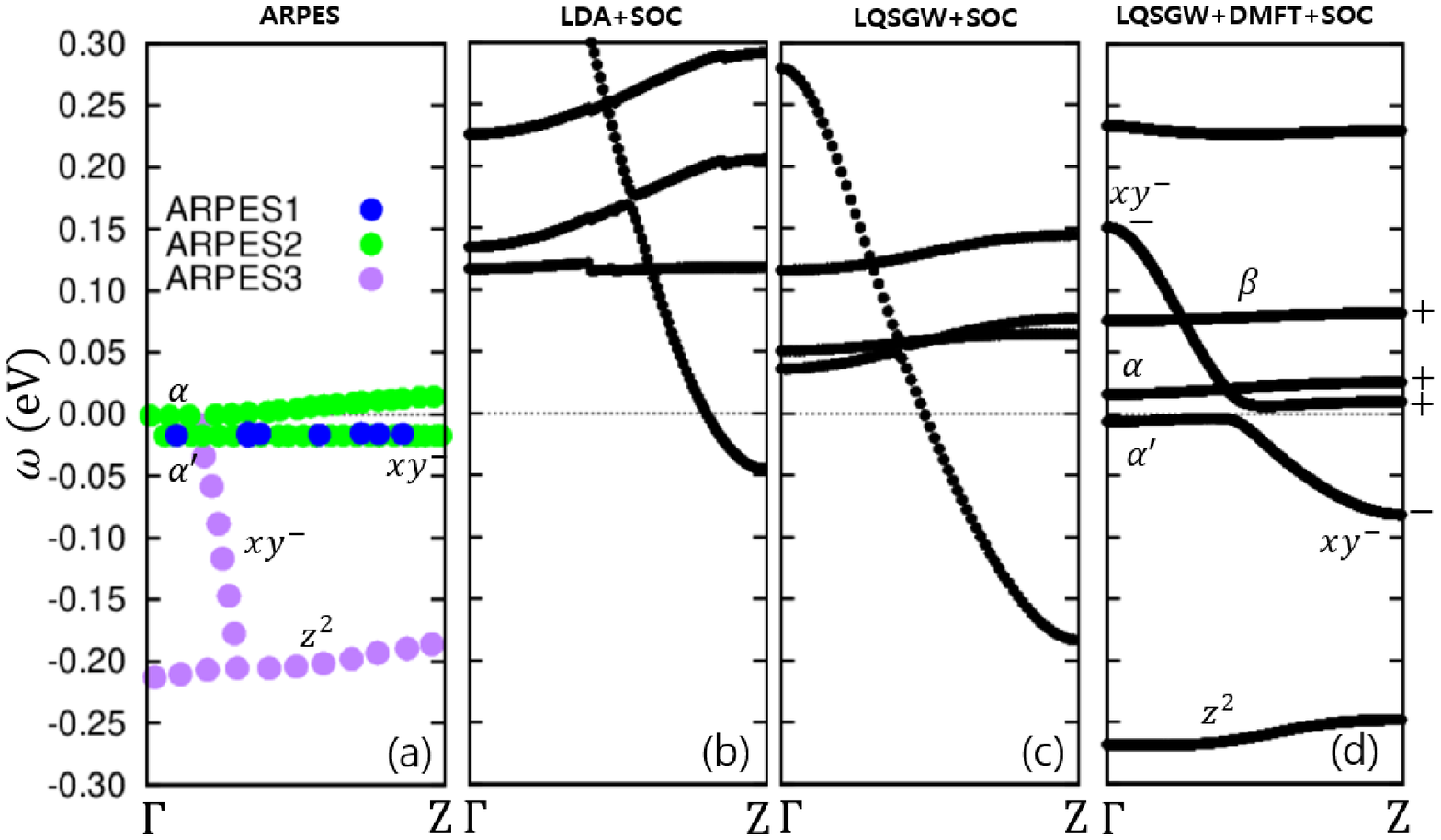}
\caption{(a) Experimental quasiparticle dispersions in the $\Gamma$-Z $k$ point line,
adapted from P. D. Johnson et al.(ARPES1 from Ref.\cite{johnson2015spin}),
H. Lohani et al.(ARPES2 from Ref.\cite{lohani2020band}), and
Z. Wang et al.(ARPES3 from Ref.\cite{wang_TopologicalNatureMathrmFeSe_2015}).
(b), (c), and (d)
Theoretical quasiparticle dispersions in the $\Gamma$-Z $k$ point line
are compared to ARPES experiments,
(b) LDA+SOC, (c) LQSGW+SOC, and (d) LQSGW+DMFT+SOC
theoretical frameworks, respectively.
The parity eigenvalue for each band is denoted in (d),
as $\alpha'$ ($+$), $\alpha$ ($+$), $\beta$ ($+$), and $xy^{-}$ ($-$).
\label{fig:outofplane}
}
\end{figure}

In Figure \ref{fig:outofplane}, we present quasiparticle dispersions from ARPES experiments~\cite{wang_TopologicalNatureMathrmFeSe_2015,lohani2020band,johnson2015spin} along the $\Gamma$-Z $k$ point line, compared to theoretical quasiparticle dispersions obtained in the LDA+SOC, the LQSGW+SOC, and the LQSGW+DMFT+SOC frameworks. Even parity bands of $\alpha'$, $\alpha$, and $\beta$ are noted. The odd parity band, $xy^{-}$, is also noted which arises from the hybridization of the $xy$ and Se($p_{z}$) orbitals (see main text). Comparison of the LQSGW+DMFT+SOC result to the LDA+SOC and LQSGW+SOC results illustrates the strong electronic correlation driven renormalization of the $xy^{-}$ band, which is dominated by the $xy$ orbital. As shown in the main text, this $xy$ orbital experiences an opening of the Mott gap in an orbital selective Mott transition. This correlated nature of the $xy^{-}$ band in FST alloy is likely the cause of the experimental variation of the $xy^{-}$ band dispersion reported in Refs.\cite{wang_TopologicalNatureMathrmFeSe_2015,lohani2020band,johnson2015spin}, as seen in Fig.\ref{fig:outofplane}(a). We note that the present LQSGW+DMFT+SOC results for the $xy^{-}$ band dispersion have an energy position at $k=Z$ that lies between the ARPES experiments of Ref.\cite{wang_TopologicalNatureMathrmFeSe_2015} and Refs.\cite{lohani2020band,johnson2015spin}.

\section{II. Computational Details}

\subsection{A. Crystal structure}

In FeSe$_{0.5}$Te$_{0.5}$,   the  isovalent substitutions Se and Te
do not form  an ordered array but a rather disordered
alloy. Furthermore, the Se and the Te chalcogen  heights are very different.
In this work, we replace this alloy by a periodic crystal   of FeSe with
effective parameters in the spirit of the virtual crystal approximation.
This has the advantage of preserving the inversion symmetry, facilitating
the analysis of the topological band.
Here, we explain the  choice of  parameters for this average crystalline structure.

The crystal structure of FeSe$_{0.5}$Te$_{0.5}$ taken from the
experiment in Refs.\cite{li2009first,tegel2010crystal}, with lattice
constant of [a=3.793 ${\AA}$, c=5.955 ${\AA}$] and Chalcogen height
of Se, Z$_{Se}$=1.48 ${\AA}$, was employed in our LQSGW, LQSGW+DMFT,
and DFT computations. This structure is also found to be the optimum
for the description of ARPES experiments as seen in
Figs.\ref{fig:148_inplane} and \ref{fig:148_outplane}. Furthermore,
the inversion symmetry of the unit cell is maintained for the
Z$_{2}$ topology resulting from the parity inversion of bands in the
presence of the time reversal symmetry, as indicated in
Refs.\cite{zhang2018observation,xu_TopologicalSuperconductivitySurface_2016,wang_TopologicalNatureMathrmFeSe_2015}.

\subsection{B. LQSGW calculation}
LQSGW calculations are performed using Flapwmbpt \cite{kutepov_linearized_2017}, which is
based on full-potential linearized augmented plane wave plus local
orbital method. The following parameters for the basis are used: muffin-tin (MT) radii (RMT ) in Bohr radius are
2.26 for Fe and 2.15 for Se. Wave functions are expanded by
spherical harmonics with l up to 4 for Fe and 4 for Se in the MT spheres, and by
plane waves with the energy cutoff determined by RMT$_{Fe}$ $\times$ Kmax = 7.7 in the interstitial
(IS) region. The Brillioun zone was sampled with 6 $\times$ 6 $\times$ 4 k-point grid. Product basis is expanded
by spherical harmonics with l up to lmax =6 in the MT spheres and RMT$_{Fe}$ × Kmax = 12.0 in
IS region. All unoccupied states are taken into
account for polarizability and self-energy calculation.

\subsection{C. Wannier function constructions}
34 Wannier functions are constructed by using Wannier90 package
\cite{mostofi_Wannier90ToolObtaining_2008} with a
frozen energy window between -10 eV to 10 eV and with a disentanglement energy window of
-10 eV to 50 eV: Fe-s, Fe-p, Fe-d, Se-p, Se-d orbitals. Initial trial
orbitals are constructed by using Muffin-tin orbitals in LAPW basis
set with well-defined angular momentum characters.

\subsection{D. Double counting energy}
The electron self-energy included in both \textit{ab initio} LQSGW and DMFT is the local Hartree term and the local GW term. They can be calculated as follows.
\begin{equation}
  \widetilde{\Sigma}_{DC,i,j}(i\omega_n)=2\sum_{k,l=m_l'}^{d\text{-orbital}} \widetilde{G}_{l,k}(\tau=0^-)\widetilde{\mathcal{U}}_{iklj}(i\nu=0)-\sum_{k,l}^{d\text{-orbital}}\int d\tau \widetilde{G}_{l,k}(\tau)\widetilde{W}_{ikjl}(\tau)e^{i \omega_n\tau}.\label{eq:dc}
\end{equation}
where $i$,$j$,$k$ and $l$ are orbital indices. $\widetilde{G}$ is the local Green's function. $\widetilde{\mathcal{U}}$ is constructed by using Slater's integrals in the
constraint random phase approximation.
\begin{equation}
  \begin{split}
    \widetilde{\mathcal{U}}_{i,j,k,l}(i\nu_n)&=\sum_{\substack{m'_1m'_2,m'_3m'_4}}S_{i,m_1}S_{j,m_2}S_{k,m_3}^{-1}S_{l,m_4}^{-1}\\
    &\sum_{k=0}^{2l,even}\frac{4\pi}{2k+1}\langle Y_{l}^{m_1'}|Y_{k}^{q}Y_{l}^{m_4'}\rangle\langle Y_{l}^{m_2'}Y_{k}^{q}|Y_{l}^{m_3'}\rangle F^{k}(i\nu_n).
    \label{eq:coulomb_so}
  \end{split}
\end{equation}

Here, we assume that the frequency-dependent interaction is of the form
\begin{equation}
  \widetilde{\mathcal{U}}_{ijkl}(i\nu_n) = \widetilde{U}_{ijkl} + F^0(i\nu_n)\delta_{il}\delta_{jk},
\end{equation}
that is, only the dynamical screening of the Slater-Condon parameter $F^0$ is taken into account. The other Slater-Condon parameters, which define $\widetilde{U}_{ijkl}$, are frequency independent and approximated by their value at $\nu_n=\infty$. $\widetilde{W}$ is the local screened Coulomb interaction given by
\begin{equation}
  \widetilde{W}_{ikjl}(i\nu_n){=}\widetilde{\mathcal{U}}_{ikjl}(i\nu_n)+\sum_{mnpq}^{d\text{-orbital}}\allowbreak \widetilde{\mathcal{U}}_{imnl}(i\nu_n) \allowbreak \widetilde{P}_{mpqn}(i\nu_n)\allowbreak \widetilde{W}_{pkjq}(i\nu_n),\label{eq:ww}
\end{equation}
where $\widetilde{P}$ is the local polarizability and it is calculated as
\begin{equation}
  \widetilde{P}_{mpqn}(i\nu_n)\allowbreak=\int d\tau
  \widetilde{G}_{n,p}(\tau)\widetilde{G}_{q,m}(-\tau)\allowbreak e^{i\nu_n\tau}.\label{eq:pi_wso}
\end{equation}

\subsection{E. DMFT self-consistent equation}
At each iteration of the fermionic DMFT self-consistent loop, the fermionic Weiss-field is constructed in the following way.
\begin{equation}
  \begin{split}
    \widetilde{\mathcal{G}}=\left(\left(\frac{1}{N_\mathbf{k}}\sum_\mathbf{k} f_\mathbf{k}^\dagger G(\mathbf{k},i\omega_n)f_\mathbf{k}\right)^{-1}+\widetilde{\Sigma}_{imp}\right)^{-1}
    \label{eq:glat}
  \end{split}
\end{equation}
Here $f_\mathbf{k}$ is the fermionic projection operator to correlation orbitals (five Fe-d orbitals) and given by $f_\mathbf{k}=\langle\mathbf{r}|W_{i\mathbf{k}}\rangle$ where $|W_{i\mathbf{k}}\rangle=\frac{1}{\sqrt{N_\mathbf{k}}}\sum_{\mathbf{R}}|W_{i\mathbf{R}}\rangle e^{i\mathbf{k}\cdot\mathbf{R}}$. $\widetilde{\Sigma}_{imp}$ is impurity self-energy from impurity solver.

Within \textit{ab initio} LQSGW+DMFT, lattice Green's function is calculated by embedding impurity self-energy into the LQSGW Green's function
\begin{equation}
  \begin{split}
    G^{-1}(\mathbf{k},i\omega_n)=i\omega_n-H_{QP}^{nl}(\mathbf{k})-f_\mathbf{k} \widetilde{\Sigma}_{imp}(i\omega_n) f_\mathbf{k}^\dagger,
    \label{eq:glat_inv}
  \end{split}
\end{equation}
where $H_{QP}^{nl}$ is non-local LQSGW Hamiltonian\cite{tomczak_MergingGWDMFT_2017}, in which double-counting self-energy is compensated up to linear order in frequency.

\begin{equation}
  \begin{split}
    H_{QP}^{nl}(\mathbf{k})=\sqrt{Z_{DC}^{-1}(\mathbf{k})} H_{QP}\sqrt{Z_{DC}^{-1}(\mathbf{k})}-f_\mathbf{k} \widetilde{\Sigma}_{DC}(\omega=0) f_\mathbf{k}^\dagger.
    \label{eq:h_qp_nl}
  \end{split}
\end{equation}
Here, $H_{QP}$ is Wannier interpolated LQSGW Hamiltonian into $15\times 15\times 10$ $k$-grid. $Z_{DC}^{-1}(\mathbf{k})=1-f_\mathbf{k}\left({\partial{\widetilde{\Sigma}_{DC}(\omega=0)}}/{\partial{i\omega_n}}\right)f_\mathbf{k}^\dagger$.

ComDMFT necessitates the solution of an impurity model action. In ComDMFT, hybridization-expansion continuous-time quantum Monte Carlo (CTQMC) is adopted. CTQMC is a stochastic approach to obtain numerically exact solutions of an impurity model. An impurity model consists of a small interacting system, the impurity, immersed in a bath of non-interacting electrons. The action of the impurity model relevant for GW+DMFT reads

\begin{equation}
  \begin{split}
    \label{equ:Action}
    S = &-\iint_0^\beta \sum_{ij} c^\dagger_i(\tau) \widetilde{\mathcal{G}}_{ij}^{-1}(\tau - \tau') c_j(\tau') d\tau d\tau' \\
    &\quad\quad+\frac{1}{2}\iint_0^\beta \sum_{ijkl} c^\dagger_i(\tau) c^\dagger_j(\tau') \widetilde{\mathcal{U}}_{ijkl}(\tau - \tau') c_k(\tau')c_l(\tau) d\tau d\tau',
  \end{split}
\end{equation}
where $c^\dagger_i$ creates an electron in the generalized orbital $i$ (which includes both spin and orbital degrees of freedom), $\beta$ is the inverse temperature, $\widetilde{\mathcal{G}}_{ij}$ is the fermionic Weiss field in eq. \eqref{eq:glat} and $\widetilde{\mathcal{U}}_{ijkl}$ in eq. \eqref{eq:coulomb_so}.

We assume that the frequency-dependent interaction is of the form
\begin{equation}
  \widetilde{\mathcal{U}}_{ijkl}(i\nu_n) = \widetilde{U}_{ijkl} + F^0(i\nu_n)\delta_{il}\delta_{jk},
\end{equation}
that is, only the dynamical screening of the Slater-Condon parameter $F^0$ is taken into account, for the simplicity in the numerical algorithm based on hybridization-expansion CTQMC. The other Slater-Condon parameters, which define $\widetilde{U}_{ijkl}$, are frequency independent and approximated by their value at $\nu_n=\infty$. DMFT self-consistent equation is solved at T=300K. 

\subsection{F. LQSGW+DMFT quasiparticle Hamiltonian construction}
By linearizing LQSGW+DMFT self-energy, we obtained
LQSGW+DMFT quasiparticle Hamiltonian ($H_{LQSGW+DMFT}(\mathbf{k})$) in
local-orbital basis in the following way.
\begin{equation}
  \begin{split}
    &H_{LQSGW+DMFT}(\mathbf{k})\\
&=\sqrt{Z_{imp}}\left(H_{QP}^{nl}(\mathbf{k})
  +f_{Fe{\text -}d}\Sigma_{imp}(\omega=0)f_{Fe{\text -}d}^\dagger\right)\sqrt{Z_{imp}},\\
\label{eq:lqsgw_dmft_Hamiltonian}
\end{split}
\end{equation}
where  $Z_{imp}=\left(1-f_{Fe{\text
      -}d}\frac{\partial\Sigma_{imp}(\omega)}{\partial\omega}|_{\omega=0}f_{Fe{\text
      -}d}^\dagger\right)^{-1}$

\subsection{G. Spin-orbit coupling for LQSGW+DMFT+SOC}
\begin{table}[t]
\caption{Parameters for the
implementation of SOC in the
LQSGW+DMFT+SOC.
The extraction of the SOC parameter
in the DFT+SOC and the LQSGW+SOC
is explained in Eq.\ref{eq:Hamiltonian},
Fig.\ref{fig:SOC}, and Fig.\ref{fig:SOC_LQSGW}.
The quasiparticle renormalization, $Z_{imp}$ of Fe($d$)
orbital is taken from the LQSGW+DMFT result for $xz/yz$ orbital ($Z_{xz/yz}$).
$Z_{imp}(\lambda_{1}$+$\Delta\lambda_{1})$ (in meV) is the renormalized quasiparticle SOC of Fe($d$)
chosen to fit ARPES experiments~\cite{wang_TopologicalNatureMathrmFeSe_2015,lohani2020band,johnson2015spin,huang2020low,miao2018universal}
by considering the orbitally off-diagonal self-energy
and the dynamical self-energy.
The Se/Te($p$) averaged SOC constant
applied in Se is $\lambda_{2}$ (in meV).
The extracted SOC constant (in meV) from the DFT
for Fe($d$), Se($p$), and Te($p$)
are $\lambda_{Fe(d),DFT}$,
$\lambda_{Se(p),DFT}$, and
$\lambda_{Te(p),DFT}$, respectively.
We use the local density approximation (LDA) for the DFT.
The extracted SOC constant (in meV) from the LQSGW
for Fe($d$), Se($p$), and Te($p$)
are $\lambda_{Fe(d),LQSGW}$,
$\lambda_{Se(p),LQSGW}$, and
$\lambda_{Te(p),LQSGW}$, respectively.}
\begin{ruledtabular}
\begin{tabular}{l c c c c c}
             & ~$Z_{imp}$
             & ~
             & ~  \\
\hline
    & 0.35 & ~ & ~ \\
\hline\hline
             & ~$Z_{imp}(\lambda_{1}$+$\Delta\lambda_{1})$
             & ~$\lambda_{Fe(d),DFT}$
             & ~$\lambda_{Fe(d),LQSGW}$ \\

\hline
   & 42 & 70 & 50 \\
\hline\hline
             & ~$\lambda_{2}$
             & ~$\lambda_{Se(p),DFT~and~LQSGW}$
             & ~$\lambda_{Te(p),DFT~and~LQSGW}$ \\
\hline
   & 450 & 200 & 700 \\
\end{tabular}
\label{Parameters}
\end{ruledtabular}
\end{table}

In Table~\ref{Parameters}, we present the parameters for the implementation of spin-orbit coupling (SOC) in the LQSGW+DMFT+SOC Hamiltonian ($H_{LQSGW+DMFT+SOC}(k)$) described in the main text. We have extracted the SOC parameter in the DFT+SOC and the LQSGW+SOC as shown in Eq.\ref{eq:Hamiltonian}, Fig.\ref{fig:SOC}, and Fig.\ref{fig:SOC_LQSGW}. The $\lambda_{1}+\Delta\lambda_{1}$ term denotes the correlation-enhanced spin-orbit coupling of Fe($d$) from the orbitally off-diagonal self-energy ($\Delta\lambda_{1}$ term). Furthermore, the quasiparticle residue of Fe($d$), $Z_{imp}$, is taken into account. We chose the corresponding $Z_{imp}(\lambda_{1}+\Delta\lambda_{1})$ value of 42 meV ($\lambda_{1}+\Delta\lambda_{1}$ = 120 meV)
to fit ARPES experiments~\cite{wang_TopologicalNatureMathrmFeSe_2015,lohani2020band,johnson2015spin,huang2020low,miao2018universal} (See Fig.\ref{fig:inplane} and Fig.\ref{fig:outofplane}). Compared to the DFT and LQSGW results on the SOC of Fe($d$) in Table~\ref{Parameters}, the present value of $Z_{imp}(\lambda_{1}+\Delta\lambda_{1})$ is consistent with previous results for other Hund metals, LiFeAs and Sr$_{2}$RuO$_{4}$\cite{kim2018spin,tamai2019high,linden2020imaginary}. Estimating $\Delta\lambda_{1}$ from DMFT computations is challenging due to (i) the large size of the quantum impurity Hilbert space constructed from 10 spin-orbitals, and (ii) the sign problem in the continuous time quantum Monte Carlo (CTQMC) impurity solver arising from the orbital off-diagonal hybridization function\cite{kim2020alleviating}.

\subsection{H. Extraction of the spin-orbit coupling constant from the DFT and the LQSGW}

\begin{figure}[t]
\includegraphics[width=\columnwidth]{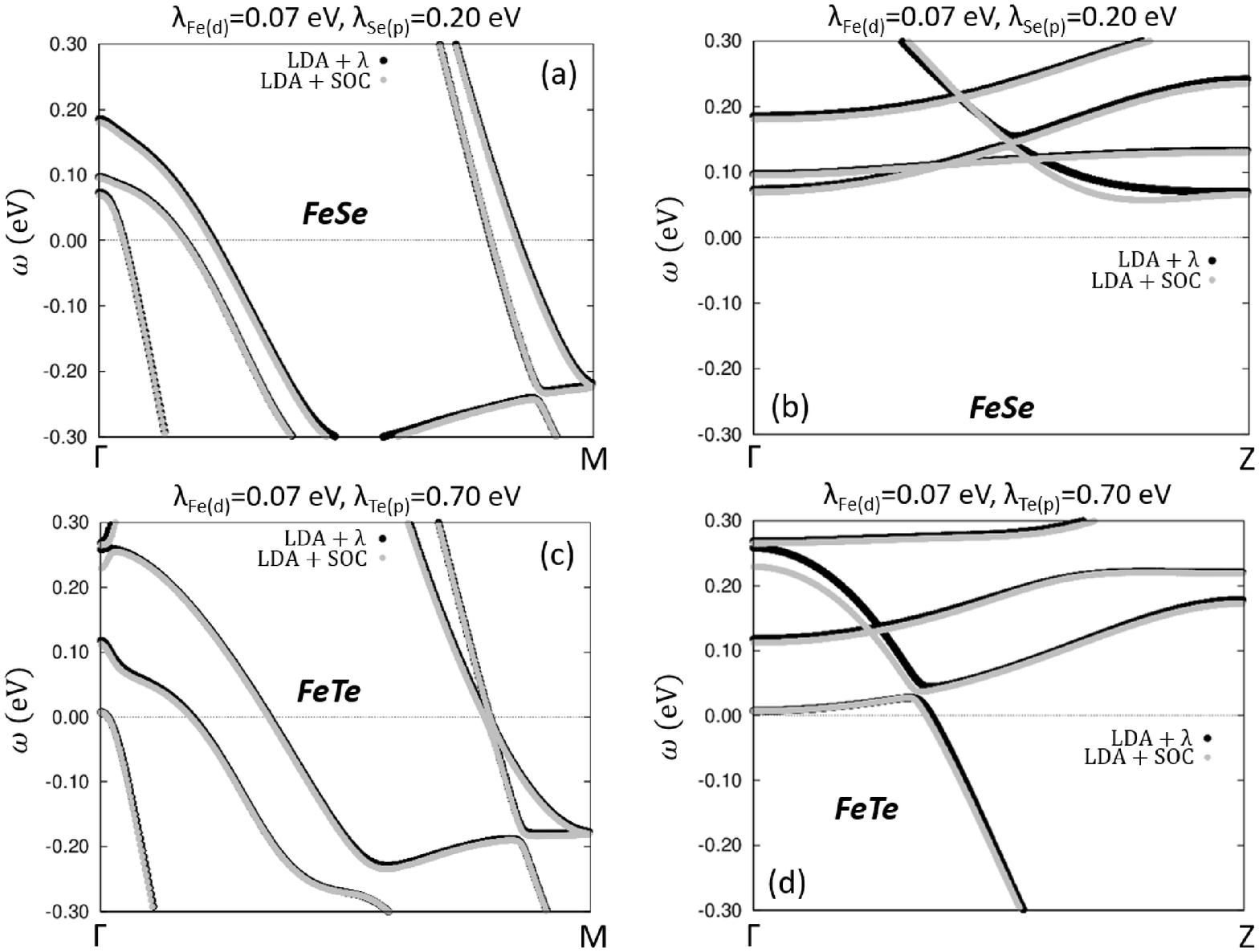}
\caption{(a) Band structure in the LDA+SOC
of FeSe is compared with the
band structure in the LDA+$\lambda$ method
in Eq.\ref{eq:Hamiltonian} with $\lambda_{Fe(d),DFT}$=0.07 eV,
and $\lambda_{Se(p),DFT}$=0.20 eV, in the $\Gamma$-M momentum path.
(b) Same as (a) in the $\Gamma$-Z momentum path.
(c) Band structure in the LDA+SOC
of FeTe is compared with the
band structure in the LDA+$\lambda$ method
in Eq.\ref{eq:Hamiltonian} with $\lambda_{Fe(d),DFT}$=0.07 eV,
and $\lambda_{Te(p),DFT}$=0.70 eV, in the $\Gamma$-M momentum path.
(d) Same as (c) in the $\Gamma$-Z momentum path.
\label{fig:SOC}
}
\end{figure}

\begin{figure}[t]
\includegraphics[width=\columnwidth]{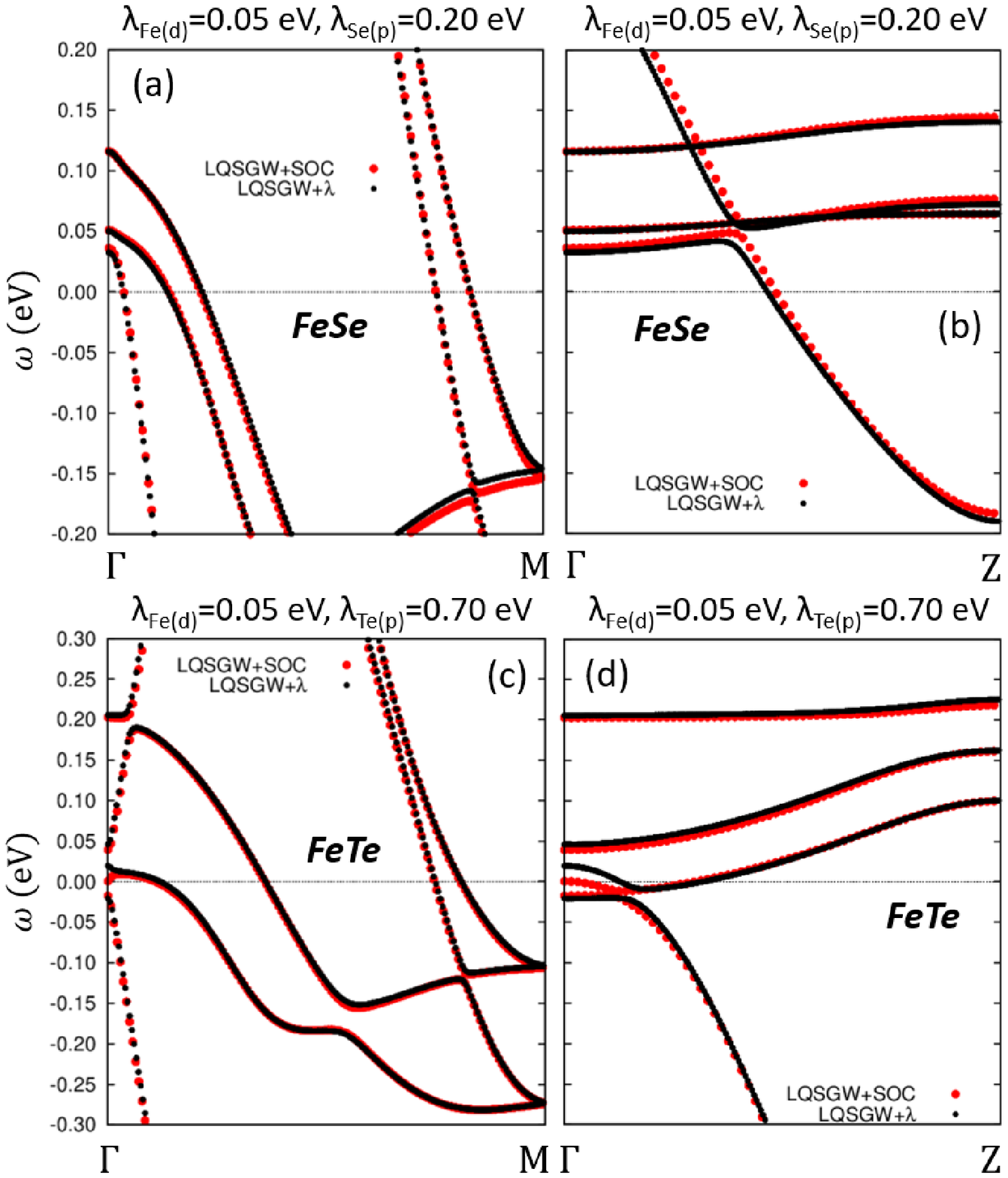}
\caption{(a) Band structure in the LQSGW+SOC
of FeSe is compared with the
band structure in the LQSGW+$\lambda$ method
in Eq.\ref{eq:Hamiltonian} with $\lambda_{Fe(d),LQSGW}$=0.05 eV,
and $\lambda_{Se(p),LQSGW}$=0.20 eV, in the $\Gamma$-M momentum path.
(b) Same as (a) in the $\Gamma$-Z momentum path.
(c) Band structure in the LQSGW+SOC
of FeTe is compared with the
band structure in the LQSGW+$\lambda$ method
in Eq.\ref{eq:Hamiltonian} with $\lambda_{Fe(d),LQSGW}$=0.05 eV,
and $\lambda_{Te(p),LQSGW}$=0.70 eV, in the $\Gamma$-M momentum path.
(d) Same as (c) in the $\Gamma$-Z momentum path.
\label{fig:SOC_LQSGW}
}
\end{figure}

In this work, we extract the spin-orbit coupling constant in the DFT+SOC and the LQSGW+SOC by constructing $H_{DFT}(k)$ and $H_{LQSGW}(k)$ (in Eq.\ref{eq:Hamiltonian}) from the DFT and the LQSGW in the local orbital basis includes Fe($d$) and Se,Te ($p$), respectively, using the maximally localized Wannier function method. For the DFT+SOC
computations in Fig.\ref{fig:SOC}, crystal structures of FeTe and FeSe are adopted from Refs.\cite{phelan2009neutron,mizuguchi2009fete}.
For the LQSGW+SOC
computations in Fig.\ref{fig:SOC_LQSGW}, the lattice constant of FeSe$_{0.5}$Te$_{0.5}$ and the Se and Te heights in FeSe$_{0.5}$Te$_{0.5}$ are adopted from Refs.\cite{li2009first,tegel2010crystal}. We then assume the local spin-orbit coupling for Fe($d$) and Se,Te ($p$), and compare the band structure of $H_{DFT+\lambda}(k)$ and $H_{LQSGW+\lambda}(k)$ to the band structure of the DFT+SOC and the LQSGW+SOC, respectively, with variation of the $\lambda$ variables. We found that in the LDA, the local $\lambda$ variables 0.07 eV (Fe($d$)), 0.20 eV (Se($p$)), and 0.70 eV (Te($p$)) are good approximations for the SOC effects on the band structure, as shown in Figure\ref{fig:SOC}. In the LQSGW, the local $\lambda$ variables 0.05 eV (Fe($d$)), 0.20 eV (Se($p$)), and 0.70 eV (Te($p$)) are good approximations for the SOC effects on the band structure, as shown in Figure\ref{fig:SOC_LQSGW}.

\begin{eqnarray} \label{eq:Hamiltonian}
H_{DFT+\lambda~or~LQSGW+\lambda}(k)= \nonumber  \\
H_{DFT~or~LQSGW}(k) \nonumber \\
+f_{Fe{\text -}d} \lambda_{Fe(d),DFT~or~LQSGW}(L\cdot S) f_{Fe{\text -}d}^\dagger  \nonumber \\
+f_{Se{\text -}p~or~Te{\text -}p} \lambda_{Se(p)~or~Te (p),DFT~or~LQSGW}(L\cdot S)f_{Se{\text -}p~or~Te{\text -}p}^\dagger
\end{eqnarray}

\subsection{I. Surface electronic structure}

\begin{figure}[t]
\includegraphics[width=6cm]{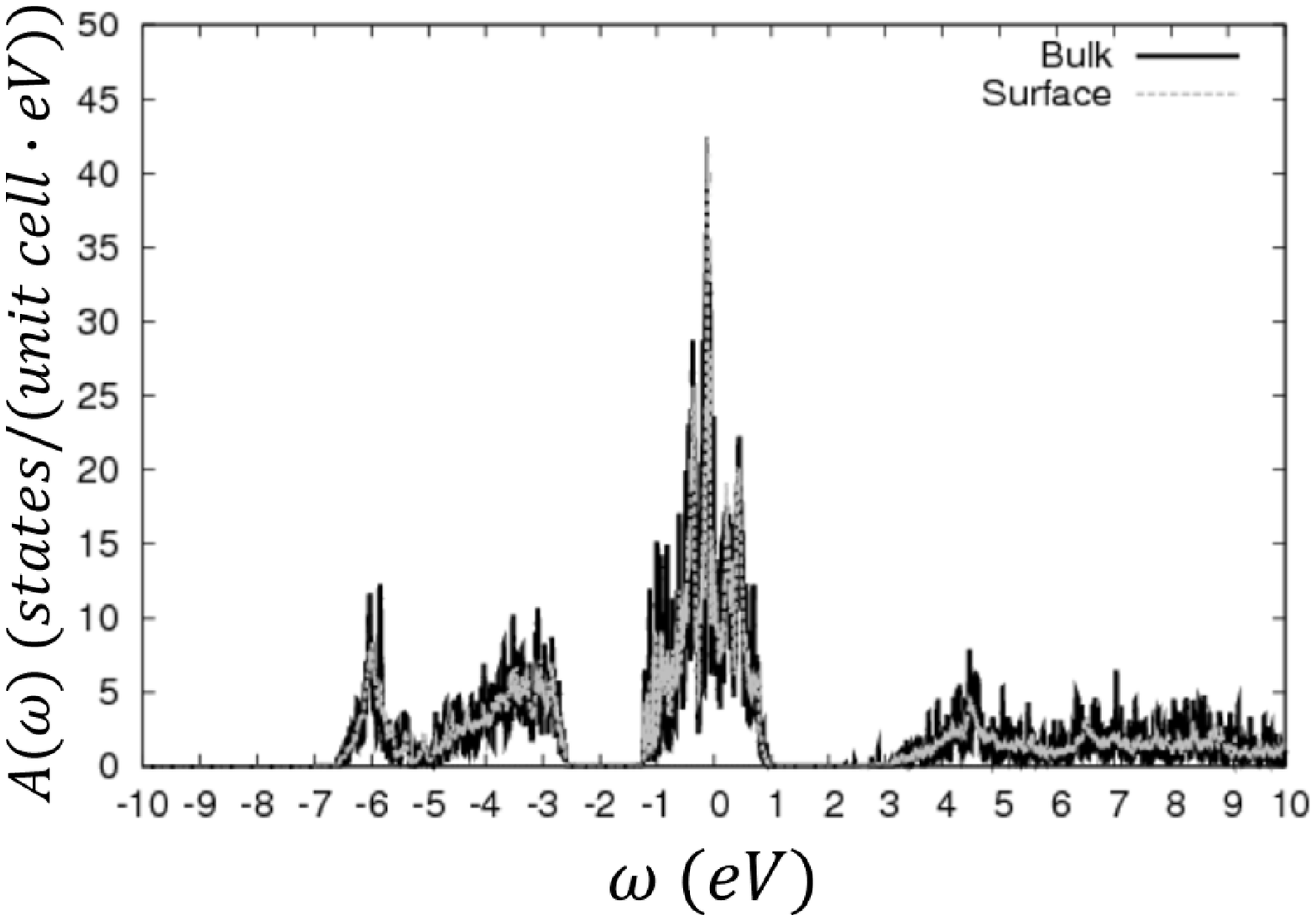}
\caption{Density of states of FST
for the bulk and the surface electronic structures
in the LQSGW+DMFT+SOC.
\label{fig:DOS}
}
\end{figure}

The electronic structure of the (001) surface of the LQSGW+DMFT+SOC was computed by converting the $H_{LQSGW+DMFT+SOC}(k)$ bulk Hamiltonian to real space via Fourier transformation, resulting in a Hamiltonian with finite hopping range of $|R_{i}|<5$ (where $i$=$x,y,z$). The Hamiltonian was then converted to $H_{LQSGW+DMFT+SOC}(k_{x},k_{y},R_{z})$ using Fourier transformation in $z$, using 99 layers for the slab. Hopping elements between the bottom and top of the slab in $R_{z}$ were set to zero, forming the slab Hamiltonian. Comparison of the density of states for both the bulk and surface of FeSe$_{0.5}$Te$_{0.5}$ in the LQSGW+DMFT+SOC (as shown in Figure \ref{fig:DOS}) implies that the slab Hamiltonian was constructed properly from the bulk Hamiltonian.



\section{III. Spin-orbit coupling enhancement and Chalcogen height}

\begin{figure}[t]
\includegraphics[width=\columnwidth]{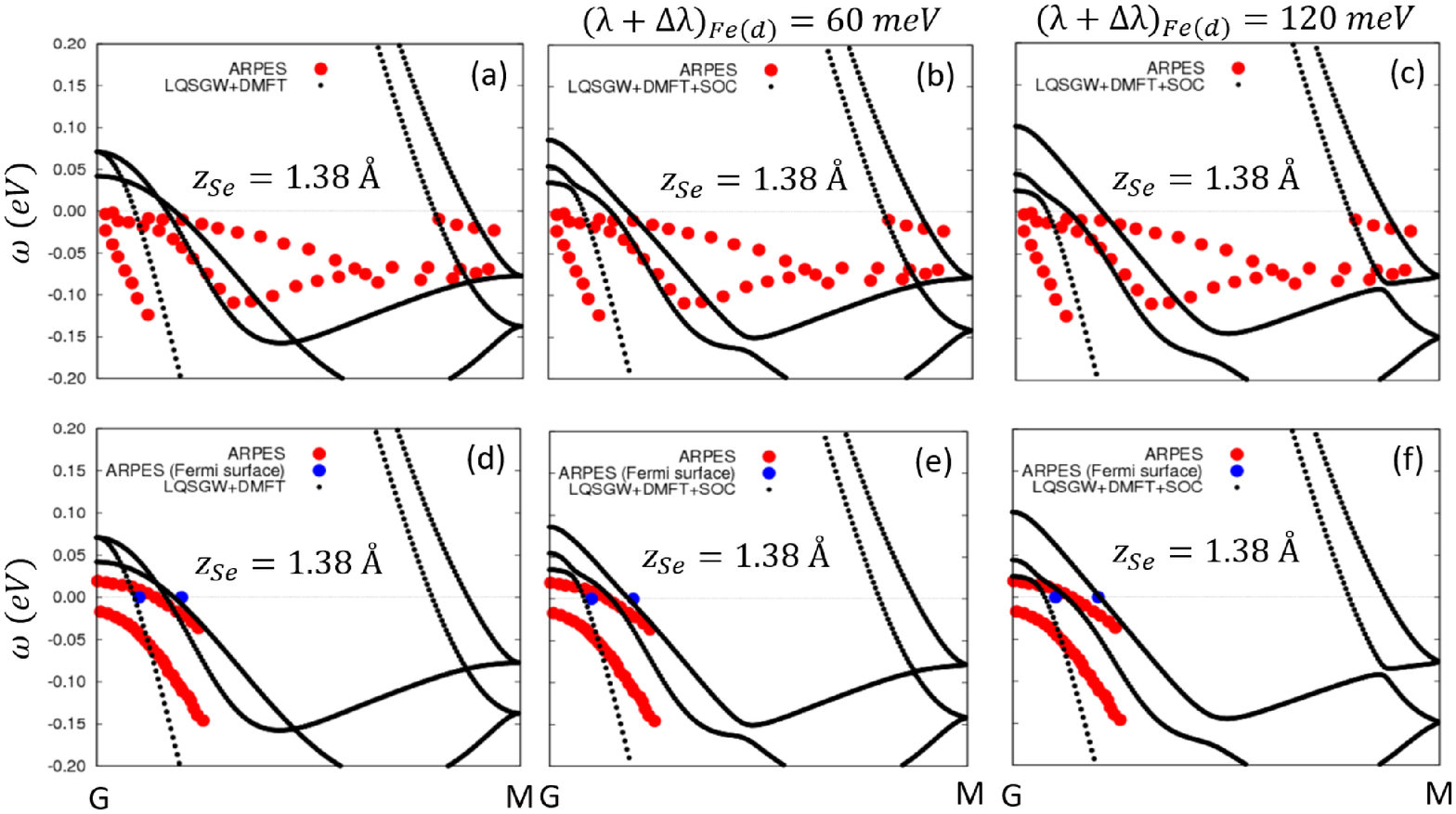}
\caption{(a) Electronic structure in the LQSGW+DMFT
with forcing Z$_{Se}$=1.38 (${\AA}$) compared with ARPES
of Ref.\cite{miao2018universal}
(b) Electronic structure in the LQSGW+DMFT+SOC
with forcing Z$_{Se}$=1.38 (${\AA}$)
and $(\lambda+\Delta \lambda)_{Fe(d)}$=($\lambda_{1}+\Delta \lambda_{1}$)=60 meV
compared with ARPES
of Ref.\cite{miao2018universal}
(c) Electronic structure in the LQSGW+DMFT+SOC
with forcing Z$_{Se}$=1.38 (${\AA}$)
and $(\lambda+\Delta \lambda)_{Fe(d)}$=($\lambda_{1}+\Delta \lambda_{1}$)=120 meV
compared with ARPES
of Ref.\cite{miao2018universal}
(d) Electronic structure in the LQSGW+DMFT
with forcing Z$_{Se}$=1.38 (${\AA}$) compared with ARPES
of Ref.\cite{johnson2015spin}
(e) Electronic structure in the LQSGW+DMFT+SOC
with forcing Z$_{Se}$=1.38 (${\AA}$)
and $(\lambda+\Delta \lambda)_{Fe(d)}$=($\lambda_{1}+\Delta \lambda_{1}$)=60 meV
compared with ARPES
of Ref.\cite{johnson2015spin}
(f) Electronic structure in the LQSGW+DMFT+SOC
with forcing Z$_{Se}$=1.38 (${\AA}$)
and $(\lambda+\Delta \lambda)_{Fe(d)}$=($\lambda_{1}+\Delta \lambda_{1}$)=120 meV
compared with ARPES
of Ref.\cite{johnson2015spin}
\label{fig:138_inplane}
}
\end{figure}

\begin{figure}[t]
\includegraphics[width=\columnwidth]{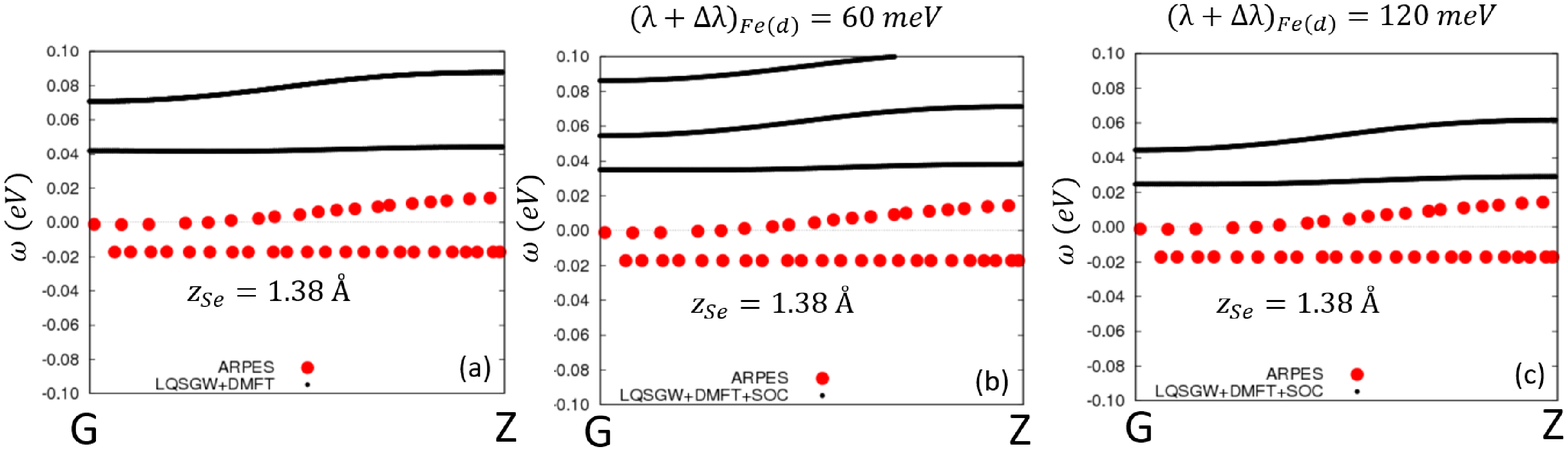}
\caption{(a) Electronic structure in the LQSGW+DMFT
with forcing Z$_{Se}$=1.38 (${\AA}$) compared with ARPES
of Ref.\cite{lohani2020band}
(b) Electronic structure in the LQSGW+DMFT+SOC
with forcing Z$_{Se}$=1.38 (${\AA}$)
and $(\lambda+\Delta \lambda)_{Fe(d)}$=($\lambda_{1}+\Delta \lambda_{1}$)=60 meV
compared with ARPES
of Ref.\cite{lohani2020band}
(c) Electronic structure in the LQSGW+DMFT+SOC
with forcing Z$_{Se}$=1.38 (${\AA}$)
and $(\lambda+\Delta \lambda)_{Fe(d)}$=($\lambda_{1}+\Delta \lambda_{1}$)=120 meV
compared with ARPES
of Ref.\cite{lohani2020band}
\label{fig:138_outplane}
}
\end{figure}

\begin{figure}[t]
\includegraphics[width=\columnwidth]{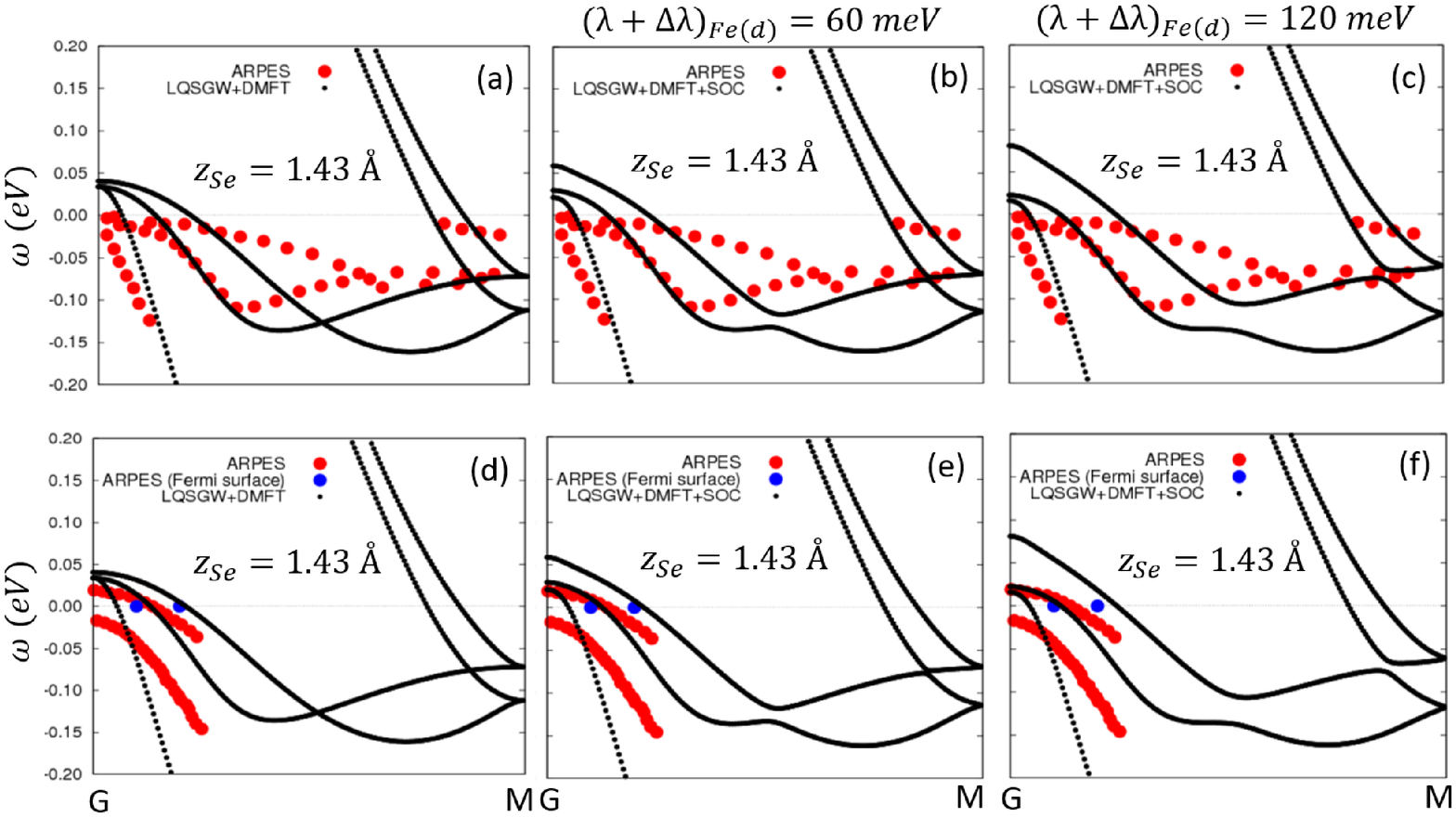}
\caption{(a) Electronic structure in the LQSGW+DMFT
with forcing Z$_{Se}$=1.43 (${\AA}$) compared with ARPES
of Ref.\cite{miao2018universal}
(b) Electronic structure in the LQSGW+DMFT+SOC
with forcing Z$_{Se}$=1.43 (${\AA}$)
and $(\lambda+\Delta \lambda)_{Fe(d)}$=($\lambda_{1}+\Delta \lambda_{1}$)=60 meV
compared with ARPES
of Ref.\cite{miao2018universal}
(c) Electronic structure in the LQSGW+DMFT+SOC
with forcing Z$_{Se}$=1.43 (${\AA}$)
and $(\lambda+\Delta \lambda)_{Fe(d)}$=($\lambda_{1}+\Delta \lambda_{1}$)=120 meV
compared with ARPES
of Ref.\cite{miao2018universal}
(d) Electronic structure in the LQSGW+DMFT
with forcing Z$_{Se}$=1.43 (${\AA}$) compared with ARPES
of Ref.\cite{johnson2015spin}
(e) Electronic structure in the LQSGW+DMFT+SOC
with forcing Z$_{Se}$=1.43 (${\AA}$)
and $(\lambda+\Delta \lambda)_{Fe(d)}$=($\lambda_{1}+\Delta \lambda_{1}$)=60 meV
compared with ARPES
of Ref.\cite{johnson2015spin}
(f) Electronic structure in the LQSGW+DMFT+SOC
with forcing Z$_{Se}$=1.43 (${\AA}$)
and $(\lambda+\Delta \lambda)_{Fe(d)}$=($\lambda_{1}+\Delta \lambda_{1}$)=120 meV
compared with ARPES
of Ref.\cite{johnson2015spin}
\label{fig:143_inplane}
}
\end{figure}

\begin{figure}[t]
\includegraphics[width=\columnwidth]{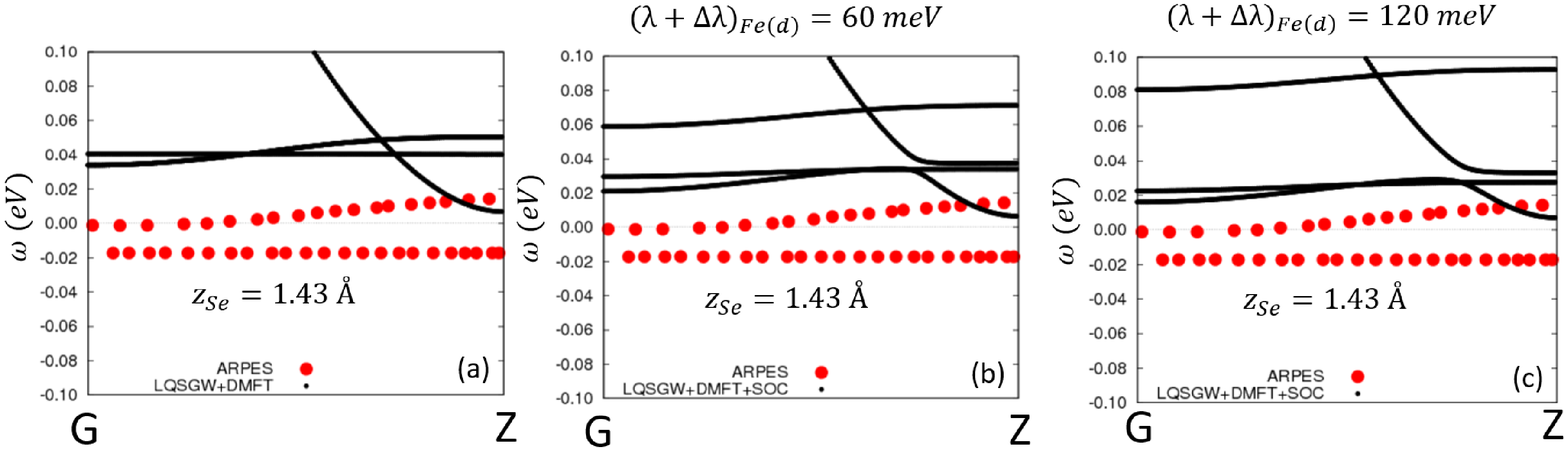}
\caption{(a) Electronic structure in the LQSGW+DMFT
with forcing Z$_{Se}$=1.43 (${\AA}$) compared with ARPES
of Ref.\cite{lohani2020band}
(b) Electronic structure in the LQSGW+DMFT+SOC
with forcing Z$_{Se}$=1.43 (${\AA}$)
and $(\lambda+\Delta \lambda)_{Fe(d)}$=($\lambda_{1}+\Delta \lambda_{1}$)=60 meV
compared with ARPES
of Ref.\cite{lohani2020band}
(c) Electronic structure in the LQSGW+DMFT+SOC
with forcing Z$_{Se}$=1.43 (${\AA}$)
and $(\lambda+\Delta \lambda)_{Fe(d)}$=($\lambda_{1}+\Delta \lambda_{1}$)=120 meV
compared with ARPES
of Ref.\cite{lohani2020band}
\label{fig:143_outplane}
}
\end{figure}

\begin{figure}[t]
\includegraphics[width=\columnwidth]{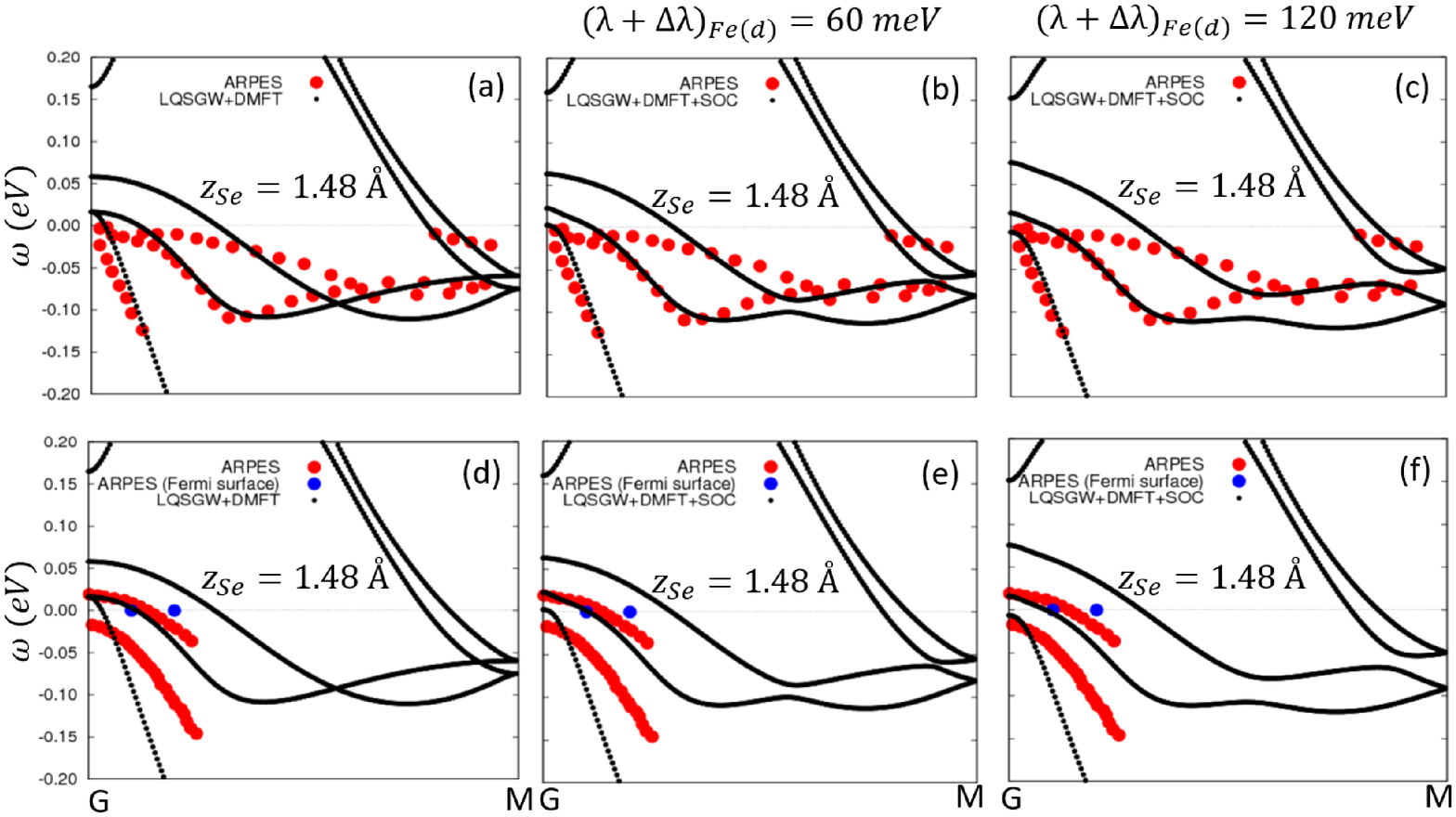}
\caption{(a) Electronic structure in the LQSGW+DMFT
with the crystal structure in the main text (Z$_{Se}$=1.48 (${\AA}$)) compared with ARPES
of Ref.\cite{miao2018universal}
(b) Electronic structure in the LQSGW+DMFT+SOC
with the crystal structure in the main text (Z$_{Se}$=1.48 (${\AA}$))
and $(\lambda+\Delta \lambda)_{Fe(d)}$=($\lambda_{1}+\Delta \lambda_{1}$)=60 meV
compared with ARPES
of Ref.\cite{miao2018universal}
(c) Electronic structure in the LQSGW+DMFT+SOC
with the crystal structure in the main text (Z$_{Se}$=1.48 (${\AA}$))
and $(\lambda+\Delta \lambda)_{Fe(d)}$=($\lambda_{1}+\Delta \lambda_{1}$)=120 meV
compared with ARPES
of Ref.\cite{miao2018universal}
(d) Electronic structure in the LQSGW+DMFT
with the crystal structure in the main text (Z$_{Se}$=1.48 (${\AA}$)) compared with ARPES
of Ref.\cite{johnson2015spin}
(e) Electronic structure in the LQSGW+DMFT+SOC
with the crystal structure in the main text (Z$_{Se}$=1.48 (${\AA}$))
and $(\lambda+\Delta \lambda)_{Fe(d)}$=($\lambda_{1}+\Delta \lambda_{1}$)=60 meV
compared with ARPES
of Ref.\cite{johnson2015spin}
(f) Electronic structure in the LQSGW+DMFT+SOC
with the crystal structure in the main text (Z$_{Se}$=1.48 (${\AA}$))
and $(\lambda+\Delta \lambda)_{Fe(d)}$=($\lambda_{1}+\Delta \lambda_{1}$)=120 meV
compared with ARPES
of Ref.\cite{johnson2015spin}
\label{fig:148_inplane}
}
\end{figure}

\begin{figure}[t]
\includegraphics[width=\columnwidth]{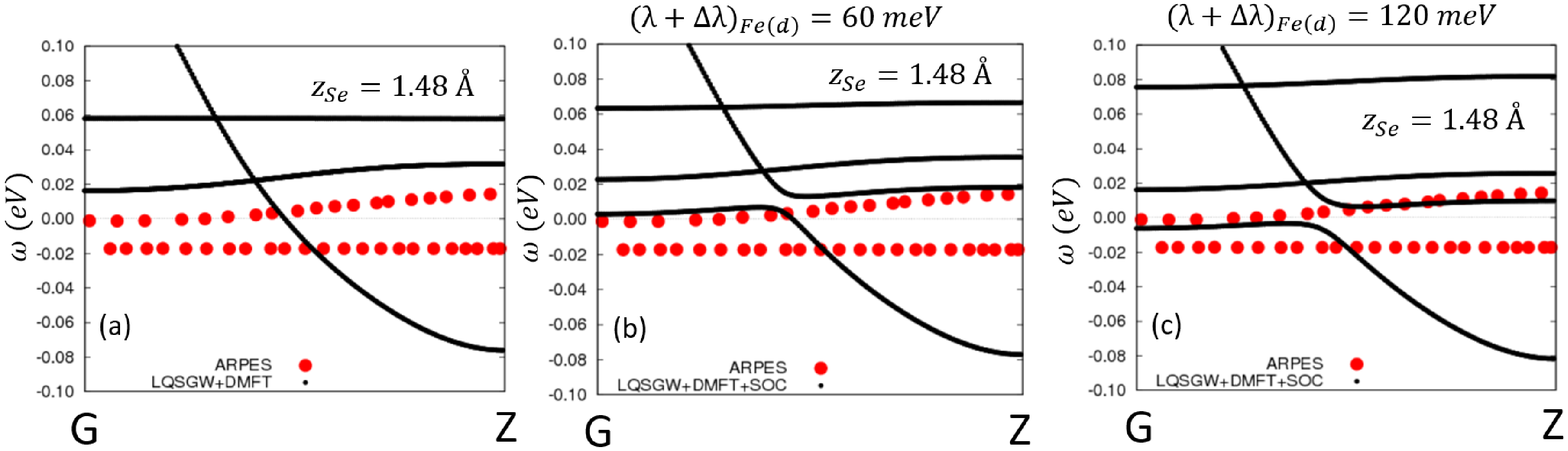}
\caption{(a) Electronic structure in the LQSGW+DMFT
with the crystal structure in the main text (Z$_{Se}$=1.48 (${\AA}$)) compared with ARPES
of Ref.\cite{lohani2020band}
(b) Electronic structure in the LQSGW+DMFT+SOC
with the crystal structure in the main text (Z$_{Se}$=1.48 (${\AA}$))
and $(\lambda+\Delta \lambda)_{Fe(d)}$=($\lambda_{1}+\Delta \lambda_{1}$)=60 meV
compared with ARPES
of Ref.\cite{lohani2020band}
(c) Electronic structure in the LQSGW+DMFT+SOC
with the crystal structure in the main text (Z$_{Se}$=1.48 (${\AA}$))
and $(\lambda+\Delta \lambda)_{Fe(d)}$=($\lambda_{1}+\Delta \lambda_{1}$)=120 meV
compared with ARPES
of Ref.\cite{lohani2020band}
\label{fig:148_outplane}
}
\end{figure}

Figure \ref{fig:138_inplane}, \ref{fig:138_outplane},
\ref{fig:143_inplane}, \ref{fig:143_outplane},
\ref{fig:148_inplane}, and \ref{fig:148_outplane}
illustrate the spin-orbit coupling (Fe($d$)) and Chalcogen height $Z_{Se}$ dependent
electronic structure of FST.
It is shown that
the agreement with the ARPES experiments
\cite{miao2018universal,johnson2015spin,lohani2020band}
is made for the condition of $\lambda_{1}$+$\Delta \lambda_{1}$=120 meV,
$Z_{Se}$=1.48 ${\AA}$, from the present LQSGW+DMFT+SOC method.
We note that this $\lambda_{1}$+$\Delta \lambda_{1}$
value of 120 meV is larger than the spin-orbit coupling
constant of Fe($d$) from LDA+SOC and LQSGW+SOC (Table\ref{Parameters}).
This enhancement of the spin-orbit coupling
realized in ARPES experiments \cite{miao2018universal,johnson2015spin,lohani2020band} demonstrates the importance of the orbitally off-diagonal
self-energy ($\Delta \lambda_{1}$) in the realistic description of the electronic structure
of FeSe$_{0.5}$Te$_{0.5}$.
This factor is further verified from the band energy levels of $\alpha$, $\alpha'$, and $\beta$
from $xz/yz$ and $xy$ orbitals at $k=\Gamma$, as shown in Table.\ref{table:SOC}.
We consider the effective Hamiltonian at $\Gamma$
for the three orbital model as shown in Eq.\ref{eq:Hamiltonian_Gamma}.

\begin{eqnarray} \label{eq:Hamiltonian_Gamma}
&H=&\lambda_{eff}(l \cdot s)_{t_{2g}} + \frac{1}{2}\varepsilon_{t,eff}(c_{xy,\uparrow}^{\dag}c_{xy,\uparrow}+c_{xy,\downarrow}^{\dag}c_{xy,\downarrow}-c_{xz,\uparrow}^{\dag}c_{xz,\uparrow}-c_{xz,\downarrow}^{\dag}c_{xz,\downarrow}-c_{yz,\uparrow}^{\dag}c_{yz,\uparrow}-c_{yz,\downarrow}^{\dag}c_{yz,\downarrow})
\end{eqnarray}

$\varepsilon_{t,eff}$ is the effective tetragonal splitting between $xz/yz$ and $xy$,
and $\lambda_{eff}$ is the effective spin-orbit coupling of the three orbital model.
The $\lambda_{eff}(l \cdot s)_{t_{2g}}$ of Eq.\ref{eq:Hamiltonian_Gamma} is given by Eq.\ref{eq:Hamiltonian_SOC},
in the order of $xz,\uparrow$, $yz,\uparrow$, $xy,\uparrow$, $xz,\downarrow$, $yz,\downarrow$, and $xy,\downarrow$ states.

\begin{eqnarray} \label{eq:Hamiltonian_SOC}
\lambda_{eff}(l \cdot s)_{t_{2g}}=\begin{pmatrix}
0 & -i\frac{\lambda_{eff}}{2} & 0 & 0 & 0 & i\frac{\lambda_{eff}}{2} \\
i\frac{\lambda_{eff}}{2} & 0 & 0 & 0 & 0 & -\frac{\lambda_{eff}}{2} \\
0 & 0 & 0 & -i\frac{\lambda_{eff}}{2} & \frac{\lambda_{eff}}{2} & 0 \\
0 & 0 & i\frac{\lambda_{eff}}{2} & 0 & i\frac{\lambda_{eff}}{2} & 0 \\
0 & 0 & \frac{\lambda_{eff}}{2} & -i\frac{\lambda_{eff}}{2} & 0 & 0 \\
-i\frac{\lambda_{eff}}{2} & -\frac{\lambda_{eff}}{2} & 0 & 0 & 0 & 0
\end{pmatrix}
\end{eqnarray}

The eigenvalue of three Kramer's doublets, A, B, and C for the Hamiltonian of Eq.\ref{eq:Hamiltonian_Gamma}
are presented in Eq.\ref{eq:eig}.

\begin{eqnarray} \label{eq:eig}
&\epsilon_{A}=&\frac{1}{4}(\sqrt{4\varepsilon^{2}_{t,eff}-4\varepsilon_{t,eff}\lambda_{eff}+9\lambda_{eff}^{2}}+2\varepsilon_{t,eff}+\lambda_{eff}) \nonumber\\
&\epsilon_{B}=&\frac{1}{4}(-\sqrt{4\varepsilon^{2}_{t,eff}-4\varepsilon_{t,eff}\lambda_{eff}+9\lambda_{eff}^{2}}+2\varepsilon_{t,eff}+\lambda_{eff}) \nonumber\\
&\epsilon_{C}=&-\frac{1}{2}\lambda_{eff}
\end{eqnarray}

In the LQSGW+DMFT+SOC study of FeSe$_{0.5}$Te$_{0.5}$, there are three Kramer's doublets at $k=\Gamma$ from the present structural model.
The effective Hamiltonian at $\Gamma$
for the three orbital model is in Eq.\ref{eq:Hamiltonian_Gamma} and Eq.\ref{eq:Hamiltonian_SOC}.
Three Kramer's doublets, A, B, and C in Eq.\ref{eq:eig}, the solution of Eq.\ref{eq:Hamiltonian_Gamma}, are identified as $\beta$, $\alpha$, and $\alpha'$ bands respectively.
The model Hamiltonian of Eq.\ref{eq:Hamiltonian_Gamma} have a validity
when those three bands are separated from other bands
in energy with respect to the energy scale of spin-orbit coupling,
which is the case for the present LQSGW+DMFT+SOC result on FST ($k$=$\Gamma$).
The splitting energy of the $\alpha$ and $\alpha'$ bands, as shown in Table.\ref{table:SOC}, is in good agreement with experimental values from Refs.\cite{lohani2020band,johnson2015spin}.
With the relation of $\lambda_{eff}$=$Z_{imp}(\lambda_{1}+\Delta \lambda_{1})$,
the consistency in Table\ref{table:SOC} implies the existence of the enhancement of the spin-orbit coupling
from the orbitally off-diagonal self-energy ($\Delta \lambda_{1}$).
Notably, the same energy scale of $\lambda_{eff}$ and $\varepsilon_{t,eff}$ in FeSe$_{0.5}$Te$_{0.5}$ indicates a sizable contribution of the $xy$ orbital. This is in contrast to the case of LiFeAs, where the energy scale of $\lambda_{eff}$ is in the perturbative regime with respect to $\varepsilon_{t,eff}$, thus having a small effect of the $xy$ orbital to the splitting energy of $\alpha$ and $\alpha'$ bands.

\begin{table*}[h!]
\caption{The effective spin-orbit coupling $\lambda_{eff}$ and tetragonal field $\varepsilon_{t,eff}$
in the LQSGW+DMFT+SOC. The difference in eigenvalues, $\epsilon_{B}-\epsilon_{C}$, for
the splitting energy of $\alpha$ and $\alpha'$ bands the LQSGW+DMFT+SOC at $k=\Gamma$ (from Eq.\ref{eq:eig}) is compared with experimental values
from ARPES\cite{lohani2020band,johnson2015spin}. (All quantities in meV unit)
}
\begin{tabular}{|| c | c | c | c ||}
\hline
 $\lambda_{eff}$(LQSGW+DMFT+SOC)=$Z_{imp}(\lambda_{1}+\Delta \lambda_{1})$  & $\varepsilon_{t,eff}$(LQSGW+DMFT) & $\epsilon_{B}-\epsilon_{C}$=$\epsilon_{\alpha}-\epsilon_{\alpha'}$ & $\epsilon_{\alpha}-\epsilon_{\alpha'}$ (ARPES)\cite{lohani2020band,johnson2015spin} \\
\hline
 42 & 42 & 21 & 17-30 \\
\hline
\end{tabular}
\label{table:SOC}
\end{table*}

\section{IV. Detailed discussion on the Z$_{2}$ bulk topology and the orbital selective Mott phase}

\begin{figure}[t]
\includegraphics[width=\columnwidth]{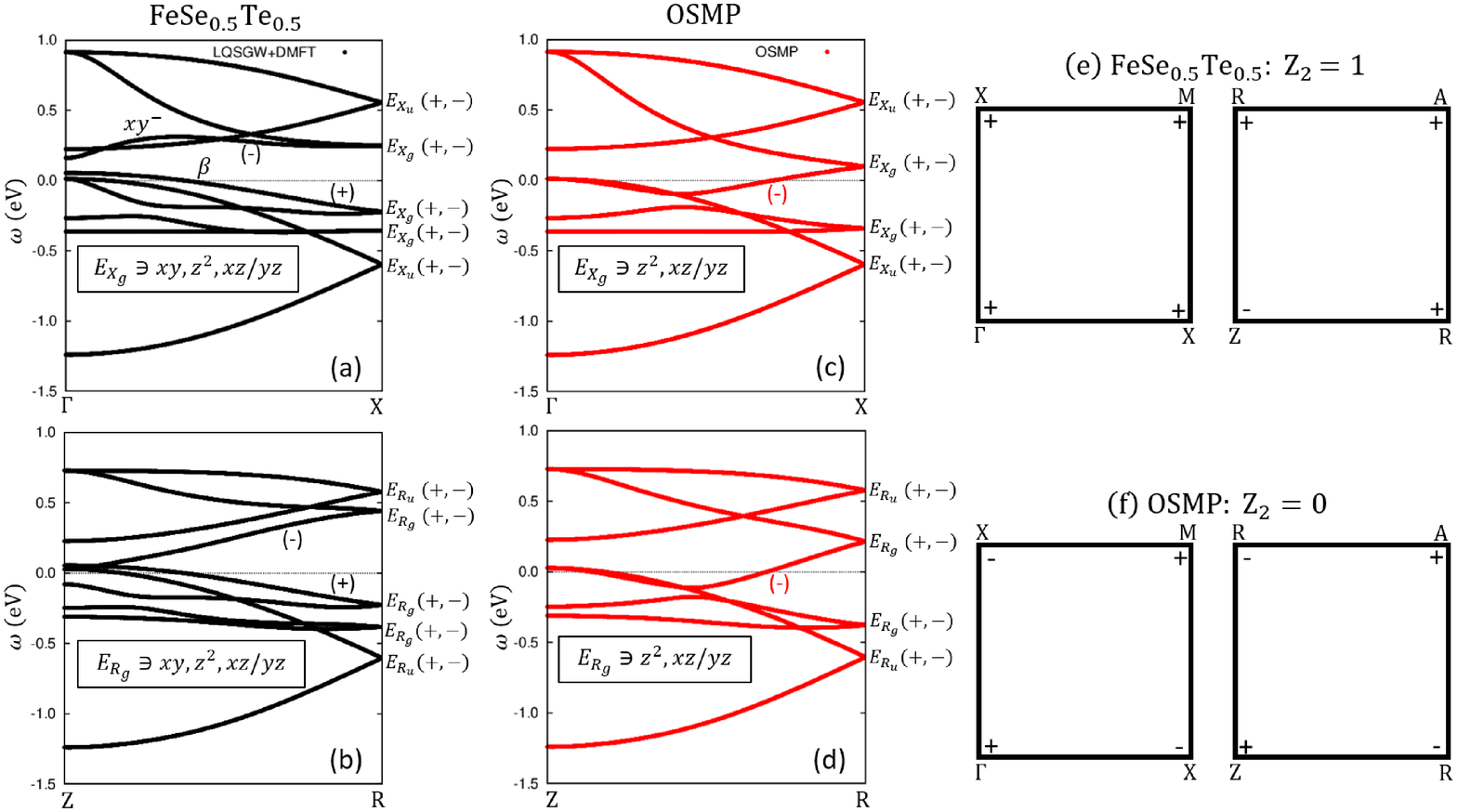}
\caption{(a) and (b) Band structure of FST from the LQSGW+DMFT framework in
the $\Gamma$-X and the Z-R momentum paths, respectively.
The band characterization from the symmetry, $E_{k_{g}}$
and $E_{k_{u}}$, is denoted at $k$=X and R.
At X and R, each of these bands has a degeneracy
of even and odd parities.
Consistent with Ref.\cite{cvetkovic2013space},
$E_{(X,R)_{g}}$ bands have Fe($d$) orbital character
of $xy$, $xz/yz$, and $z^{2}$.
The $\beta$ (even parity) and $xy^{-}$ (odd parity)
bands are denoted.
(c) and (d) Band structure in the orbital selective Mott phase (OSMP)
(by forcing $Z_{xy}=0$ in the LQSGW+DMFT results)
in the $\Gamma$-X and the Z-R momentum paths, respectively.
We removed isolated $xy$ orbital driven bands,
$\beta$ and $xy^{-}$, merged to the flat band
at the chemical potential.
The band characterization from the symmetry, $E_{k_{g}}$
and $E_{k_{u}}$, is denoted at $k$=X and R.
At X and R, each of these bands has a degeneracy
of even and odd parities.
From the orbital selective Mott phase,
$E_{(X,R)_{g}}$ bands have Fe($d$) orbital character
of $xz/yz$ and $z^{2}$ without $xy$ component.
The odd parity band with the emergence of
the new Fermi surface (denoted as (-)) at X point gives rise
to the switching of the parity in $\Gamma$-X
and Z-R momentum paths, respectively.
(e) and (f) Parity at the time reversal
invariant momentum and the Z$_{2}$ invariant
for (i) the FST in the LQSGW+DMFT+SOC
and (ii) the orbital selective Mott phase, respectively.
\label{fig:OSMP_Z2}
}
\end{figure}

Fig.\ref{fig:OSMP_Z2} illustrates the relationship between
the orbital selective Mott phase and Z$_{2}$ bulk topology in FST.
As demonstrated in Ref.\cite{huang2020low}, an increase in the Te ratio in FST leads to an orbital selective Mott transition, resulting in the emergence of a new Fermi surface sheet at the X point, with a hybridized orbital character of $xz/yz$ and $z^{2}$. Our calculations of the electronic structure in the orbital selective Mott phase corroborate the results of Ref.\cite{huang2020low}, showing that the switching of the parity in $\Gamma$-X and Z-R momentum paths corresponds to the emergence of the new Fermi surface sheet. This transition results in a trivial Z$_{2}$ topology.

Fig.\ref{fig:OSMP_Z2}(a and b) present the electronic structure of FST obtained from the LQSGW+DMFT framework in the $\Gamma$-X and Z-R momentum paths. Consistent with Ref.\cite{cvetkovic2013space}, five Fe($d$) dominant bands are characterized as three $E_{(X,R)_{g}}$ bands and two $E_{(X,R)_{u}}$ bands at $k$=X and R. Ref.\cite{cvetkovic2013space} further revealed that the three $E_{(X,R)_{g}}$ bands at $k$=X and R have a hybridized orbital character of $xy$, $xz/yz$, and $z^{2}$, which suggests that band characterization from the orbital character is problematic at these momentums. Additionally, the even parity $\beta$ band is connected to the topmost occupied even parity $E_{(X)_{g}}$ band at X point, while the odd parity $xy^{-}$ band is connected to the lowermost unoccupied odd parity $E_{(X)_{g}}$ band at X point.

Fig.~\ref{fig:OSMP_Z2}(c and d) present the electronic structure of FST in the orbital selective Mott phase, wherein $Z_{xy}$ (the quasiparticle residue of the $xy$ orbital) is set to zero. As a result of this transition, the $\beta$ and the $xy^{-}$ bands become incoherent and isolated, leading to the removal of one of the $E_{(X,R)_{g}}$ bands. Furthermore, the topmost occupied odd parity $E_{(X,R)_{g}}$ band merges with the lowermost unoccupied even parity $E_{(X,R)_{g}}$ band at X and R points, thereby preserving the inversion symmetry driven degeneracy of parity and isolating the $xy$ driven bands. This switch in parity is the cause of the new Fermi surface sheet at $k$=X observed in Ref.~\cite{huang2020low}. The $xz/yz$ and the $z^{2}$ hybridized orbital character of this band is consistent with the space group analysis in Ref.~\cite{cvetkovic2013space}.

The parity of band at the time reversal invariant momentum is shown in Fig.\ref{fig:OSMP_Z2}(e and f) for the FST ($x\approx0.5$) and the orbital selective Mott phase, respectively. As discussed in the main text, the FST ($x\approx0.5$) exhibits a non-trivial Z$_{2}$ invariant of bulk, due to the band inversion in the $\Gamma$-Z momentum path (Fig.\ref{fig:OSMP_Z2}(e)). With the orbital selective Mott transition, the band inversion in the $\Gamma$-Z momentum path is removed and the parity at X and R point is odd (Fig.\ref{fig:OSMP_Z2}(c and d)). However, due to the tetragonal symmetry, the bulk Z$_{2}$ invariant of this system in the orbital selective Mott phase is trivial (Fig.\ref{fig:OSMP_Z2}(f)).

\section{V. Electrons Quasiparticle Effective  Hamiltonian}

\begin{figure}[t]
\includegraphics[width=\columnwidth]{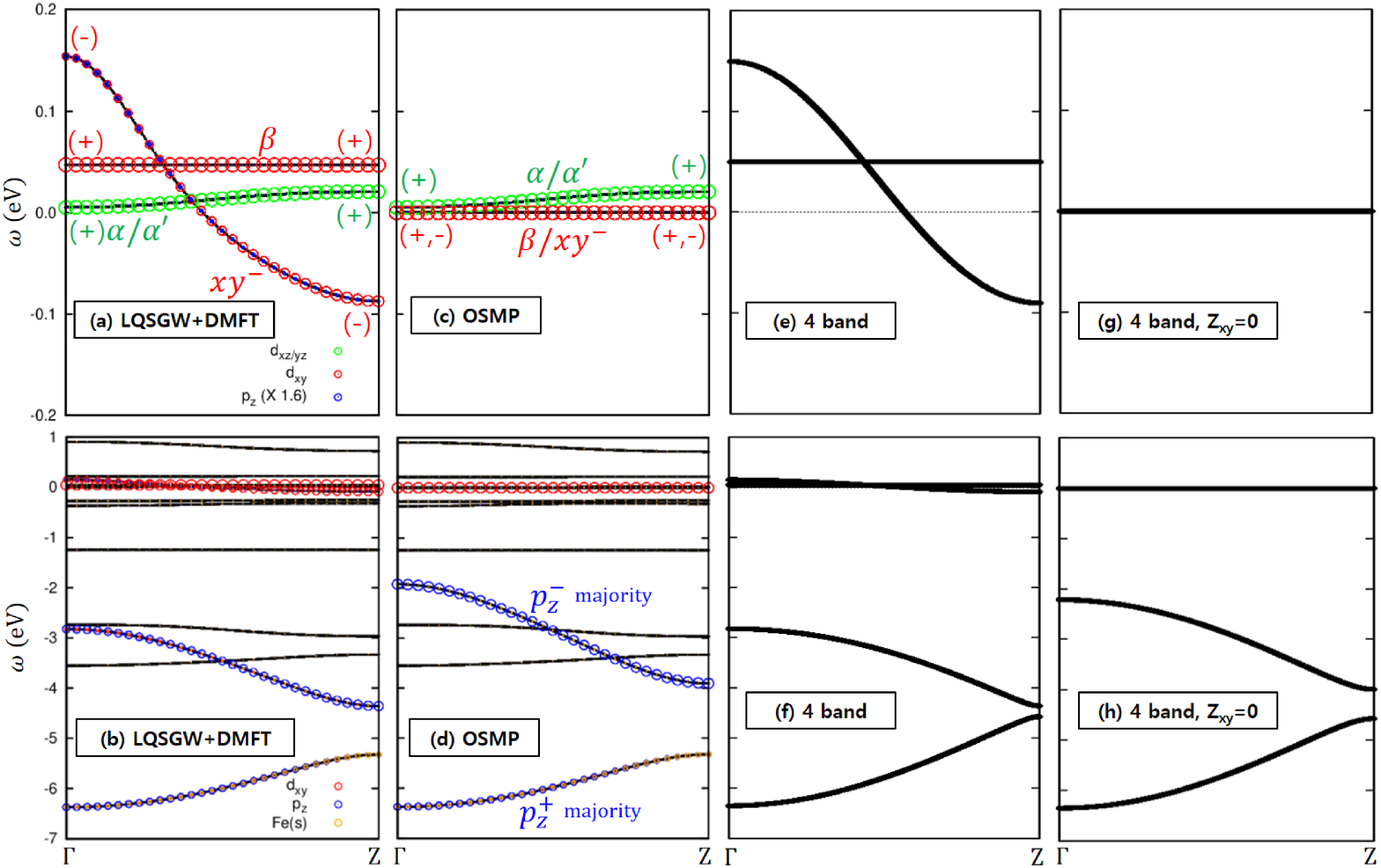}
\caption{(a) Electronic structures in the LQSGW+DMFT framework
, with the narrow energy range near the Fermi level.
The size of red, green, and blue circles indicates
Fe($d_{xy}$), Fe($d_{xz/yz}$), and Se($p_{z}$) orbitals contributions.
The size of blue circles for the Se($p_{z}$) orbital
is multiplied by the factor of 1.6.
(b) Electronic structures in the LQSGW+DMFT framework
, with the wide energy range.
The size of red, blue, and orange circles indicates
Fe($d_{xy}$), Se($p_{z}$), and Fe($s$) orbitals contributions.
(c) Electronic structures in the orbital selective Mott phase
(by forcing $Z_{xy}$ to zero for the LQSGW+DMFT Hamiltonian)
, with the narrow energy range near the Fermi level.
The size of red, green, and blue circles indicates
Fe($d_{xy}$), Fe($d_{xz/yz}$), and Se($p_{z}$) orbitals contributions.
The size of blue circles for the Se($p_{z}$) orbital
is multiplied by factor of 1.6.
(d) Electronic structures in the orbital selective Mott phase
(by forcing $Z_{xy}$ to zero for the LQSGW+DMFT Hamiltonian)
, with the wide energy range.
The size of red, blue, and orange circles indicates
Fe($d_{xy}$), Se($p_{z}$), and Fe($s$) orbitals contributions.
(e) Electronic structures from the effective quasiparticle
Hamiltonian in the main text with parameters in Table~\ref{table:4band}
, with the narrow energy range near the Fermi level.
(f) Electronic structures from the effective quasiparticle
Hamiltonian in the main text with parameters in Table~\ref{table:4band}
, with the wide energy range.
(g) Electronic structures from the effective quasiparticle
Hamiltonian forcing $Z_{xy}=0$ in the main text with parameters in Table~\ref{table:4band}
, with the narrow energy range near the Fermi level.
(h) Electronic structures from the effective quasiparticle
Hamiltonian forcing $Z_{xy}=0$ in the main text with parameters in Table~\ref{table:4band}
, with the wide energy range.
\label{fig:4band}
}
\end{figure}

Fig.~\ref{fig:4band} compares the electronic structure obtained from the LQSGW+DMFT framework to that derived from a four-band quasiparticle effective Hamiltonian
(see main text), which has dominant Fe($d_{xy}$) and Se($p_{z}$) orbital character. The comparison implies that the model captures both the essential electronic structure of the Fe($d_{xy}$) and Se($p_{z}$) bands, as well as the topological transition from the non-trivial to trivial Z$_{2}$ phase in the orbital selective Mott phase by forcing $Z_{xy}$ to zero. It is important to note that the tight-binding parameters used in the effective Hamiltonian in Table~\ref{table:4band} differ from those in Fig.~\ref{fig:tightbinding_projection}, owing to contributions from longer range hopping and the hybridization with other dispersive orbitals.

\begin{table*}[h!]
\caption{The tight-binding parameters of the electrons quasiparticle  effective Hamiltonian
in the main text and the Fig.\ref{fig:4band} (All quantities in eV unit).
}
\begin{tabular}{|| c | c | c | c | c | c ||}
\hline
 $\tilde{\epsilon}_{xy}$ & $\tilde{\epsilon}_{p_{z}}$ & $\tilde{t}_{p_{z}}$ & $\tilde{t}_{xy}$ & $\tilde{t}_{1}$ & $\tilde{t}_{2}$ \\
\hline
 -0.204 & -4.282 & 0.295 & 0.064 & 0.300 & 0.223 \\
\hline
\end{tabular}
\label{table:4band}
\end{table*}

\section{VI. Crystal structure dependent electronic structures and tight-binding parameters}

\subsection{A. Crystal structure model dependency in the DFT}

\begin{figure}[t]
\includegraphics[width=\columnwidth]{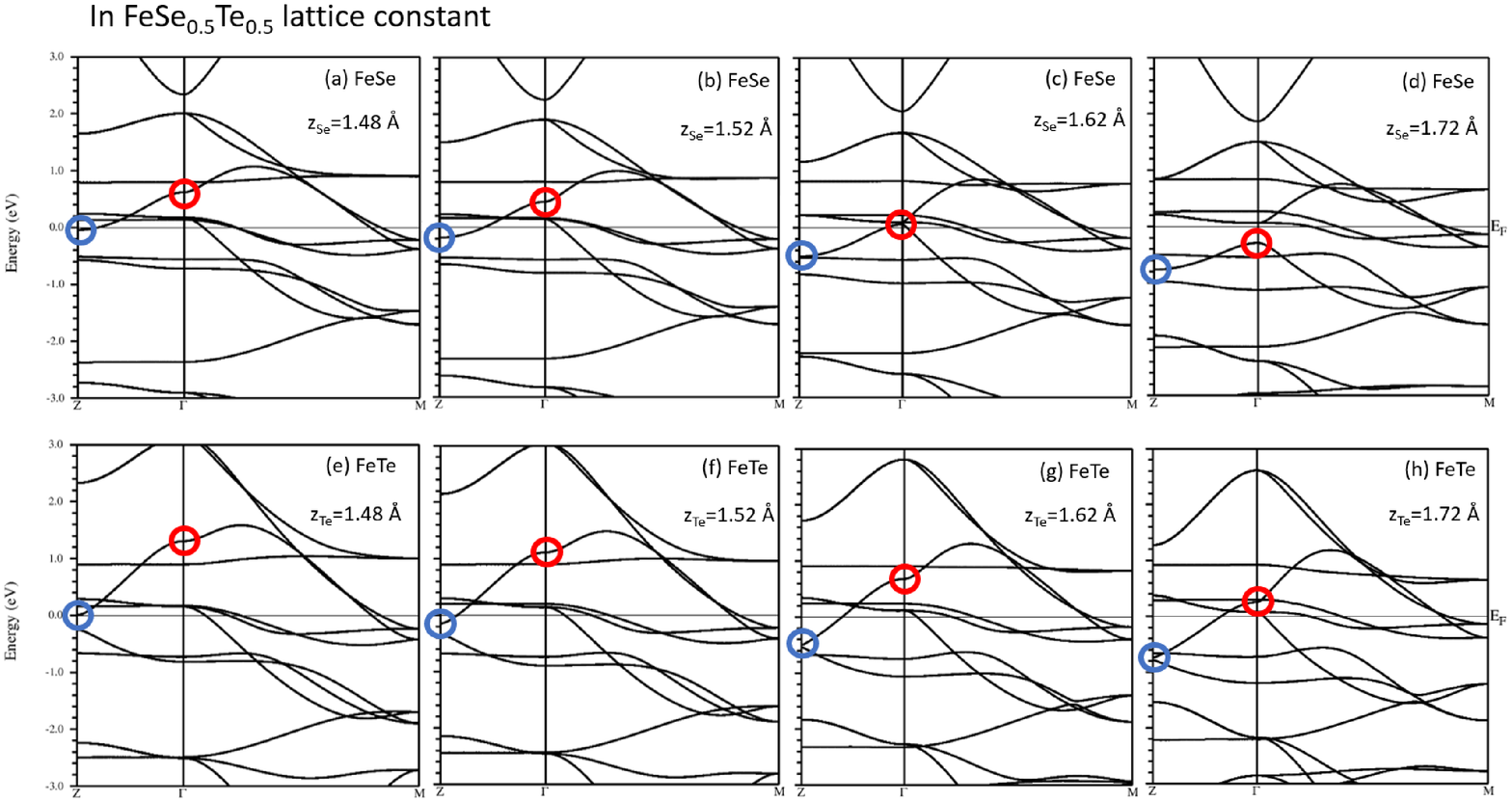}
\caption{Comparison of LDA band structures along the Z-$\Gamma$-M  point line
using the fixed lattice  parameters (taken from the experiment) of  FST.  We  vary the chalcogen type (Se or Te) and its height.  The center of the $xy^-$ band strongly varies with the chalcogen height. The type of chalcogen modifies the  $xy^-$  bandwidth.
(a), (b), (c), and (d) are
FeSe chemical formula
with the lattice constant of FeSe$_{0.5}$Te$_{0.5}$ \cite{li2009first,tegel2010crystal},
(a) Z$_{Se}$=1.48 ${\AA}$, (b) Z$_{Se}$=1.52 ${\AA}$, (c) Z$_{Se}$=1.62 ${\AA}$, and (d) Z$_{Se}$=1.72 ${\AA}$.
(e), (f), (g), and (h) are
FeTe chemical formula
with the lattice constant of FeSe$_{0.5}$Te$_{0.5}$ \cite{li2009first,tegel2010crystal},
(e) Z$_{Te}$=1.48 ${\AA}$, (f) Z$_{Te}$=1.52 ${\AA}$, (g) Z$_{Te}$=1.62 ${\AA}$, and (h) Z$_{Te}$=1.72 ${\AA}$.
Red and blue circles indicate the top of the $xy^{-}$ band at $k$=$\Gamma$ and the bottom of the $xy^{-}$ band
at $k$=Z, respectively.
\label{fig:chemical}
}
\end{figure}

Fig.\ref{fig:chemical} depicts a comparison of the local density approximation (LDA) electronic structures of the FeTe and FeSe chemical formulas in the lattice constant of FeSe$_{0.5}$Te$_{0.5}$ with varying chalcogen heights from 1.48 ${\AA}$ to 1.72 ${\AA}$, the chalcogen heights range observed in FeSe$_{0.5}$Te$_{0.5}$.\cite{li2009first,tegel2010crystal}. It was observed that with the enhancement of chalcogen heights, the $xy^{-}$ band in the $\Gamma$-Z momentum path was shifted down for both FeTe and FeSe chemical formulas. In addition, the bottom energies of the $xy^{-}$ band at Z were similar in value for both FeTe and FeSe chemical formulas with the same chalcogen height. However, the top energy of the $xy^{-}$ band at $\Gamma$ was found to be higher in FeTe than in FeSe in the same chalcogen height, thereby indicating that the substitution of Te for Se increases the $xy^{-}$ band dispersion in the $\Gamma$-Z momentum path.

\begin{figure}[t]
\includegraphics[width=12cm]{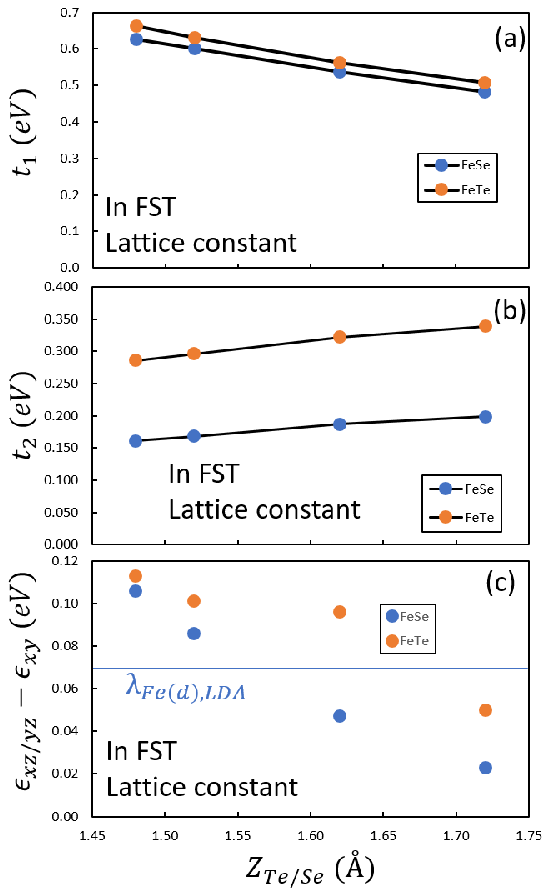}
\caption{
\label{fig:parameters}
(a) The Te/Se heights versus
the nearest neighboring hopping between $xy$ and $p_{z}$
from maximally localized Wannier orbitals in the LDA level.
(b) The Te/Se heights versus
the nearest out-of-plane hopping of chalcogen $p_{z}$ orbitals,
from maximally localized Wannier orbitals in the LDA level.
(c) The Te/Se heights
versus the difference in energy levels of $xz/yz$ and $xy$
maximally localized Wannier orbitals in the LDA level.
The energy of the spin-orbit coupling in the Fe($d$)
orbital for the LDA+SOC is compared.
We use the FeSe and FeTe chemical formulas with the lattice constant of
FeSe$_{0.5}$Te$_{0.5}$\cite{li2009first,tegel2010crystal}.
}
\end{figure}

Fig.\ref{fig:parameters} presents the chalcogen heights dependence of
(i) the nearest neighboring hopping between $xy$ and $p_{z}$, and (ii) the nearest out-of-plane hopping between chalcogen $p_{z}$ orbitals, and (iii) the difference of $xz/yz$ and $xy$ orbital energy level, for the FeSe and FeTe chemical formula model with the lattice constant of FeSe$_{0.5}$Te$_{0.5}$ \cite{li2009first,tegel2010crystal}.
It is revealed that the strong $xy$-$p_{z}$ hopping
from the reduction of the chalcogen heights gives rise to the upshift of the center of the dispersive $xy^{-}$ band in $\Gamma$ - Z from Ref.\cite{wang_TopologicalNatureMathrmFeSe_2015},
as shown in Fig.\ref{fig:chemical}.
The substitution of Se for Te leads to the reduction of the out-of-plane hopping of the chalcogen $p_{z}$ orbital and the corresponding reduction of the $xy^{-}$ band dispersion in $\Gamma$-$Z$, as shown in Fig.\ref{fig:chemical}.
We emphasize that this substitution brings a minor variation in the $xy$-$p_{z}$ hopping.

We also show that the enhancement of the chalcogen heights from 1.48 (${\AA}$) to 1.72 (${\AA}$) gives rise to a slight reduction of the difference in energy levels of the axial orbital-$xz/yz$ and planar orbital-$xy$, around 63 meV for the FeTe model and 83 meV for the FeSe model. The scale of the difference in energy levels is approximately 100 meV, which is comparable to the energy scale of the spin-orbit coupling in the LDA level, 70 meV. This comparable energy scale implies that in the LDA, $xz/yz$ and $xy$ orbitals should be considered together when including the spin-orbit coupling.

\subsection{B. Chalcogen height dependency in the LQSGW+DMFT}

\begin{figure}[t]
\includegraphics[width=\columnwidth]{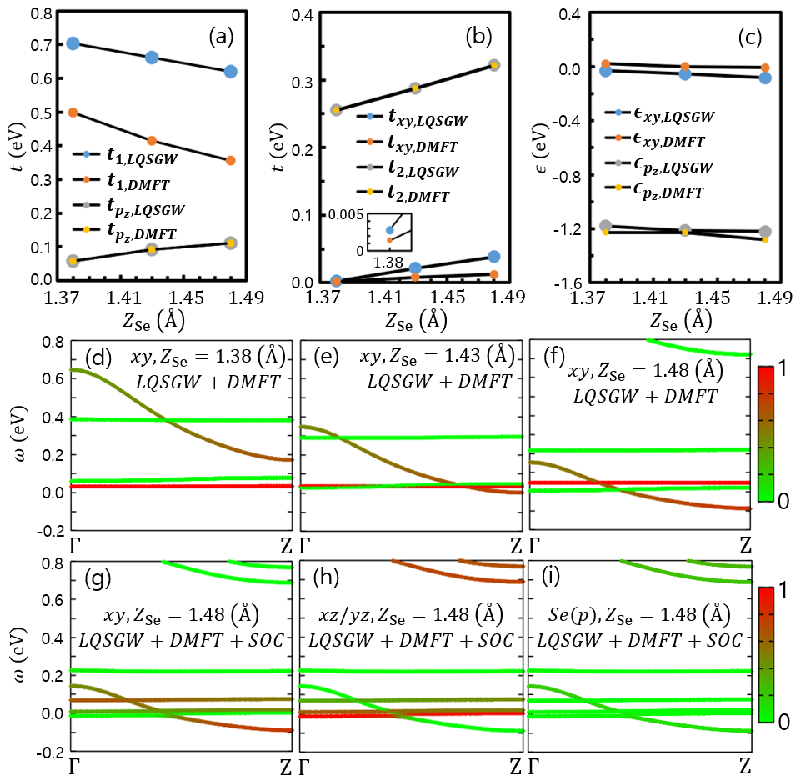}
\caption{(a), (b), and (c) Elements of Hamiltonian from the
maximally localized Wannier function in the LQSGW (LQSGW)
and LQSGW+DMFT (DMFT).
(d), (e), and (f) Se height ($Z_{Se}$) dependent
quasiparticle dispersions in the $\Gamma$-Z $k$ point line
from the LQSGW+DMFT,
(d) $Z_{Se}$=1.38 ($\AA$),
(e) $Z_{Se}$=1.43 ($\AA$), and
(f) $Z_{Se}$=1.48 ($\AA$), respectively.
The color scale from green to red
indicates the $xy$ orbital contribution to each band from 0 to 1.
(g), (h), and (i)
Quasiparticle dispersion in the $\Gamma$-Z $k$ point line
from the LQSGW+DMFT+SOC at $Z_{Se}$=1.48 ($\AA$)
for (g) $xy$ orbital, (h) $xz/yz$ orbitals, and (i) Se($p$) orbitals
contributions in the color scale, respectively.
\label{fig:tightbinding_projection}
}
\end{figure}

Fig. \ref{fig:tightbinding_projection} (a, b, and c) present the chalcogen height ($Z_{Se}$) dependent elements of Hamiltonian in the maximally localized Wannier function (MLWF) from the LQSGW and LQSGW+DMFT frameworks, starting from the average structure for FST. It is shown that the dynamical correlation in the LQSGW+DMFT substantially renormalizes the $t_{1}$ term in comparison to the LQSGW. Additionally, three major $Z_{Se}$ dependent variations of the tight-binding parameter are recognized for the LQSGW+DMFT result: (i) a reduction of the $t_{1}$ term, (ii) an enhancement of the $t_{2}$ term, and (iii) an enhancement of the $t_{p_{z}}$ term due to the enhancement of $Z_{Se}$. Fig. \ref{fig:tightbinding_projection} (d, e, and f) present the $Z_{Se}$ dependent band structure in the LQSGW+DMFT; due to the large $\epsilon_{xy}-\epsilon_{p_{z}}$ value in Fig. \ref{fig:tightbinding_projection} (c), the major feature for the $Z_{Se}$ dependence of the $xy^{-}$ band is explained from the variation of the $t_{1}$ term. This reduction of the $t_{1}$ term results in (i) a down shifted energy level of the $xy^{-}$ band and (ii) a less dispersive $xy^{-}$ band in $\Gamma-$Z, as shown in Fig. \ref{fig:tightbinding_projection} (d, e, and f). Furthermore, Fig. \ref{fig:tightbinding_projection} (g, h, and i) presents the band dispersion in the LQSGW+DMFT+SOC in the present structure of FST ($Z_{Se}$=1.48 ($\AA$)), with orbital weights for $xy$, $xz/yz$, and $p_{z}$. It is demonstrated that the $p_{z}$ orbital has a minor contribution to the $xy^{-}$ band.
We note that the tight-binding parameter ($t$,$\epsilon$) in Fig. \ref{fig:tightbinding_projection} (a, b, and c) is different from the effective parameter ($\tilde{t}$,$\tilde{\epsilon}$)  in Table~\ref{table:4band},
due to the factor that the effective parameter in Table~\ref{table:4band}
absorbs longer range hopping and other dispersive orbitals contribution.

%


\bibliography{refs_FeSeTe,zotero}

\begin{thebibliography}{60}%
\makeatletter
\providecommand \@ifxundefined [1]{%
 \@ifx{#1\undefined}
}%
\providecommand \@ifnum [1]{%
 \ifnum #1\expandafter \@firstoftwo
 \else \expandafter \@secondoftwo
 \fi
}%
\providecommand \@ifx [1]{%
 \ifx #1\expandafter \@firstoftwo
 \else \expandafter \@secondoftwo
 \fi
}%
\providecommand \natexlab [1]{#1}%
\providecommand \enquote  [1]{``#1''}%
\providecommand \bibnamefont  [1]{#1}%
\providecommand \bibfnamefont [1]{#1}%
\providecommand \citenamefont [1]{#1}%
\providecommand \href@noop [0]{\@secondoftwo}%
\providecommand \href [0]{\begingroup \@sanitize@url \@href}%
\providecommand \@href[1]{\@@startlink{#1}\@@href}%
\providecommand \@@href[1]{\endgroup#1\@@endlink}%
\providecommand \@sanitize@url [0]{\catcode `\\12\catcode `\$12\catcode
  `\&12\catcode `\#12\catcode `\^12\catcode `\_12\catcode `\%12\relax}%
\providecommand \@@startlink[1]{}%
\providecommand \@@endlink[0]{}%
\providecommand \url  [0]{\begingroup\@sanitize@url \@url }%
\providecommand \@url [1]{\endgroup\@href {#1}{\urlprefix }}%
\providecommand \urlprefix  [0]{URL }%
\providecommand \Eprint [0]{\href }%
\providecommand \doibase [0]{https://doi.org/}%
\providecommand \selectlanguage [0]{\@gobble}%
\providecommand \bibinfo  [0]{\@secondoftwo}%
\providecommand \bibfield  [0]{\@secondoftwo}%
\providecommand \translation [1]{[#1]}%
\providecommand \BibitemOpen [0]{}%
\providecommand \bibitemStop [0]{}%
\providecommand \bibitemNoStop [0]{.\EOS\space}%
\providecommand \EOS [0]{\spacefactor3000\relax}%
\providecommand \BibitemShut  [1]{\csname bibitem#1\endcsname}%
\let\auto@bib@innerbib\@empty
\bibitem [{\citenamefont
  {DiVincenzo}(2000)}]{divincenzo_PhysicalImplementationQuantum_2000}%
  \BibitemOpen
  \bibfield  {author} {\bibinfo {author} {\bibfnamefont {D.~P.}\ \bibnamefont
  {DiVincenzo}},\ }\bibfield  {title} {\bibinfo {title} {The {{Physical
  Implementation}} of {{Quantum Computation}}},\ }\href
  {https://doi.org/10.1002/1521-3978(200009)48:9/11<771::AID-PROP771>3.0.CO;2-E}
  {\bibfield  {journal} {\bibinfo  {journal} {Fortschritte der Physik}\
  }\textbf {\bibinfo {volume} {48}},\ \bibinfo {pages} {771} (\bibinfo {year}
  {2000})}\BibitemShut {NoStop}%
\bibitem [{\citenamefont {Dowling}\ and\ \citenamefont
  {Milburn}(2003)}]{dowling_QuantumTechnologySecond_2003}%
  \BibitemOpen
  \bibfield  {author} {\bibinfo {author} {\bibfnamefont {J.~P.}\ \bibnamefont
  {Dowling}}\ and\ \bibinfo {author} {\bibfnamefont {G.~J.}\ \bibnamefont
  {Milburn}},\ }\bibfield  {title} {\bibinfo {title} {Quantum technology: The
  second quantum revolution},\ }\href@noop {} {\bibfield  {journal} {\bibinfo
  {journal} {Series A: Mathematical, Physical and Engineering Sciences}\ }
  (\bibinfo {year} {2003})}\BibitemShut {NoStop}%
\bibitem [{\citenamefont {Atzori}\ and\ \citenamefont
  {Sessoli}(2019)}]{ohira_ZeroCorrelationEntanglement_2019}%
  \BibitemOpen
  \bibfield  {author} {\bibinfo {author} {\bibfnamefont {M.}~\bibnamefont
  {Atzori}}\ and\ \bibinfo {author} {\bibfnamefont {R.}~\bibnamefont
  {Sessoli}},\ }\bibfield  {title} {\bibinfo {title} {The {{Second Quantum
  Revolution}}: {{Role}} and {{Challenges}} of {{Molecular Chemistry}}},\
  }\href {https://doi.org/10.1021/jacs.9b00984} {\bibfield  {journal} {\bibinfo
   {journal} {J. Am. Chem. Soc.}\ }\textbf {\bibinfo {volume} {141}},\ \bibinfo
  {pages} {11339} (\bibinfo {year} {2019})}\BibitemShut {NoStop}%
\bibitem [{\citenamefont {Preskill}(2018)}]{preskillQuantumComputingNISQ2018}%
  \BibitemOpen
  \bibfield  {author} {\bibinfo {author} {\bibfnamefont {J.}~\bibnamefont
  {Preskill}},\ }\bibfield  {title} {\bibinfo {title} {Quantum {{Computing}} in
  the {{NISQ}} era and beyond},\ }\href
  {https://doi.org/10.22331/q-2018-08-06-79} {\bibfield  {journal} {\bibinfo
  {journal} {Quantum}\ }\textbf {\bibinfo {volume} {2}},\ \bibinfo {pages} {79}
  (\bibinfo {year} {2018})}\BibitemShut {NoStop}%
\bibitem [{\citenamefont
  {Kitaev}(2001)}]{kitaev_UnpairedMajoranaFermions_2001}%
  \BibitemOpen
  \bibfield  {author} {\bibinfo {author} {\bibfnamefont {A.~Y.}\ \bibnamefont
  {Kitaev}},\ }\bibfield  {title} {\bibinfo {title} {Unpaired {{Majorana}}
  fermions in quantum wires},\ }\href
  {https://doi.org/10.1070/1063-7869/44/10S/S29} {\bibfield  {journal}
  {\bibinfo  {journal} {Phys.-Usp.}\ }\textbf {\bibinfo {volume} {44}},\
  \bibinfo {pages} {131} (\bibinfo {year} {2001})}\BibitemShut {NoStop}%
\bibitem [{\citenamefont
  {Kitaev}(2003)}]{kitaev_FaulttolerantQuantumComputation_2003}%
  \BibitemOpen
  \bibfield  {author} {\bibinfo {author} {\bibfnamefont {A.~Y.}\ \bibnamefont
  {Kitaev}},\ }\bibfield  {title} {\bibinfo {title} {Fault-tolerant quantum
  computation by anyons},\ }\href
  {https://doi.org/10.1016/S0003-4916(02)00018-0} {\bibfield  {journal}
  {\bibinfo  {journal} {Annals of Physics}\ }\textbf {\bibinfo {volume}
  {303}},\ \bibinfo {pages} {2} (\bibinfo {year} {2003})}\BibitemShut {NoStop}%
\bibitem [{\citenamefont {Read}\ and\ \citenamefont
  {Green}(2000)}]{read_PairedStatesFermions_2000}%
  \BibitemOpen
  \bibfield  {author} {\bibinfo {author} {\bibfnamefont {N.}~\bibnamefont
  {Read}}\ and\ \bibinfo {author} {\bibfnamefont {D.}~\bibnamefont {Green}},\
  }\bibfield  {title} {\bibinfo {title} {Paired states of fermions in two
  dimensions with breaking of parity and time-reversal symmetries and the
  fractional quantum {{Hall}} effect},\ }\href
  {https://doi.org/10.1103/PhysRevB.61.10267} {\bibfield  {journal} {\bibinfo
  {journal} {Phys. Rev. B}\ }\textbf {\bibinfo {volume} {61}},\ \bibinfo
  {pages} {10267} (\bibinfo {year} {2000})}\BibitemShut {NoStop}%
\bibitem [{\citenamefont {Wang}\ \emph {et~al.}(2015)\citenamefont {Wang},
  \citenamefont {Zhang}, \citenamefont {Xu}, \citenamefont {Zeng},
  \citenamefont {Miao}, \citenamefont {Xu}, \citenamefont {Qian}, \citenamefont
  {Weng}, \citenamefont {Richard}, \citenamefont {Fedorov}, \citenamefont
  {Ding}, \citenamefont {Dai},\ and\ \citenamefont
  {Fang}}]{wang_TopologicalNatureMathrmFeSe_2015}%
  \BibitemOpen
  \bibfield  {author} {\bibinfo {author} {\bibfnamefont {Z.}~\bibnamefont
  {Wang}}, \bibinfo {author} {\bibfnamefont {P.}~\bibnamefont {Zhang}},
  \bibinfo {author} {\bibfnamefont {G.}~\bibnamefont {Xu}}, \bibinfo {author}
  {\bibfnamefont {L.~K.}\ \bibnamefont {Zeng}}, \bibinfo {author}
  {\bibfnamefont {H.}~\bibnamefont {Miao}}, \bibinfo {author} {\bibfnamefont
  {X.}~\bibnamefont {Xu}}, \bibinfo {author} {\bibfnamefont {T.}~\bibnamefont
  {Qian}}, \bibinfo {author} {\bibfnamefont {H.}~\bibnamefont {Weng}}, \bibinfo
  {author} {\bibfnamefont {P.}~\bibnamefont {Richard}}, \bibinfo {author}
  {\bibfnamefont {A.~V.}\ \bibnamefont {Fedorov}}, \bibinfo {author}
  {\bibfnamefont {H.}~\bibnamefont {Ding}}, \bibinfo {author} {\bibfnamefont
  {X.}~\bibnamefont {Dai}},\ and\ \bibinfo {author} {\bibfnamefont
  {Z.}~\bibnamefont {Fang}},\ }\bibfield  {title} {\bibinfo {title}
  {Topological nature of the \$\{\textbackslash
  mathrm\{\vphantom{\}\}}{{FeSe}}\vphantom\{\}\vphantom\{\}\_\{0.5\}\{\textbackslash
  mathrm\{\vphantom{\}\}}{{Te}}\vphantom\{\}\vphantom\{\}\_\{0.5\}\$
  superconductor},\ }\href {https://doi.org/10.1103/PhysRevB.92.115119}
  {\bibfield  {journal} {\bibinfo  {journal} {Phys. Rev. B}\ }\textbf {\bibinfo
  {volume} {92}},\ \bibinfo {pages} {115119} (\bibinfo {year}
  {2015})}\BibitemShut {NoStop}%
\bibitem [{\citenamefont {Xu}\ \emph {et~al.}(2016)\citenamefont {Xu},
  \citenamefont {Lian}, \citenamefont {Tang}, \citenamefont {Qi},\ and\
  \citenamefont {Zhang}}]{xu_TopologicalSuperconductivitySurface_2016}%
  \BibitemOpen
  \bibfield  {author} {\bibinfo {author} {\bibfnamefont {G.}~\bibnamefont
  {Xu}}, \bibinfo {author} {\bibfnamefont {B.}~\bibnamefont {Lian}}, \bibinfo
  {author} {\bibfnamefont {P.}~\bibnamefont {Tang}}, \bibinfo {author}
  {\bibfnamefont {X.-L.}\ \bibnamefont {Qi}},\ and\ \bibinfo {author}
  {\bibfnamefont {S.-C.}\ \bibnamefont {Zhang}},\ }\bibfield  {title} {\bibinfo
  {title} {Topological {{Superconductivity}} on the {{Surface}} of
  {{Fe}}-{{Based Superconductors}}},\ }\href
  {https://doi.org/10.1103/PhysRevLett.117.047001} {\bibfield  {journal}
  {\bibinfo  {journal} {Phys. Rev. Lett.}\ }\textbf {\bibinfo {volume} {117}},\
  \bibinfo {pages} {047001} (\bibinfo {year} {2016})}\BibitemShut {NoStop}%
\bibitem [{\citenamefont {Farhang}\ \emph {et~al.}(2023)\citenamefont
  {Farhang}, \citenamefont {Zaki}, \citenamefont {Wang}, \citenamefont {Gu},
  \citenamefont {Johnson},\ and\ \citenamefont {Xia}}]{TRS_surface}%
  \BibitemOpen
  \bibfield  {author} {\bibinfo {author} {\bibfnamefont {C.}~\bibnamefont
  {Farhang}}, \bibinfo {author} {\bibfnamefont {N.}~\bibnamefont {Zaki}},
  \bibinfo {author} {\bibfnamefont {J.}~\bibnamefont {Wang}}, \bibinfo {author}
  {\bibfnamefont {G.}~\bibnamefont {Gu}}, \bibinfo {author} {\bibfnamefont
  {P.~D.}\ \bibnamefont {Johnson}},\ and\ \bibinfo {author} {\bibfnamefont
  {J.}~\bibnamefont {Xia}},\ }\bibfield  {title} {\bibinfo {title} {Revealing
  the origin of time-reversal symmetry breaking in fe-chalcogenide
  superconductor ${\mathrm{fete}}_{1\ensuremath{-}x}{\mathrm{se}}_{x}$},\
  }\href {https://doi.org/10.1103/PhysRevLett.130.046702} {\bibfield  {journal}
  {\bibinfo  {journal} {Phys. Rev. Lett.}\ }\textbf {\bibinfo {volume} {130}},\
  \bibinfo {pages} {046702} (\bibinfo {year} {2023})}\BibitemShut {NoStop}%
\bibitem [{\citenamefont {Yeh}\ \emph {et~al.}(2008)\citenamefont {Yeh},
  \citenamefont {Huang}, \citenamefont {Huang}, \citenamefont {Chen},
  \citenamefont {Hsu}, \citenamefont {Wu}, \citenamefont {Lee}, \citenamefont
  {Chu}, \citenamefont {Chen}, \citenamefont {Luo}, \citenamefont {Yan},\ and\
  \citenamefont {Wu}}]{yehTelluriumSubstitutionEffect2008}%
  \BibitemOpen
  \bibfield  {author} {\bibinfo {author} {\bibfnamefont {K.-W.}\ \bibnamefont
  {Yeh}}, \bibinfo {author} {\bibfnamefont {T.-W.}\ \bibnamefont {Huang}},
  \bibinfo {author} {\bibfnamefont {Y.-l.}\ \bibnamefont {Huang}}, \bibinfo
  {author} {\bibfnamefont {T.-K.}\ \bibnamefont {Chen}}, \bibinfo {author}
  {\bibfnamefont {F.-C.}\ \bibnamefont {Hsu}}, \bibinfo {author} {\bibfnamefont
  {P.~M.}\ \bibnamefont {Wu}}, \bibinfo {author} {\bibfnamefont {Y.-C.}\
  \bibnamefont {Lee}}, \bibinfo {author} {\bibfnamefont {Y.-Y.}\ \bibnamefont
  {Chu}}, \bibinfo {author} {\bibfnamefont {C.-L.}\ \bibnamefont {Chen}},
  \bibinfo {author} {\bibfnamefont {J.-Y.}\ \bibnamefont {Luo}}, \bibinfo
  {author} {\bibfnamefont {D.-C.}\ \bibnamefont {Yan}},\ and\ \bibinfo {author}
  {\bibfnamefont {M.-K.}\ \bibnamefont {Wu}},\ }\bibfield  {title} {\bibinfo
  {title} {Tellurium substitution effect on superconductivity of the
  \$\textbackslash upalpha\$-phase iron selenide},\ }\href
  {https://doi.org/10.1209/0295-5075/84/37002} {\bibfield  {journal} {\bibinfo
  {journal} {EPL}\ }\textbf {\bibinfo {volume} {84}},\ \bibinfo {pages} {37002}
  (\bibinfo {year} {2008})}\BibitemShut {NoStop}%
\bibitem [{\citenamefont {Fang}\ \emph {et~al.}(2008)\citenamefont {Fang},
  \citenamefont {Pham}, \citenamefont {Qian}, \citenamefont {Liu},
  \citenamefont {Vehstedt}, \citenamefont {Liu}, \citenamefont {Spinu},\ and\
  \citenamefont {Mao}}]{fangSuperconductivityCloseMagnetic2008}%
  \BibitemOpen
  \bibfield  {author} {\bibinfo {author} {\bibfnamefont {M.~H.}\ \bibnamefont
  {Fang}}, \bibinfo {author} {\bibfnamefont {H.~M.}\ \bibnamefont {Pham}},
  \bibinfo {author} {\bibfnamefont {B.}~\bibnamefont {Qian}}, \bibinfo {author}
  {\bibfnamefont {T.~J.}\ \bibnamefont {Liu}}, \bibinfo {author} {\bibfnamefont
  {E.~K.}\ \bibnamefont {Vehstedt}}, \bibinfo {author} {\bibfnamefont
  {Y.}~\bibnamefont {Liu}}, \bibinfo {author} {\bibfnamefont {L.}~\bibnamefont
  {Spinu}},\ and\ \bibinfo {author} {\bibfnamefont {Z.~Q.}\ \bibnamefont
  {Mao}},\ }\bibfield  {title} {\bibinfo {title} {Superconductivity close to
  magnetic instability in {{Fe}} ( {{Se}} 1 - x {{Te}} x ) 0.82},\ }\href
  {https://doi.org/10.1103/PhysRevB.78.224503} {\bibfield  {journal} {\bibinfo
  {journal} {Phys. Rev. B}\ }\textbf {\bibinfo {volume} {78}},\ \bibinfo
  {pages} {224503} (\bibinfo {year} {2008})}\BibitemShut {NoStop}%
\bibitem [{\citenamefont {Sales}\ \emph {et~al.}(2009)\citenamefont {Sales},
  \citenamefont {Sefat}, \citenamefont {McGuire}, \citenamefont {Jin},
  \citenamefont {Mandrus},\ and\ \citenamefont
  {Mozharivskyj}}]{salesBulkSuperconductivity142009}%
  \BibitemOpen
  \bibfield  {author} {\bibinfo {author} {\bibfnamefont {B.~C.}\ \bibnamefont
  {Sales}}, \bibinfo {author} {\bibfnamefont {A.~S.}\ \bibnamefont {Sefat}},
  \bibinfo {author} {\bibfnamefont {M.~A.}\ \bibnamefont {McGuire}}, \bibinfo
  {author} {\bibfnamefont {R.~Y.}\ \bibnamefont {Jin}}, \bibinfo {author}
  {\bibfnamefont {D.}~\bibnamefont {Mandrus}},\ and\ \bibinfo {author}
  {\bibfnamefont {Y.}~\bibnamefont {Mozharivskyj}},\ }\bibfield  {title}
  {\bibinfo {title} {Bulk superconductivity at 14 {{K}} in single crystals of
  {{Fe}} 1 + y {{Te}} x {{Se}} 1 - x},\ }\href
  {https://doi.org/10.1103/PhysRevB.79.094521} {\bibfield  {journal} {\bibinfo
  {journal} {Phys. Rev. B}\ }\textbf {\bibinfo {volume} {79}},\ \bibinfo
  {pages} {094521} (\bibinfo {year} {2009})}\BibitemShut {NoStop}%
\bibitem [{\citenamefont {Zhang}\ \emph {et~al.}(2018)\citenamefont {Zhang},
  \citenamefont {Yaji}, \citenamefont {Hashimoto}, \citenamefont {Ota},
  \citenamefont {Kondo}, \citenamefont {Okazaki}, \citenamefont {Wang},
  \citenamefont {Wen}, \citenamefont {Gu}, \citenamefont {Ding} \emph
  {et~al.}}]{zhang2018observation}%
  \BibitemOpen
  \bibfield  {author} {\bibinfo {author} {\bibfnamefont {P.}~\bibnamefont
  {Zhang}}, \bibinfo {author} {\bibfnamefont {K.}~\bibnamefont {Yaji}},
  \bibinfo {author} {\bibfnamefont {T.}~\bibnamefont {Hashimoto}}, \bibinfo
  {author} {\bibfnamefont {Y.}~\bibnamefont {Ota}}, \bibinfo {author}
  {\bibfnamefont {T.}~\bibnamefont {Kondo}}, \bibinfo {author} {\bibfnamefont
  {K.}~\bibnamefont {Okazaki}}, \bibinfo {author} {\bibfnamefont
  {Z.}~\bibnamefont {Wang}}, \bibinfo {author} {\bibfnamefont {J.}~\bibnamefont
  {Wen}}, \bibinfo {author} {\bibfnamefont {G.}~\bibnamefont {Gu}}, \bibinfo
  {author} {\bibfnamefont {H.}~\bibnamefont {Ding}}, \emph {et~al.},\
  }\href@noop {} {\bibfield  {journal} {\bibinfo  {journal} {Science}\ }\textbf
  {\bibinfo {volume} {360}},\ \bibinfo {pages} {182} (\bibinfo {year}
  {2018})}\BibitemShut {NoStop}%
\bibitem [{\citenamefont {Rameau}\ \emph {et~al.}(2019)\citenamefont {Rameau},
  \citenamefont {Zaki}, \citenamefont {Gu}, \citenamefont {Johnson},\ and\
  \citenamefont {Weinert}}]{rameauInterplayParamagnetismTopology2019}%
  \BibitemOpen
  \bibfield  {author} {\bibinfo {author} {\bibfnamefont {J.~D.}\ \bibnamefont
  {Rameau}}, \bibinfo {author} {\bibfnamefont {N.}~\bibnamefont {Zaki}},
  \bibinfo {author} {\bibfnamefont {G.~D.}\ \bibnamefont {Gu}}, \bibinfo
  {author} {\bibfnamefont {P.~D.}\ \bibnamefont {Johnson}},\ and\ \bibinfo
  {author} {\bibfnamefont {M.}~\bibnamefont {Weinert}},\ }\bibfield  {title}
  {\bibinfo {title} {Interplay of paramagnetism and topology in the
  {{Fe}}-chalcogenide high-\$\{\vphantom\}{{T}}\vphantom\{\}\_\{c\}\$
  superconductors},\ }\href {https://doi.org/10.1103/PhysRevB.99.205117}
  {\bibfield  {journal} {\bibinfo  {journal} {Phys. Rev. B}\ }\textbf {\bibinfo
  {volume} {99}},\ \bibinfo {pages} {205117} (\bibinfo {year}
  {2019})}\BibitemShut {NoStop}%
\bibitem [{\citenamefont {Zaki}\ \emph {et~al.}(2021)\citenamefont {Zaki},
  \citenamefont {Gu}, \citenamefont {Tsvelik}, \citenamefont {Wu},\ and\
  \citenamefont {Johnson}}]{zakiTimereversalSymmetryBreaking2021}%
  \BibitemOpen
  \bibfield  {author} {\bibinfo {author} {\bibfnamefont {N.}~\bibnamefont
  {Zaki}}, \bibinfo {author} {\bibfnamefont {G.}~\bibnamefont {Gu}}, \bibinfo
  {author} {\bibfnamefont {A.}~\bibnamefont {Tsvelik}}, \bibinfo {author}
  {\bibfnamefont {C.}~\bibnamefont {Wu}},\ and\ \bibinfo {author}
  {\bibfnamefont {P.~D.}\ \bibnamefont {Johnson}},\ }\bibfield  {title}
  {\bibinfo {title} {Time-reversal symmetry breaking in the {{Fe}}-chalcogenide
  superconductors},\ }\bibfield  {journal} {\bibinfo  {journal} {PNAS}\
  }\textbf {\bibinfo {volume} {118}},\ \href
  {https://doi.org/10.1073/pnas.2007241118} {10.1073/pnas.2007241118} (\bibinfo
  {year} {2021})\BibitemShut {NoStop}%
\bibitem [{\citenamefont {Fu}\ and\ \citenamefont
  {Kane}(2008)}]{fu_SuperconductingProximityEffect_2008}%
  \BibitemOpen
  \bibfield  {author} {\bibinfo {author} {\bibfnamefont {L.}~\bibnamefont
  {Fu}}\ and\ \bibinfo {author} {\bibfnamefont {C.~L.}\ \bibnamefont {Kane}},\
  }\bibfield  {title} {\bibinfo {title} {Superconducting {{Proximity Effect}}
  and {{Majorana Fermions}} at the {{Surface}} of a {{Topological
  Insulator}}},\ }\href {https://doi.org/10.1103/PhysRevLett.100.096407}
  {\bibfield  {journal} {\bibinfo  {journal} {Phys. Rev. Lett.}\ }\textbf
  {\bibinfo {volume} {100}},\ \bibinfo {pages} {096407} (\bibinfo {year}
  {2008})}\BibitemShut {NoStop}%
\bibitem [{\citenamefont {Wang}\ \emph {et~al.}(2018)\citenamefont {Wang},
  \citenamefont {Kong}, \citenamefont {Fan}, \citenamefont {Chen},
  \citenamefont {Zhu}, \citenamefont {Liu}, \citenamefont {Cao}, \citenamefont
  {Sun}, \citenamefont {Du}, \citenamefont {Schneeloch} \emph
  {et~al.}}]{wang2018evidence}%
  \BibitemOpen
  \bibfield  {author} {\bibinfo {author} {\bibfnamefont {D.}~\bibnamefont
  {Wang}}, \bibinfo {author} {\bibfnamefont {L.}~\bibnamefont {Kong}}, \bibinfo
  {author} {\bibfnamefont {P.}~\bibnamefont {Fan}}, \bibinfo {author}
  {\bibfnamefont {H.}~\bibnamefont {Chen}}, \bibinfo {author} {\bibfnamefont
  {S.}~\bibnamefont {Zhu}}, \bibinfo {author} {\bibfnamefont {W.}~\bibnamefont
  {Liu}}, \bibinfo {author} {\bibfnamefont {L.}~\bibnamefont {Cao}}, \bibinfo
  {author} {\bibfnamefont {Y.}~\bibnamefont {Sun}}, \bibinfo {author}
  {\bibfnamefont {S.}~\bibnamefont {Du}}, \bibinfo {author} {\bibfnamefont
  {J.}~\bibnamefont {Schneeloch}}, \emph {et~al.},\ }\href@noop {} {\bibfield
  {journal} {\bibinfo  {journal} {Science}\ }\textbf {\bibinfo {volume}
  {362}},\ \bibinfo {pages} {333} (\bibinfo {year} {2018})}\BibitemShut
  {NoStop}%
\bibitem [{\citenamefont {Wang}\ \emph {et~al.}(2020)\citenamefont {Wang},
  \citenamefont {Rodriguez}, \citenamefont {Jiao}, \citenamefont {Howard},
  \citenamefont {Graham}, \citenamefont {Gu}, \citenamefont {Hughes},
  \citenamefont {Morr},\ and\ \citenamefont {Madhavan}}]{wang2020evidence}%
  \BibitemOpen
  \bibfield  {author} {\bibinfo {author} {\bibfnamefont {Z.}~\bibnamefont
  {Wang}}, \bibinfo {author} {\bibfnamefont {J.~O.}\ \bibnamefont {Rodriguez}},
  \bibinfo {author} {\bibfnamefont {L.}~\bibnamefont {Jiao}}, \bibinfo {author}
  {\bibfnamefont {S.}~\bibnamefont {Howard}}, \bibinfo {author} {\bibfnamefont
  {M.}~\bibnamefont {Graham}}, \bibinfo {author} {\bibfnamefont
  {G.}~\bibnamefont {Gu}}, \bibinfo {author} {\bibfnamefont {T.~L.}\
  \bibnamefont {Hughes}}, \bibinfo {author} {\bibfnamefont {D.~K.}\
  \bibnamefont {Morr}},\ and\ \bibinfo {author} {\bibfnamefont
  {V.}~\bibnamefont {Madhavan}},\ }\bibfield  {title} {\bibinfo {title}
  {Evidence for dispersing 1d majorana channels in an iron-based
  superconductor},\ }\href@noop {} {\bibfield  {journal} {\bibinfo  {journal}
  {Science}\ }\textbf {\bibinfo {volume} {367}},\ \bibinfo {pages} {104}
  (\bibinfo {year} {2020})}\BibitemShut {NoStop}%
\bibitem [{\citenamefont {Yi}\ \emph {et~al.}(2015)\citenamefont {Yi},
  \citenamefont {Liu}, \citenamefont {Zhang}, \citenamefont {Yu}, \citenamefont
  {Zhu}, \citenamefont {Lee}, \citenamefont {Moore}, \citenamefont {Schmitt},
  \citenamefont {Li}, \citenamefont {Riggs} \emph
  {et~al.}}]{yi2015observation}%
  \BibitemOpen
  \bibfield  {author} {\bibinfo {author} {\bibfnamefont {M.}~\bibnamefont
  {Yi}}, \bibinfo {author} {\bibfnamefont {Z.}~\bibnamefont {Liu}}, \bibinfo
  {author} {\bibfnamefont {Y.}~\bibnamefont {Zhang}}, \bibinfo {author}
  {\bibfnamefont {R.}~\bibnamefont {Yu}}, \bibinfo {author} {\bibfnamefont
  {J.-X.}\ \bibnamefont {Zhu}}, \bibinfo {author} {\bibfnamefont
  {J.}~\bibnamefont {Lee}}, \bibinfo {author} {\bibfnamefont {R.}~\bibnamefont
  {Moore}}, \bibinfo {author} {\bibfnamefont {F.}~\bibnamefont {Schmitt}},
  \bibinfo {author} {\bibfnamefont {W.}~\bibnamefont {Li}}, \bibinfo {author}
  {\bibfnamefont {S.}~\bibnamefont {Riggs}}, \emph {et~al.},\ }\bibfield
  {title} {\bibinfo {title} {Observation of universal strong orbital-dependent
  correlation effects in iron chalcogenides},\ }\href@noop {} {\bibfield
  {journal} {\bibinfo  {journal} {Nature communications}\ }\textbf {\bibinfo
  {volume} {6}},\ \bibinfo {pages} {1} (\bibinfo {year} {2015})}\BibitemShut
  {NoStop}%
\bibitem [{\citenamefont {Huang}\ \emph {et~al.}(2022)\citenamefont {Huang},
  \citenamefont {Yu}, \citenamefont {Xu}, \citenamefont {Zhu}, \citenamefont
  {Oh}, \citenamefont {Jiang}, \citenamefont {Wang}, \citenamefont {Wu},
  \citenamefont {Chen}, \citenamefont {Denlinger} \emph
  {et~al.}}]{huang2020low}%
  \BibitemOpen
  \bibfield  {author} {\bibinfo {author} {\bibfnamefont {J.}~\bibnamefont
  {Huang}}, \bibinfo {author} {\bibfnamefont {R.}~\bibnamefont {Yu}}, \bibinfo
  {author} {\bibfnamefont {Z.}~\bibnamefont {Xu}}, \bibinfo {author}
  {\bibfnamefont {J.-X.}\ \bibnamefont {Zhu}}, \bibinfo {author} {\bibfnamefont
  {J.~S.}\ \bibnamefont {Oh}}, \bibinfo {author} {\bibfnamefont
  {Q.}~\bibnamefont {Jiang}}, \bibinfo {author} {\bibfnamefont
  {M.}~\bibnamefont {Wang}}, \bibinfo {author} {\bibfnamefont {H.}~\bibnamefont
  {Wu}}, \bibinfo {author} {\bibfnamefont {T.}~\bibnamefont {Chen}}, \bibinfo
  {author} {\bibfnamefont {J.~D.}\ \bibnamefont {Denlinger}}, \emph {et~al.},\
  }\bibfield  {title} {\bibinfo {title} {Correlation-driven electronic
  reconstruction in fete1- xsex},\ }\href@noop {} {\bibfield  {journal}
  {\bibinfo  {journal} {Communications Physics}\ }\textbf {\bibinfo {volume}
  {5}},\ \bibinfo {pages} {1} (\bibinfo {year} {2022})}\BibitemShut {NoStop}%
\bibitem [{\citenamefont {Miao}\ \emph {et~al.}(2018)\citenamefont {Miao},
  \citenamefont {Brito}, \citenamefont {Yin}, \citenamefont {Zhong},
  \citenamefont {Gu}, \citenamefont {Johnson}, \citenamefont {Dean},
  \citenamefont {Choi}, \citenamefont {Kotliar}, \citenamefont {Ku} \emph
  {et~al.}}]{miao2018universal}%
  \BibitemOpen
  \bibfield  {author} {\bibinfo {author} {\bibfnamefont {H.}~\bibnamefont
  {Miao}}, \bibinfo {author} {\bibfnamefont {W.}~\bibnamefont {Brito}},
  \bibinfo {author} {\bibfnamefont {Z.}~\bibnamefont {Yin}}, \bibinfo {author}
  {\bibfnamefont {R.}~\bibnamefont {Zhong}}, \bibinfo {author} {\bibfnamefont
  {G.}~\bibnamefont {Gu}}, \bibinfo {author} {\bibfnamefont {P.}~\bibnamefont
  {Johnson}}, \bibinfo {author} {\bibfnamefont {M.}~\bibnamefont {Dean}},
  \bibinfo {author} {\bibfnamefont {S.}~\bibnamefont {Choi}}, \bibinfo {author}
  {\bibfnamefont {G.}~\bibnamefont {Kotliar}}, \bibinfo {author} {\bibfnamefont
  {W.}~\bibnamefont {Ku}}, \emph {et~al.},\ }\href@noop {} {\bibfield
  {journal} {\bibinfo  {journal} {Physical Review B}\ }\textbf {\bibinfo
  {volume} {98}},\ \bibinfo {pages} {020502} (\bibinfo {year}
  {2018})}\BibitemShut {NoStop}%
\bibitem [{\citenamefont {Hohenberg}\ and\ \citenamefont
  {Kohn}(1964)}]{hohenberg_InhomogeneousElectronGas_1964}%
  \BibitemOpen
  \bibfield  {author} {\bibinfo {author} {\bibfnamefont {P.}~\bibnamefont
  {Hohenberg}}\ and\ \bibinfo {author} {\bibfnamefont {W.}~\bibnamefont
  {Kohn}},\ }\bibfield  {title} {\bibinfo {title} {Inhomogeneous {{Electron
  Gas}}},\ }\href {https://doi.org/10.1103/PhysRev.136.B864} {\bibfield
  {journal} {\bibinfo  {journal} {Phys. Rev.}\ }\textbf {\bibinfo {volume}
  {136}},\ \bibinfo {pages} {B864} (\bibinfo {year} {1964})}\BibitemShut
  {NoStop}%
\bibitem [{\citenamefont {Kohn}\ and\ \citenamefont
  {Sham}(1965)}]{kohn_SelfConsistentEquationsIncluding_1965}%
  \BibitemOpen
  \bibfield  {author} {\bibinfo {author} {\bibfnamefont {W.}~\bibnamefont
  {Kohn}}\ and\ \bibinfo {author} {\bibfnamefont {L.~J.}\ \bibnamefont
  {Sham}},\ }\bibfield  {title} {\bibinfo {title} {Self-{{Consistent Equations
  Including Exchange}} and {{Correlation Effects}}},\ }\href
  {https://doi.org/10.1103/PhysRev.140.A1133} {\bibfield  {journal} {\bibinfo
  {journal} {Phys. Rev.}\ }\textbf {\bibinfo {volume} {140}},\ \bibinfo {pages}
  {A1133} (\bibinfo {year} {1965})}\BibitemShut {NoStop}%
\bibitem [{\citenamefont {Li}\ \emph {et~al.}(2021)\citenamefont {Li},
  \citenamefont {Zaki}, \citenamefont {Garlea}, \citenamefont {Savici},
  \citenamefont {Fobes}, \citenamefont {Xu}, \citenamefont {Camino},
  \citenamefont {Petrovic}, \citenamefont {Gu}, \citenamefont {Johnson} \emph
  {et~al.}}]{liElectronicPropertiesBulk2021}%
  \BibitemOpen
  \bibfield  {author} {\bibinfo {author} {\bibfnamefont {Y.}~\bibnamefont
  {Li}}, \bibinfo {author} {\bibfnamefont {N.}~\bibnamefont {Zaki}}, \bibinfo
  {author} {\bibfnamefont {V.~O.}\ \bibnamefont {Garlea}}, \bibinfo {author}
  {\bibfnamefont {A.~T.}\ \bibnamefont {Savici}}, \bibinfo {author}
  {\bibfnamefont {D.}~\bibnamefont {Fobes}}, \bibinfo {author} {\bibfnamefont
  {Z.}~\bibnamefont {Xu}}, \bibinfo {author} {\bibfnamefont {F.}~\bibnamefont
  {Camino}}, \bibinfo {author} {\bibfnamefont {C.}~\bibnamefont {Petrovic}},
  \bibinfo {author} {\bibfnamefont {G.}~\bibnamefont {Gu}}, \bibinfo {author}
  {\bibfnamefont {P.~D.}\ \bibnamefont {Johnson}}, \emph {et~al.},\ }\bibfield
  {title} {\bibinfo {title} {Electronic properties of the bulk and surface
  states of fe1+ yte1- xsex},\ }\href@noop {} {\bibfield  {journal} {\bibinfo
  {journal} {Nature Materials}\ }\textbf {\bibinfo {volume} {20}},\ \bibinfo
  {pages} {1221} (\bibinfo {year} {2021})}\BibitemShut {NoStop}%
\bibitem [{\citenamefont {Ma}\ \emph {et~al.}(2022)\citenamefont {Ma},
  \citenamefont {Wang}, \citenamefont {Liu}, \citenamefont {Yu}, \citenamefont
  {Peng}, \citenamefont {Zheng},\ and\ \citenamefont {Yin}}]{Yin2022}%
  \BibitemOpen
  \bibfield  {author} {\bibinfo {author} {\bibfnamefont {X.}~\bibnamefont
  {Ma}}, \bibinfo {author} {\bibfnamefont {G.}~\bibnamefont {Wang}}, \bibinfo
  {author} {\bibfnamefont {R.}~\bibnamefont {Liu}}, \bibinfo {author}
  {\bibfnamefont {T.}~\bibnamefont {Yu}}, \bibinfo {author} {\bibfnamefont
  {Y.}~\bibnamefont {Peng}}, \bibinfo {author} {\bibfnamefont {P.}~\bibnamefont
  {Zheng}},\ and\ \bibinfo {author} {\bibfnamefont {Z.}~\bibnamefont {Yin}},\
  }\bibfield  {title} {\bibinfo {title} {Correlation-corrected band topology
  and topological surface states in iron-based superconductors},\ }\href
  {https://doi.org/10.1103/PhysRevB.106.115114} {\bibfield  {journal} {\bibinfo
   {journal} {Phys. Rev. B}\ }\textbf {\bibinfo {volume} {106}},\ \bibinfo
  {pages} {115114} (\bibinfo {year} {2022})}\BibitemShut {NoStop}%
\bibitem [{\citenamefont {Kutepov}\ \emph {et~al.}(2012)\citenamefont
  {Kutepov}, \citenamefont {Haule}, \citenamefont {Savrasov},\ and\
  \citenamefont {Kotliar}}]{kutepov_ElectronicStructurePu_2012}%
  \BibitemOpen
  \bibfield  {author} {\bibinfo {author} {\bibfnamefont {A.}~\bibnamefont
  {Kutepov}}, \bibinfo {author} {\bibfnamefont {K.}~\bibnamefont {Haule}},
  \bibinfo {author} {\bibfnamefont {S.~Y.}\ \bibnamefont {Savrasov}},\ and\
  \bibinfo {author} {\bibfnamefont {G.}~\bibnamefont {Kotliar}},\ }\bibfield
  {title} {\bibinfo {title} {Electronic structure of {{Pu}} and {{Am}} metals
  by self-consistent relativistic {{GW}} method},\ }\href
  {https://doi.org/10.1103/PhysRevB.85.155129} {\bibfield  {journal} {\bibinfo
  {journal} {Phys. Rev. B}\ }\textbf {\bibinfo {volume} {85}},\ \bibinfo
  {pages} {155129} (\bibinfo {year} {2012})}\BibitemShut {NoStop}%
\bibitem [{\citenamefont {Kutepov}\ \emph {et~al.}(2017)\citenamefont
  {Kutepov}, \citenamefont {Oudovenko},\ and\ \citenamefont
  {Kotliar}}]{kutepov_linearized_2017}%
  \BibitemOpen
  \bibfield  {author} {\bibinfo {author} {\bibfnamefont {A.}~\bibnamefont
  {Kutepov}}, \bibinfo {author} {\bibfnamefont {V.}~\bibnamefont {Oudovenko}},\
  and\ \bibinfo {author} {\bibfnamefont {G.}~\bibnamefont {Kotliar}},\ }\href
  {https://doi.org/10.1016/j.cpc.2017.06.012} {\bibfield  {journal} {\bibinfo
  {journal} {Computer Physics Communications}\ }\textbf {\bibinfo {volume}
  {219}},\ \bibinfo {pages} {407} (\bibinfo {year} {2017})}\BibitemShut
  {NoStop}%
\bibitem [{\citenamefont {Georges}\ \emph {et~al.}(1996)\citenamefont
  {Georges}, \citenamefont {Kotliar}, \citenamefont {Krauth},\ and\
  \citenamefont {Rozenberg}}]{georges_DynamicalMeanfieldTheory_1996}%
  \BibitemOpen
  \bibfield  {author} {\bibinfo {author} {\bibfnamefont {A.}~\bibnamefont
  {Georges}}, \bibinfo {author} {\bibfnamefont {G.}~\bibnamefont {Kotliar}},
  \bibinfo {author} {\bibfnamefont {W.}~\bibnamefont {Krauth}},\ and\ \bibinfo
  {author} {\bibfnamefont {M.~J.}\ \bibnamefont {Rozenberg}},\ }\bibfield
  {title} {\bibinfo {title} {Dynamical mean-field theory of strongly correlated
  fermion systems and the limit of infinite dimensions},\ }\href
  {https://doi.org/10.1103/RevModPhys.68.13} {\bibfield  {journal} {\bibinfo
  {journal} {Rev. Mod. Phys.}\ }\textbf {\bibinfo {volume} {68}},\ \bibinfo
  {pages} {13} (\bibinfo {year} {1996})}\BibitemShut {NoStop}%
\bibitem [{\citenamefont {Metzner}\ and\ \citenamefont
  {Vollhardt}(1989)}]{metzner_CorrelatedLatticeFermions_1989}%
  \BibitemOpen
  \bibfield  {author} {\bibinfo {author} {\bibfnamefont {W.}~\bibnamefont
  {Metzner}}\ and\ \bibinfo {author} {\bibfnamefont {D.}~\bibnamefont
  {Vollhardt}},\ }\bibfield  {title} {\bibinfo {title} {Correlated {{Lattice
  Fermions}} in \$d=\textbackslash ensuremath\{\textbackslash infty\}\$
  {{Dimensions}}},\ }\href {https://doi.org/10.1103/PhysRevLett.62.324}
  {\bibfield  {journal} {\bibinfo  {journal} {Phys. Rev. Lett.}\ }\textbf
  {\bibinfo {volume} {62}},\ \bibinfo {pages} {324} (\bibinfo {year}
  {1989})}\BibitemShut {NoStop}%
\bibitem [{\citenamefont
  {{M{\"u}ller-Hartmann}}(1989)}]{muller-hartmann_CorrelatedFermionsLattice_1989}%
  \BibitemOpen
  \bibfield  {author} {\bibinfo {author} {\bibfnamefont {E.}~\bibnamefont
  {{M{\"u}ller-Hartmann}}},\ }\bibfield  {title} {\bibinfo {title} {Correlated
  fermions on a lattice in high dimensions},\ }\href
  {https://doi.org/10.1007/BF01311397} {\bibfield  {journal} {\bibinfo
  {journal} {Z. Physik B - Condensed Matter}\ }\textbf {\bibinfo {volume}
  {74}},\ \bibinfo {pages} {507} (\bibinfo {year} {1989})}\BibitemShut
  {NoStop}%
\bibitem [{\citenamefont {Brandt}\ and\ \citenamefont
  {Mielsch}(1989)}]{brandt_ThermodynamicsCorrelationFunctions_1989}%
  \BibitemOpen
  \bibfield  {author} {\bibinfo {author} {\bibfnamefont {U.}~\bibnamefont
  {Brandt}}\ and\ \bibinfo {author} {\bibfnamefont {C.}~\bibnamefont
  {Mielsch}},\ }\bibfield  {title} {\bibinfo {title} {Thermodynamics and
  correlation functions of the {{Falicov}}-{{Kimball}} model in large
  dimensions},\ }\href {https://doi.org/10.1007/BF01321824} {\bibfield
  {journal} {\bibinfo  {journal} {Z. Physik B - Condensed Matter}\ }\textbf
  {\bibinfo {volume} {75}},\ \bibinfo {pages} {365} (\bibinfo {year}
  {1989})}\BibitemShut {NoStop}%
\bibitem [{\citenamefont {Jani{\v
  s}}(1991)}]{janis_NewConstructionThermodynamic_1991}%
  \BibitemOpen
  \bibfield  {author} {\bibinfo {author} {\bibfnamefont {V.}~\bibnamefont
  {Jani{\v s}}},\ }\bibfield  {title} {\bibinfo {title} {A new construction of
  thermodynamic mean-field theories of itinerant fermions: Application to the
  {{Falicov}}-{{Kimball}} model},\ }\href {https://doi.org/10.1007/BF01309423}
  {\bibfield  {journal} {\bibinfo  {journal} {Z. Physik B - Condensed Matter}\
  }\textbf {\bibinfo {volume} {83}},\ \bibinfo {pages} {227} (\bibinfo {year}
  {1991})}\BibitemShut {NoStop}%
\bibitem [{\citenamefont {Georges}\ and\ \citenamefont
  {Kotliar}(1992)}]{georges_HubbardModelInfinite_1992}%
  \BibitemOpen
  \bibfield  {author} {\bibinfo {author} {\bibfnamefont {A.}~\bibnamefont
  {Georges}}\ and\ \bibinfo {author} {\bibfnamefont {G.}~\bibnamefont
  {Kotliar}},\ }\bibfield  {title} {\bibinfo {title} {Hubbard model in infinite
  dimensions},\ }\href {https://doi.org/10.1103/PhysRevB.45.6479} {\bibfield
  {journal} {\bibinfo  {journal} {Phys. Rev. B}\ }\textbf {\bibinfo {volume}
  {45}},\ \bibinfo {pages} {6479} (\bibinfo {year} {1992})}\BibitemShut
  {NoStop}%
\bibitem [{\citenamefont {Jarrell}(1992)}]{jarrell_HubbardModelInfinite_1992}%
  \BibitemOpen
  \bibfield  {author} {\bibinfo {author} {\bibfnamefont {M.}~\bibnamefont
  {Jarrell}},\ }\bibfield  {title} {\bibinfo {title} {Hubbard model in infinite
  dimensions: {{A}} quantum {{Monte Carlo}} study},\ }\href
  {https://doi.org/10.1103/PhysRevLett.69.168} {\bibfield  {journal} {\bibinfo
  {journal} {Phys. Rev. Lett.}\ }\textbf {\bibinfo {volume} {69}},\ \bibinfo
  {pages} {168} (\bibinfo {year} {1992})}\BibitemShut {NoStop}%
\bibitem [{\citenamefont {Rozenberg}\ \emph {et~al.}(1992)\citenamefont
  {Rozenberg}, \citenamefont {Zhang},\ and\ \citenamefont
  {Kotliar}}]{rozenberg_MottHubbardTransitionInfinite_1992}%
  \BibitemOpen
  \bibfield  {author} {\bibinfo {author} {\bibfnamefont {M.~J.}\ \bibnamefont
  {Rozenberg}}, \bibinfo {author} {\bibfnamefont {X.~Y.}\ \bibnamefont
  {Zhang}},\ and\ \bibinfo {author} {\bibfnamefont {G.}~\bibnamefont
  {Kotliar}},\ }\bibfield  {title} {\bibinfo {title} {Mott-{{Hubbard}}
  transition in infinite dimensions},\ }\href
  {https://doi.org/10.1103/PhysRevLett.69.1236} {\bibfield  {journal} {\bibinfo
   {journal} {Phys. Rev. Lett.}\ }\textbf {\bibinfo {volume} {69}},\ \bibinfo
  {pages} {1236} (\bibinfo {year} {1992})}\BibitemShut {NoStop}%
\bibitem [{\citenamefont {Georges}\ and\ \citenamefont
  {Krauth}(1992)}]{georges_NumericalSolutionEnsuremath_1992}%
  \BibitemOpen
  \bibfield  {author} {\bibinfo {author} {\bibfnamefont {A.}~\bibnamefont
  {Georges}}\ and\ \bibinfo {author} {\bibfnamefont {W.}~\bibnamefont
  {Krauth}},\ }\bibfield  {title} {\bibinfo {title} {Numerical solution of the
  d=\textbackslash ensuremath\{\textbackslash infty\} {{Hubbard}} model:
  {{Evidence}} for a {{Mott}} transition},\ }\href
  {https://doi.org/10.1103/PhysRevLett.69.1240} {\bibfield  {journal} {\bibinfo
   {journal} {Phys. Rev. Lett.}\ }\textbf {\bibinfo {volume} {69}},\ \bibinfo
  {pages} {1240} (\bibinfo {year} {1992})}\BibitemShut {NoStop}%
\bibitem [{\citenamefont {Choi}\ \emph {et~al.}(2016)\citenamefont {Choi},
  \citenamefont {Kutepov}, \citenamefont {Haule}, \citenamefont {van
  Schilfgaarde},\ and\ \citenamefont {Kotliar}}]{choi_first-principles_2016}%
  \BibitemOpen
  \bibfield  {author} {\bibinfo {author} {\bibfnamefont {S.}~\bibnamefont
  {Choi}}, \bibinfo {author} {\bibfnamefont {A.}~\bibnamefont {Kutepov}},
  \bibinfo {author} {\bibfnamefont {K.}~\bibnamefont {Haule}}, \bibinfo
  {author} {\bibfnamefont {M.}~\bibnamefont {van Schilfgaarde}},\ and\ \bibinfo
  {author} {\bibfnamefont {G.}~\bibnamefont {Kotliar}},\ }\href
  {https://doi.org/10.1038/npjquantmats.2016.1} {\bibfield  {journal} {\bibinfo
   {journal} {npj Quantum Materials}\ }\textbf {\bibinfo {volume} {1}},\
  \bibinfo {pages} {16001} (\bibinfo {year} {2016})}\BibitemShut {NoStop}%
\bibitem [{\citenamefont {Choi}\ \emph {et~al.}(2019)\citenamefont {Choi},
  \citenamefont {Semon}, \citenamefont {Kang}, \citenamefont {Kutepov},\ and\
  \citenamefont {Kotliar}}]{choi_comdmft:_2019}%
  \BibitemOpen
  \bibfield  {author} {\bibinfo {author} {\bibfnamefont {S.}~\bibnamefont
  {Choi}}, \bibinfo {author} {\bibfnamefont {P.}~\bibnamefont {Semon}},
  \bibinfo {author} {\bibfnamefont {B.}~\bibnamefont {Kang}}, \bibinfo {author}
  {\bibfnamefont {A.}~\bibnamefont {Kutepov}},\ and\ \bibinfo {author}
  {\bibfnamefont {G.}~\bibnamefont {Kotliar}},\ }\href
  {https://doi.org/10.1016/j.cpc.2019.07.003} {\bibfield  {journal} {\bibinfo
  {journal} {Computer Physics Communications}\ }\textbf {\bibinfo {volume}
  {244}},\ \bibinfo {pages} {277} (\bibinfo {year} {2019})}\BibitemShut
  {NoStop}%
\bibitem [{\citenamefont {Fernandes}\ \emph {et~al.}(2022)\citenamefont
  {Fernandes}, \citenamefont {Coldea}, \citenamefont {Ding}, \citenamefont
  {Fisher}, \citenamefont {Hirschfeld},\ and\ \citenamefont
  {Kotliar}}]{fernandes2022iron}%
  \BibitemOpen
  \bibfield  {author} {\bibinfo {author} {\bibfnamefont {R.~M.}\ \bibnamefont
  {Fernandes}}, \bibinfo {author} {\bibfnamefont {A.~I.}\ \bibnamefont
  {Coldea}}, \bibinfo {author} {\bibfnamefont {H.}~\bibnamefont {Ding}},
  \bibinfo {author} {\bibfnamefont {I.~R.}\ \bibnamefont {Fisher}}, \bibinfo
  {author} {\bibfnamefont {P.}~\bibnamefont {Hirschfeld}},\ and\ \bibinfo
  {author} {\bibfnamefont {G.}~\bibnamefont {Kotliar}},\ }\bibfield  {title}
  {\bibinfo {title} {Iron pnictides and chalcogenides: a new paradigm for
  superconductivity},\ }\href@noop {} {\bibfield  {journal} {\bibinfo
  {journal} {Nature}\ }\textbf {\bibinfo {volume} {601}},\ \bibinfo {pages}
  {35} (\bibinfo {year} {2022})}\BibitemShut {NoStop}%
\bibitem [{\citenamefont {Li}\ \emph {et~al.}(2009)\citenamefont {Li},
  \citenamefont {de~La~Cruz}, \citenamefont {Huang}, \citenamefont {Chen},
  \citenamefont {Lynn}, \citenamefont {Hu}, \citenamefont {Huang},
  \citenamefont {Hsu}, \citenamefont {Yeh}, \citenamefont {Wu} \emph
  {et~al.}}]{li2009first}%
  \BibitemOpen
  \bibfield  {author} {\bibinfo {author} {\bibfnamefont {S.}~\bibnamefont
  {Li}}, \bibinfo {author} {\bibfnamefont {C.}~\bibnamefont {de~La~Cruz}},
  \bibinfo {author} {\bibfnamefont {Q.}~\bibnamefont {Huang}}, \bibinfo
  {author} {\bibfnamefont {Y.}~\bibnamefont {Chen}}, \bibinfo {author}
  {\bibfnamefont {J.}~\bibnamefont {Lynn}}, \bibinfo {author} {\bibfnamefont
  {J.}~\bibnamefont {Hu}}, \bibinfo {author} {\bibfnamefont {Y.-L.}\
  \bibnamefont {Huang}}, \bibinfo {author} {\bibfnamefont {F.-C.}\ \bibnamefont
  {Hsu}}, \bibinfo {author} {\bibfnamefont {K.-W.}\ \bibnamefont {Yeh}},
  \bibinfo {author} {\bibfnamefont {M.-K.}\ \bibnamefont {Wu}}, \emph
  {et~al.},\ }\href@noop {} {\bibfield  {journal} {\bibinfo  {journal}
  {Physical Review B}\ }\textbf {\bibinfo {volume} {79}},\ \bibinfo {pages}
  {054503} (\bibinfo {year} {2009})}\BibitemShut {NoStop}%
\bibitem [{\citenamefont {Tegel}\ \emph {et~al.}(2010)\citenamefont {Tegel},
  \citenamefont {L{\"o}hnert},\ and\ \citenamefont
  {Johrendt}}]{tegel2010crystal}%
  \BibitemOpen
  \bibfield  {author} {\bibinfo {author} {\bibfnamefont {M.}~\bibnamefont
  {Tegel}}, \bibinfo {author} {\bibfnamefont {C.}~\bibnamefont {L{\"o}hnert}},\
  and\ \bibinfo {author} {\bibfnamefont {D.}~\bibnamefont {Johrendt}},\
  }\bibfield  {title} {\bibinfo {title} {The crystal structure of fese0. 44te0.
  56},\ }\href@noop {} {\bibfield  {journal} {\bibinfo  {journal} {Solid state
  communications}\ }\textbf {\bibinfo {volume} {150}},\ \bibinfo {pages} {383}
  (\bibinfo {year} {2010})}\BibitemShut {NoStop}%
\bibitem [{Note1()}]{Note1}%
  \BibitemOpen
  \bibinfo {note} {The Supplementary Material includes (i) a comparison of
  ARPES and the present theory, (ii) computational details, (iii) the
  definition of the projector $f_{Fe-d~or~Se/Te-p}$, (iv) the extraction of the
  spin-orbit coupling (SOC) constants, (v) the computation of surface
  electronic structures, (vi) the SOC enhancement and chalcogen heights, (vii)
  a comparison of the local density approximation (LDA) electronic structures
  and essential low energy parameters with the FeTe and FeSe chemical formula
  in the lattice constant of FeSe$_{0.5}$Te$_{0.5}$ with the consideration of
  the variation in the chalcogen heights, (viii) tight-binding parameters and
  electronic structures from the local quasi-particle self-consistent GW
  (LQSGW) and the LQSGW plus dynamical mean-field theory (LQSGW+DMFT), (ix) a
  detailed discussion on the Z$_{2}$ topology and OSMP, and (x) the effective
  tight-binding parameters in the Hamiltonian of Eq.\ref {eq:Hamiltonian_rot}
  and its comparison to the LQSGW+DMFT result in Fig.\ref
  {fig:OSMP}.}\BibitemShut {Stop}%
\bibitem [{\citenamefont {Johnson}\ \emph {et~al.}(2015)\citenamefont
  {Johnson}, \citenamefont {Yang}, \citenamefont {Rameau}, \citenamefont {Gu},
  \citenamefont {Pan}, \citenamefont {Valla}, \citenamefont {Weinert},\ and\
  \citenamefont {Fedorov}}]{johnson2015spin}%
  \BibitemOpen
  \bibfield  {author} {\bibinfo {author} {\bibfnamefont {P.}~\bibnamefont
  {Johnson}}, \bibinfo {author} {\bibfnamefont {H.-B.}\ \bibnamefont {Yang}},
  \bibinfo {author} {\bibfnamefont {J.}~\bibnamefont {Rameau}}, \bibinfo
  {author} {\bibfnamefont {G.}~\bibnamefont {Gu}}, \bibinfo {author}
  {\bibfnamefont {Z.-H.}\ \bibnamefont {Pan}}, \bibinfo {author} {\bibfnamefont
  {T.}~\bibnamefont {Valla}}, \bibinfo {author} {\bibfnamefont
  {M.}~\bibnamefont {Weinert}},\ and\ \bibinfo {author} {\bibfnamefont
  {A.}~\bibnamefont {Fedorov}},\ }\href@noop {} {\bibfield  {journal} {\bibinfo
   {journal} {Physical review letters}\ }\textbf {\bibinfo {volume} {114}},\
  \bibinfo {pages} {167001} (\bibinfo {year} {2015})}\BibitemShut {NoStop}%
\bibitem [{\citenamefont {Lohani}\ \emph {et~al.}(2020)\citenamefont {Lohani},
  \citenamefont {Hazra}, \citenamefont {Ribak}, \citenamefont {Nitzav},
  \citenamefont {Fu}, \citenamefont {Yan}, \citenamefont {Randeria},\ and\
  \citenamefont {Kanigel}}]{lohani2020band}%
  \BibitemOpen
  \bibfield  {author} {\bibinfo {author} {\bibfnamefont {H.}~\bibnamefont
  {Lohani}}, \bibinfo {author} {\bibfnamefont {T.}~\bibnamefont {Hazra}},
  \bibinfo {author} {\bibfnamefont {A.}~\bibnamefont {Ribak}}, \bibinfo
  {author} {\bibfnamefont {Y.}~\bibnamefont {Nitzav}}, \bibinfo {author}
  {\bibfnamefont {H.}~\bibnamefont {Fu}}, \bibinfo {author} {\bibfnamefont
  {B.}~\bibnamefont {Yan}}, \bibinfo {author} {\bibfnamefont {M.}~\bibnamefont
  {Randeria}},\ and\ \bibinfo {author} {\bibfnamefont {A.}~\bibnamefont
  {Kanigel}},\ }\href@noop {} {\bibfield  {journal} {\bibinfo  {journal}
  {Physical Review B}\ }\textbf {\bibinfo {volume} {101}},\ \bibinfo {pages}
  {245146} (\bibinfo {year} {2020})}\BibitemShut {NoStop}%
\bibitem [{\citenamefont {Kim}\ \emph {et~al.}(2021)\citenamefont {Kim},
  \citenamefont {Miao}, \citenamefont {Choi}, \citenamefont {Zingl},
  \citenamefont {Georges},\ and\ \citenamefont {Kotliar}}]{kim2021spatial}%
  \BibitemOpen
  \bibfield  {author} {\bibinfo {author} {\bibfnamefont {M.}~\bibnamefont
  {Kim}}, \bibinfo {author} {\bibfnamefont {H.}~\bibnamefont {Miao}}, \bibinfo
  {author} {\bibfnamefont {S.}~\bibnamefont {Choi}}, \bibinfo {author}
  {\bibfnamefont {M.}~\bibnamefont {Zingl}}, \bibinfo {author} {\bibfnamefont
  {A.}~\bibnamefont {Georges}},\ and\ \bibinfo {author} {\bibfnamefont
  {G.}~\bibnamefont {Kotliar}},\ }\href@noop {} {\bibfield  {journal} {\bibinfo
   {journal} {Physical Review B}\ }\textbf {\bibinfo {volume} {103}},\ \bibinfo
  {pages} {155107} (\bibinfo {year} {2021})}\BibitemShut {NoStop}%
\bibitem [{\citenamefont {Kim}\ \emph {et~al.}(2018)\citenamefont {Kim},
  \citenamefont {Mravlje}, \citenamefont {Ferrero}, \citenamefont {Parcollet},\
  and\ \citenamefont {Georges}}]{kim2018spin}%
  \BibitemOpen
  \bibfield  {author} {\bibinfo {author} {\bibfnamefont {M.}~\bibnamefont
  {Kim}}, \bibinfo {author} {\bibfnamefont {J.}~\bibnamefont {Mravlje}},
  \bibinfo {author} {\bibfnamefont {M.}~\bibnamefont {Ferrero}}, \bibinfo
  {author} {\bibfnamefont {O.}~\bibnamefont {Parcollet}},\ and\ \bibinfo
  {author} {\bibfnamefont {A.}~\bibnamefont {Georges}},\ }\href@noop {}
  {\bibfield  {journal} {\bibinfo  {journal} {Physical Review Letters}\
  }\textbf {\bibinfo {volume} {120}},\ \bibinfo {pages} {126401} (\bibinfo
  {year} {2018})}\BibitemShut {NoStop}%
\bibitem [{\citenamefont {Tamai}\ \emph {et~al.}(2019)\citenamefont {Tamai},
  \citenamefont {Zingl}, \citenamefont {Rozbicki}, \citenamefont {Cappelli},
  \citenamefont {Ricco}, \citenamefont {de~la Torre}, \citenamefont {Walker},
  \citenamefont {Bruno}, \citenamefont {King}, \citenamefont {Meevasana} \emph
  {et~al.}}]{tamai2019high}%
  \BibitemOpen
  \bibfield  {author} {\bibinfo {author} {\bibfnamefont {A.}~\bibnamefont
  {Tamai}}, \bibinfo {author} {\bibfnamefont {M.}~\bibnamefont {Zingl}},
  \bibinfo {author} {\bibfnamefont {E.}~\bibnamefont {Rozbicki}}, \bibinfo
  {author} {\bibfnamefont {E.}~\bibnamefont {Cappelli}}, \bibinfo {author}
  {\bibfnamefont {S.}~\bibnamefont {Ricco}}, \bibinfo {author} {\bibfnamefont
  {A.}~\bibnamefont {de~la Torre}}, \bibinfo {author} {\bibfnamefont {S.~M.}\
  \bibnamefont {Walker}}, \bibinfo {author} {\bibfnamefont {F.}~\bibnamefont
  {Bruno}}, \bibinfo {author} {\bibfnamefont {P.}~\bibnamefont {King}},
  \bibinfo {author} {\bibfnamefont {W.}~\bibnamefont {Meevasana}}, \emph
  {et~al.},\ }\href@noop {} {\bibfield  {journal} {\bibinfo  {journal}
  {Physical Review X}\ }\textbf {\bibinfo {volume} {9}},\ \bibinfo {pages}
  {021048} (\bibinfo {year} {2019})}\BibitemShut {NoStop}%
\bibitem [{\citenamefont {Linden}\ \emph {et~al.}(2020)\citenamefont {Linden},
  \citenamefont {Zingl}, \citenamefont {Hubig}, \citenamefont {Parcollet},\
  and\ \citenamefont {Schollw{\"o}ck}}]{linden2020imaginary}%
  \BibitemOpen
  \bibfield  {author} {\bibinfo {author} {\bibfnamefont {N.-O.}\ \bibnamefont
  {Linden}}, \bibinfo {author} {\bibfnamefont {M.}~\bibnamefont {Zingl}},
  \bibinfo {author} {\bibfnamefont {C.}~\bibnamefont {Hubig}}, \bibinfo
  {author} {\bibfnamefont {O.}~\bibnamefont {Parcollet}},\ and\ \bibinfo
  {author} {\bibfnamefont {U.}~\bibnamefont {Schollw{\"o}ck}},\ }\href@noop {}
  {\bibfield  {journal} {\bibinfo  {journal} {Physical Review B}\ }\textbf
  {\bibinfo {volume} {101}},\ \bibinfo {pages} {041101} (\bibinfo {year}
  {2020})}\BibitemShut {NoStop}%
\bibitem [{Note2()}]{Note2}%
  \BibitemOpen
  \bibinfo {note} {ComDMFT is built on top of FlapwMBPT\cite
  {kutepov_linearized_2017} for the LQSGW part, and ComCTQMC for the quantum
  impurity problem solution \cite {melnickAcceleratedImpuritySolver2021}. We
  employed Wien2k~\cite {blaha2001wien2k} to calculate the DFT band
  structures.}\BibitemShut {Stop}%
\bibitem [{\citenamefont {Richler}\ \emph {et~al.}(2018)\citenamefont
  {Richler}, \citenamefont {Fratini}, \citenamefont {Ciuchi},\ and\
  \citenamefont {Mayou}}]{richler2018inhomogeneous}%
  \BibitemOpen
  \bibfield  {author} {\bibinfo {author} {\bibfnamefont {K.-D.}\ \bibnamefont
  {Richler}}, \bibinfo {author} {\bibfnamefont {S.}~\bibnamefont {Fratini}},
  \bibinfo {author} {\bibfnamefont {S.}~\bibnamefont {Ciuchi}},\ and\ \bibinfo
  {author} {\bibfnamefont {D.}~\bibnamefont {Mayou}},\ }\bibfield  {title}
  {\bibinfo {title} {Inhomogeneous dynamical mean-field theory of the small
  polaron problem},\ }\href@noop {} {\bibfield  {journal} {\bibinfo  {journal}
  {Journal of Physics: Condensed Matter}\ }\textbf {\bibinfo {volume} {30}},\
  \bibinfo {pages} {465902} (\bibinfo {year} {2018})}\BibitemShut {NoStop}%
\bibitem [{\citenamefont {Li}\ \emph {et~al.}(2022)\citenamefont {Li},
  \citenamefont {Li}, \citenamefont {Cao}, \citenamefont {Zhou}, \citenamefont
  {Wang}, \citenamefont {Jin}, \citenamefont {Chiu}, \citenamefont {Pennycook},
  \citenamefont {Wang},\ and\ \citenamefont {Gao}}]{li2022ordered}%
  \BibitemOpen
  \bibfield  {author} {\bibinfo {author} {\bibfnamefont {M.}~\bibnamefont
  {Li}}, \bibinfo {author} {\bibfnamefont {G.}~\bibnamefont {Li}}, \bibinfo
  {author} {\bibfnamefont {L.}~\bibnamefont {Cao}}, \bibinfo {author}
  {\bibfnamefont {X.}~\bibnamefont {Zhou}}, \bibinfo {author} {\bibfnamefont
  {X.}~\bibnamefont {Wang}}, \bibinfo {author} {\bibfnamefont {C.}~\bibnamefont
  {Jin}}, \bibinfo {author} {\bibfnamefont {C.-K.}\ \bibnamefont {Chiu}},
  \bibinfo {author} {\bibfnamefont {S.~J.}\ \bibnamefont {Pennycook}}, \bibinfo
  {author} {\bibfnamefont {Z.}~\bibnamefont {Wang}},\ and\ \bibinfo {author}
  {\bibfnamefont {H.-J.}\ \bibnamefont {Gao}},\ }\bibfield  {title} {\bibinfo
  {title} {Ordered and tunable majorana-zero-mode lattice in naturally strained
  lifeas},\ }\href@noop {} {\bibfield  {journal} {\bibinfo  {journal} {Nature}\
  }\textbf {\bibinfo {volume} {606}},\ \bibinfo {pages} {890} (\bibinfo {year}
  {2022})}\BibitemShut {NoStop}%
\bibitem [{\citenamefont {Melnick}\ \emph {et~al.}(2021)\citenamefont
  {Melnick}, \citenamefont {S{\'e}mon}, \citenamefont {Yu}, \citenamefont
  {D'Imperio}, \citenamefont {Tremblay},\ and\ \citenamefont
  {Kotliar}}]{melnickAcceleratedImpuritySolver2021}%
  \BibitemOpen
  \bibfield  {author} {\bibinfo {author} {\bibfnamefont {C.}~\bibnamefont
  {Melnick}}, \bibinfo {author} {\bibfnamefont {P.}~\bibnamefont {S{\'e}mon}},
  \bibinfo {author} {\bibfnamefont {K.}~\bibnamefont {Yu}}, \bibinfo {author}
  {\bibfnamefont {N.}~\bibnamefont {D'Imperio}}, \bibinfo {author}
  {\bibfnamefont {A.-M.}\ \bibnamefont {Tremblay}},\ and\ \bibinfo {author}
  {\bibfnamefont {G.}~\bibnamefont {Kotliar}},\ }\bibfield  {title} {\bibinfo
  {title} {Accelerated impurity solver for {{DMFT}} and its diagrammatic
  extensions},\ }\href {https://doi.org/10.1016/j.cpc.2021.108075} {\bibfield
  {journal} {\bibinfo  {journal} {Computer Physics Communications}\ }\textbf
  {\bibinfo {volume} {267}},\ \bibinfo {pages} {108075} (\bibinfo {year}
  {2021})}\BibitemShut {NoStop}%
\bibitem [{\citenamefont {Blaha}\ \emph {et~al.}(2001)\citenamefont {Blaha},
  \citenamefont {Schwarz}, \citenamefont {Madsen}, \citenamefont {Kvasnicka},\
  and\ \citenamefont {Luitz}}]{blaha2001wien2k}%
  \BibitemOpen
  \bibfield  {author} {\bibinfo {author} {\bibfnamefont {P.}~\bibnamefont
  {Blaha}}, \bibinfo {author} {\bibfnamefont {K.}~\bibnamefont {Schwarz}},
  \bibinfo {author} {\bibfnamefont {G.~K.}\ \bibnamefont {Madsen}}, \bibinfo
  {author} {\bibfnamefont {D.}~\bibnamefont {Kvasnicka}},\ and\ \bibinfo
  {author} {\bibfnamefont {J.}~\bibnamefont {Luitz}},\ }\bibfield  {title}
  {\bibinfo {title} {wien2k},\ }\href@noop {} {\bibfield  {journal} {\bibinfo
  {journal} {An augmented plane wave+ local orbitals program for calculating
  crystal properties}\ } (\bibinfo {year} {2001})}\BibitemShut {NoStop}%
\bibitem [{\citenamefont {Mostofi}\ \emph {et~al.}(2008)\citenamefont
  {Mostofi}, \citenamefont {Yates}, \citenamefont {Lee}, \citenamefont {Souza},
  \citenamefont {Vanderbilt},\ and\ \citenamefont
  {Marzari}}]{mostofi_Wannier90ToolObtaining_2008}%
  \BibitemOpen
  \bibfield  {author} {\bibinfo {author} {\bibfnamefont {A.~A.}\ \bibnamefont
  {Mostofi}}, \bibinfo {author} {\bibfnamefont {J.~R.}\ \bibnamefont {Yates}},
  \bibinfo {author} {\bibfnamefont {Y.-S.}\ \bibnamefont {Lee}}, \bibinfo
  {author} {\bibfnamefont {I.}~\bibnamefont {Souza}}, \bibinfo {author}
  {\bibfnamefont {D.}~\bibnamefont {Vanderbilt}},\ and\ \bibinfo {author}
  {\bibfnamefont {N.}~\bibnamefont {Marzari}},\ }\bibfield  {title} {\bibinfo
  {title} {Wannier90: {{A}} tool for obtaining maximally-localised {{Wannier}}
  functions},\ }\href {https://doi.org/10.1016/j.cpc.2007.11.016} {\bibfield
  {journal} {\bibinfo  {journal} {Computer Physics Communications}\ }\textbf
  {\bibinfo {volume} {178}},\ \bibinfo {pages} {685} (\bibinfo {year}
  {2008})}\BibitemShut {NoStop}%
\bibitem [{\citenamefont {Tomczak}\ \emph {et~al.}(2017)\citenamefont
  {Tomczak}, \citenamefont {Liu}, \citenamefont {Toschi}, \citenamefont
  {Kresse},\ and\ \citenamefont {Held}}]{tomczak_MergingGWDMFT_2017}%
  \BibitemOpen
  \bibfield  {author} {\bibinfo {author} {\bibfnamefont {J.~M.}\ \bibnamefont
  {Tomczak}}, \bibinfo {author} {\bibfnamefont {P.}~\bibnamefont {Liu}},
  \bibinfo {author} {\bibfnamefont {A.}~\bibnamefont {Toschi}}, \bibinfo
  {author} {\bibfnamefont {G.}~\bibnamefont {Kresse}},\ and\ \bibinfo {author}
  {\bibfnamefont {K.}~\bibnamefont {Held}},\ }\bibfield  {title} {\bibinfo
  {title} {Merging {{{\emph{GW}}}} with {{DMFT}} and non-local correlations
  beyond},\ }\href {https://doi.org/10.1140/epjst/e2017-70053-1} {\bibfield
  {journal} {\bibinfo  {journal} {Eur. Phys. J. Spec. Top.}\ }\textbf {\bibinfo
  {volume} {226}},\ \bibinfo {pages} {2565} (\bibinfo {year}
  {2017})}\BibitemShut {NoStop}%
\bibitem [{\citenamefont {Kim}\ \emph {et~al.}(2020)\citenamefont {Kim},
  \citenamefont {Werner},\ and\ \citenamefont
  {Valent{\'\i}}}]{kim2020alleviating}%
  \BibitemOpen
  \bibfield  {author} {\bibinfo {author} {\bibfnamefont {A.~J.}\ \bibnamefont
  {Kim}}, \bibinfo {author} {\bibfnamefont {P.}~\bibnamefont {Werner}},\ and\
  \bibinfo {author} {\bibfnamefont {R.}~\bibnamefont {Valent{\'\i}}},\
  }\href@noop {} {\bibfield  {journal} {\bibinfo  {journal} {Physical Review
  B}\ }\textbf {\bibinfo {volume} {101}},\ \bibinfo {pages} {045108} (\bibinfo
  {year} {2020})}\BibitemShut {NoStop}%
\bibitem [{\citenamefont {Phelan}\ \emph {et~al.}(2009)\citenamefont {Phelan},
  \citenamefont {Millican}, \citenamefont {Thomas}, \citenamefont {Leao},
  \citenamefont {Qiu},\ and\ \citenamefont {Paul}}]{phelan2009neutron}%
  \BibitemOpen
  \bibfield  {author} {\bibinfo {author} {\bibfnamefont {D.}~\bibnamefont
  {Phelan}}, \bibinfo {author} {\bibfnamefont {J.}~\bibnamefont {Millican}},
  \bibinfo {author} {\bibfnamefont {E.}~\bibnamefont {Thomas}}, \bibinfo
  {author} {\bibfnamefont {J.}~\bibnamefont {Leao}}, \bibinfo {author}
  {\bibfnamefont {Y.}~\bibnamefont {Qiu}},\ and\ \bibinfo {author}
  {\bibfnamefont {R.}~\bibnamefont {Paul}},\ }\bibfield  {title} {\bibinfo
  {title} {Neutron scattering measurements of the phonon density of states of
  fese 1- x superconductors},\ }\href@noop {} {\bibfield  {journal} {\bibinfo
  {journal} {Physical Review B}\ }\textbf {\bibinfo {volume} {79}},\ \bibinfo
  {pages} {014519} (\bibinfo {year} {2009})}\BibitemShut {NoStop}%
\bibitem [{\citenamefont {Mizuguchi}\ \emph {et~al.}(2009)\citenamefont
  {Mizuguchi}, \citenamefont {Tomioka}, \citenamefont {Tsuda}, \citenamefont
  {Yamaguchi},\ and\ \citenamefont {Takano}}]{mizuguchi2009fete}%
  \BibitemOpen
  \bibfield  {author} {\bibinfo {author} {\bibfnamefont {Y.}~\bibnamefont
  {Mizuguchi}}, \bibinfo {author} {\bibfnamefont {F.}~\bibnamefont {Tomioka}},
  \bibinfo {author} {\bibfnamefont {S.}~\bibnamefont {Tsuda}}, \bibinfo
  {author} {\bibfnamefont {T.}~\bibnamefont {Yamaguchi}},\ and\ \bibinfo
  {author} {\bibfnamefont {Y.}~\bibnamefont {Takano}},\ }\bibfield  {title}
  {\bibinfo {title} {Fete as a candidate material for new iron-based
  superconductor},\ }\href@noop {} {\bibfield  {journal} {\bibinfo  {journal}
  {Physica C: Superconductivity}\ }\textbf {\bibinfo {volume} {469}},\ \bibinfo
  {pages} {1027} (\bibinfo {year} {2009})}\BibitemShut {NoStop}%
\bibitem [{\citenamefont {Cvetkovic}\ and\ \citenamefont
  {Vafek}(2013)}]{cvetkovic2013space}%
  \BibitemOpen
  \bibfield  {author} {\bibinfo {author} {\bibfnamefont {V.}~\bibnamefont
  {Cvetkovic}}\ and\ \bibinfo {author} {\bibfnamefont {O.}~\bibnamefont
  {Vafek}},\ }\bibfield  {title} {\bibinfo {title} {Space group symmetry,
  spin-orbit coupling, and the low-energy effective hamiltonian for iron-based
  superconductors},\ }\href@noop {} {\bibfield  {journal} {\bibinfo  {journal}
  {Physical Review B}\ }\textbf {\bibinfo {volume} {88}},\ \bibinfo {pages}
  {134510} (\bibinfo {year} {2013})}\BibitemShut {NoStop}%
\end{thebibliography}%

\end{document}